\documentclass{article}

\usepackage[numbers,sort&compress]{natbib}
\usepackage[final]{neurips_2024} % final

\makeatletter\renewcommand{\@noticestring}{}\renewcommand{\@notice}{}\makeatother
\AtBeginDocument{%  runs after the sty's own \newgeometry 
  \newgeometry{textwidth=16.2cm, textheight=23.5cm, top=2.2cm,
               headheight=12pt, headsep=20pt, footskip=30pt}}
\usepackage{amsmath,amssymb}
\setcitestyle{numbers,square,comma}

\usepackage[utf8]{inputenc}
\usepackage[T1]{fontenc}

\usepackage{mathtools}

\usepackage{graphicx}
\usepackage{overpic}
\usepackage{wrapfig}
\graphicspath{{figures/}}
\usepackage{caption}
\usepackage{subcaption}
\usepackage{booktabs}
\usepackage{tabularx}
\usepackage{multirow}
\usepackage[dvipsnames]{xcolor}
\usepackage{colortbl}   % \cellcolor in the input-variable table
\usepackage{microtype}
\usepackage{xspace}

\usepackage{float}
\usepackage{listings}

\usepackage[section]{placeins}

\usepackage[colorlinks=true,linkcolor=blue,citecolor=blue,urlcolor=blue]{hyperref}
\usepackage{xurl}   % allow line breaks anywhere in long bibliography URLs
\usepackage[nameinlink,capitalise]{cleveref}

\newcommand{\pt}{\ensuremath{p_{\text{T}}}\xspace}
\newcommand{\mN}{m_N}
\newcommand{\mL}{m_l}
\newcommand{\etabase}{\eta_{\mathrm{base}}}
\newcommand{\etastar}{\eta^{\star}}
\newcommand{\Lstar}{L^{\star}}
\newcommand{\ARstar}{\mathrm{AR}^{\star}}
\newcommand{\bstar}{b^{\star}}
\newcommand{\bref}{b_{\mathrm{ref}}}
\newcommand{\Tref}{T_{\mathrm{ref}}}
\newcommand{\Nref}{N_{\mathrm{ref}}}
\newcommand{\Lref}{l_{\mathrm{ref}}}

\newcommand{\WL}{w{/}l}
\newcommand{\salt}{\mathtt{salt}}

\definecolor{colFull}{HTML}{BDBDBD}     % full set      (light grey backbone)
\definecolor{colGN2}{HTML}{9467BD}      % GN2 set       (violet)
\definecolor{colFourmom}{HTML}{2CA02C}  % 4-momenta only(green)
\definecolor{colNoflow}{HTML}{FFD700}   % no-flows      (gold)
\definecolor{colGN2e}{HTML}{D62728}     % GN2+electrons (red)

\providecommand{\yF}{\cellcolor{colFull}}
\providecommand{\yG}{\cellcolor{colGN2}}
\providecommand{\yM}{\cellcolor{colFourmom}}
\providecommand{\yN}{\cellcolor{colNoflow}}
\providecommand{\yE}{\cellcolor{colGN2e}}

\begin{document}

\title{How to scale your HEP ML models: \\ A recipe for robust architecture comparisons at scale}

 \author{%
   \textbf{Matthias Vigl}$^{\,1,2}$\thanks{\href{mailto:mvigl@slac.stanford.edu}{\texttt{mvigl@slac.stanford.edu}}} \quad
   \textbf{Nikita Pond}$^{\,3}$ \quad
   \textbf{Jackson Barr}$^{\,3}$ \quad
   \textbf{Alexander Froch}$^{\,4}$ \\[0.3em]
   \textbf{Dan Guest}$^{\,5}$ \quad
   \textbf{Nicole Hartman}$^{\,6}$ \quad
   \textbf{Michael Kagan}$^{\,2}$ \quad
   \textbf{Lukas Heinrich}$^{\,1,7}$ \\[0.7em]
   $^{1}$Technical University of Munich \quad
   $^{2}$SLAC National Accelerator Laboratory \\
   $^{3}$University College London \quad
   $^{4}$University of Geneva \quad
   $^{5}$Humboldt-Universit\"at zu Berlin \\
   $^{6}$NSF AI Institute for Artificial Intelligence and Fundamental Interactions\\ 
   $^{7}$Munich Center for Machine Learning \quad
 }

\maketitle

\begin{abstract}
Much of the recent progress in machine learning domains such as language models has come from \emph{scaling laws} that predict performance as a function of training effort. In high-energy physics (HEP) similar behavior has now been observed. To aid further study, we present a systematic procedure to derive robust scaling laws and compare design choices on the relevant budget axes for HEP tasks. We first validate the full scaling trajectory on toy problems and then apply the procedure to multi-task transformers on the $\sim$11 billion-jet ATLAS JetSet2 dataset, in both the compute- and data-constrained regimes. For the latter, we predict, to the best of our knowledge for the first time, the jointly optimal model size, training horizon, learning rate and batch size under early stopping. At compute-optimal scaling, we recover a near-equal $\sqrt{C}$ dependence of model and dataset size, and find that auxiliary objectives lower the primary jet-classification loss at equal compute budget. Expanding the inputs toward lower-level data systematically lowers the loss while leaving the scaling exponent nearly unchanged. The onset of the power-law regime is itself set by scale: below a threshold in dataset size the loss carries little information about high-compute scaling, underscoring the value of large, high-quality full-simulation datasets as a foundation for scaling studies and the development of foundation models in HEP.
\end{abstract}

% .............................................................................
\section{Introduction and Related work}
\label{sec:intro}
% .............................................................................

Jet flavor tagging, the identification of the flavor of jets, including those
originating from $b$- and $c$-quarks, $\tau$-leptons, and light quarks or gluons,
is a central component of the LHC~\cite{Lyndon_Evans_2008} physics program.
ATLAS~\cite{The_ATLAS_Collaboration_2008} has progressively moved toward increasingly powerful
neural-network-based taggers, from recurrent and deep-sets architectures over
tracks~\cite{ATL-PHYS-PUB-2020-014} to graph and transformer
models~\cite{ATL-PHYS-PUB-2022-027,Aad:2933089}, with the latest GN
family~\cite{ATL-PHYS-PUB-2026-001,ATL-PHYS-PUB-2026-005,ATL-PHYS-PUB-2026-013} using transformers trained directly on low-level tracks and
particle-flow constituents. A similar progression has taken place at
CMS~\cite{The_CMS_Collaboration_2008} with UParT and GloParT~\cite{CMS-DP-2022-050,CMS-DP-2024-066,CMS-DP-2025-081,CMS-DP-2026-104}, mirroring a broader move in jet physics toward constituent-level
architectures~\cite{Cogan:2014oua,deOliveira:2015xxd,Qu_2020,qu2024particletransformerjettagging,Gong:2022lye,Bogatskiy_2024,Brehmer:2024yqw,Favaro:2025pgz,Favaro:2026amj,Mondal:2024nsa}.

In machine learning, increasing model size, training data, and compute has driven
much of the recent progress in
language~\cite{kaplan2020scalinglawsneurallanguage,hoffmann2022trainingcomputeoptimallargelanguage,openai2024gpt4technicalreport}
and vision~\cite{zhai2022scalingvisiontransformers}. Neural scaling
laws~\cite{kaplan2020scalinglawsneurallanguage,hoffmann2022trainingcomputeoptimallargelanguage}
quantify this dependence through power-law relations between achievable loss and
model size $N$, dataset size $D$, and compute $C$. They can be used both to predict
performance at a larger training budget and to decide how a given budget should be
divided between $N$ and $D$. General-purpose foundation models in industry have
continued to grow in both model size and training data, reaching trillions of
parameters~\cite{kimiteam2026kimik3openfrontier,openai2026gpt6astra,anthropic_mythos5}. This paradigm is now entering HEP, where foundation models trained on increasingly
large samples are being explored with a variety of pretraining
strategies~\cite{Amram_2025,leigh2024tokenizationneededmaskedparticle,Golling_2024,Vigl_2024,Birk_2024,Birk:2025wai,Birk_2026,Mikuni_2025,Bhimji_2026,hsu2026evenetfoundationmodelparticle,kakati2026promptingparticlephysicstokenized}.
For heavy-flavor tagging, ATLAS recently demonstrated the first
Chinchilla-style~\cite{hoffmann2022trainingcomputeoptimallargelanguage} scaling
laws for transformer-based taggers trained on billions of simulated
jets~\cite{ATL-SOFT-PUB-2026-002}, showing that large gains are possible from
scaling compute alone, and CMS has studied transformer scaling for the same
task~\cite{CMS-DP-2026-133}. Scaling behavior has also been studied across a
growing range of other HEP problems, including data scaling of jet
classifiers~\cite{batson2023scalinglawsjetclassification}, compute-optimal scaling
of boosted-jet taggers~\cite{vigl2026neuralscalinglawsboosted}, jet~\cite{amram2026neuralscalinglawsjet} and full-event~\cite{godi2026scalingcollidereventgeneration}
generation, amplitude
surrogates~\cite{bahl2026scalinglawsamplitudesurrogates}, foundation-model
pretraining~\cite{park2025fm4nppscalingfoundationmodel,Elsharkawy:2026mev,uslu2026engineeringscalinglawspretraining},
and over-parameterization effects in HEP
data~\cite{vigl2025doubledescentoverparameterizationparticle}. Taken together,
these studies establish that a wide range of HEP tasks exhibit systematic scaling
behavior.

Scaling models, however, requires more than observing performance as a function
of model or dataset size, as the scaling behavior depends on whether each run is
trained under appropriately tuned
dynamics~\cite{porian2025resolvingdiscrepanciescomputeoptimalscaling}. As scaling
is explored more deeply in HEP, we offer this paper as a pedagogical guide to
deriving robust scaling laws. We also study how scaling depends on choices within
the machine-learning setup, such as auxiliary objectives and input features, and
how these design choices can be compared.

\begin{figure}[!htb]
\centering
\includegraphics[width=0.8\textwidth]{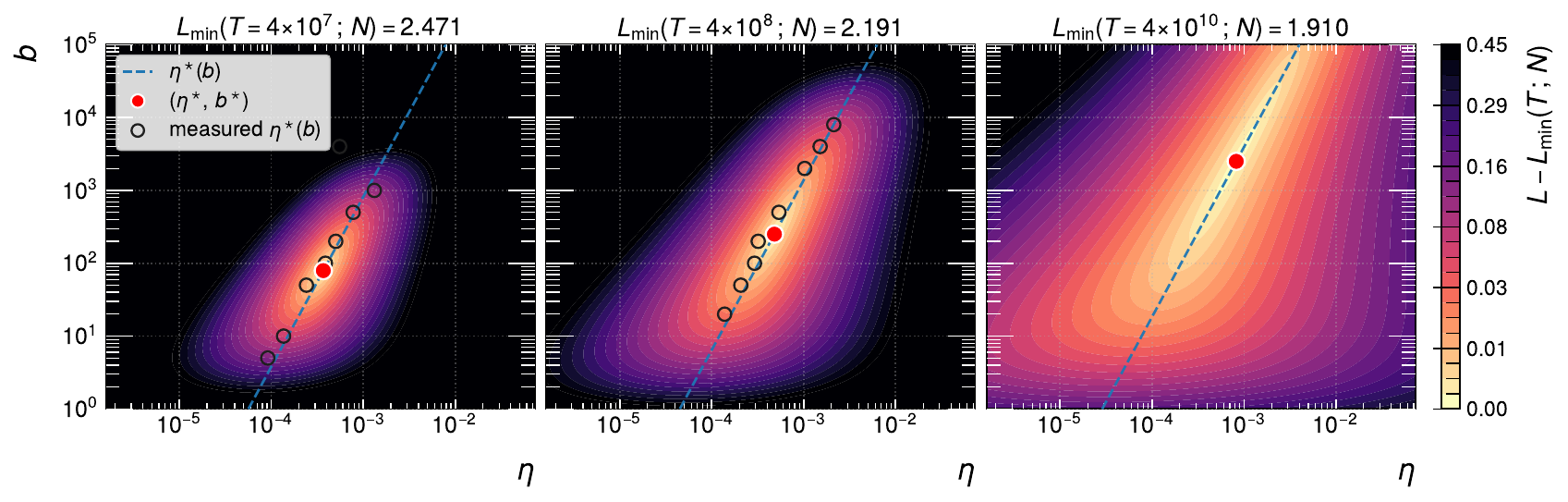}
\caption{Validation-loss landscape $L(\eta, b)$ of the
flavor tagger trained with all auxiliary tasks included, at fixed model size $N\sim6.7$M, for training-token
budgets $T = 4\times10^7$, $4\times10^8$, and $4\times10^{10}$. The dashed line traces the optimal
learning rate $\eta^\star(b;T)$ at each batch size for a given token budget, and the red
marker the joint optimum $(\eta^\star(T), b^\star(T))$. As the token budget
grows, the optimal learning rate decreases at fixed batch size, while at fixed learning rate the optimal batch size grows. The joint optimum migrates toward larger batch sizes and learning
rates. Scaling these quantities optimally and transferring them across model sizes and shapes
is a prerequisite for constructing robust scaling laws.}
\label{fig:intro:apex2x2}
\end{figure}

The central issue is illustrated in \cref{fig:intro:apex2x2}. At fixed
model size and shape, the optimal learning rate $\eta$ and batch size $b$ change systematically
with the training budget~\cite{mccandlish2018empiricalmodellargebatchtraining,filatov2025timetransferoptimallearning,li2025predictablescaleistep,zhang2025doescriticalbatchsize,deepseekai2024deepseekllmscalingopensource}. Fixing hyperparameters at a reference budget can therefore mix genuine
scaling behavior with the movement of the training optimum. A robust scaling analysis
must instead account for this movement explicitly across all scalable axes.

\begin{wrapfigure}{r}{0.5\textwidth}
\centering
%\vspace{-10pt}
\includegraphics[width=\linewidth]{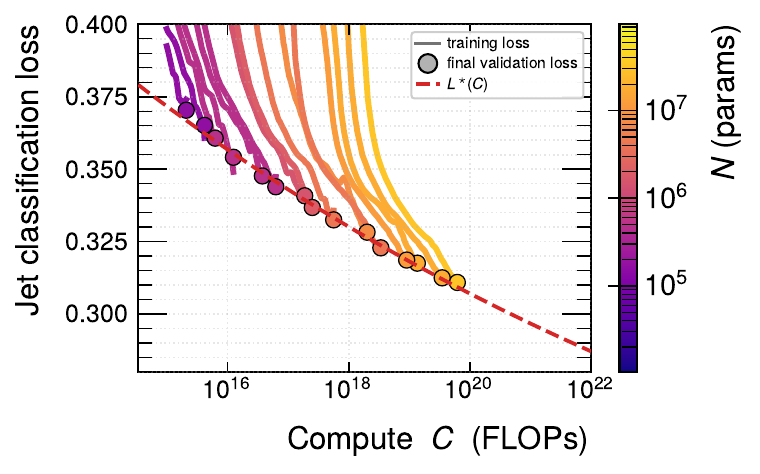}
\caption{Jet-classification loss against training compute, with every run
trained at its predicted optimal $(\eta^\star, b^\star)$ and at the
compute-optimal allocation $(N^\star, D^\star)$ at every budget $C$.}
\label{fig:intro:jclsfrontier}
%\vspace{-8pt}
\end{wrapfigure}

In this work, we show that in the practical setting of jet flavor tagging the
optimal training hyperparameters are themselves predictable through their own
scaling laws. Building on theoretical descriptions of the joint $(\eta, b, T)$
optimum~\cite{shulgin2026derivinghyperparameterscalinglaws} and on
parameterizations that make hyperparameters transferable across width and
depth~\cite{yang2022tensorprogramsvtuning,dey2026dontlazycompletepenables,everett2024scalingexponentsparameterizationsoptimizers,mlodozeniec2025completedhyperparametertransfermodules,dezoort2026hyperparametertransfergraphneural,qiu2026hyperparametertransferenablesconsistent}, we fit how the joint optimal learning rate, batch size and model shape move with the token budget. When
this recipe is followed, the loss as a function of compute becomes a well-behaved
scaling law (\cref{fig:intro:jclsfrontier}). The resulting compute-optimal
frontier is a property of the architecture rather than of hyperparameter
choices, and provides the basis for reliable comparisons of architectures,
objectives, and input representations at scale.

The paper is structured as follows. \cref{sec:tagging:gn3} introduces the
GN family tagger, the dataset, and the multi-task training setup.
\cref{sec:scaling-laws} reviews compute-optimal scaling laws, the logic of
hyperparameter tuning under a compute constraint, and how architectures
should be compared at scale.
\cref{sec:warmup} follows a scaling law from onset to the irreducible floor, on
a teacher--student problem with a known floor and on MLP probes over compressed
jets. \cref{sec:completep} derives the hyperparameter recipe: a width- and
depth-transferable parameterization of the joint optimum
$(\etastar(T),\bstar(T))$, and the optimal aspect ratio. \cref{sec:final}
applies the recipe to the full tagger and extracts the compute-optimal
allocation, and \cref{sec:repetition_main} extends it to repeated data under early stopping, predicting the jointly optimal model size, training horizon, learning rate and batch size. \cref{sec:ablations} compares design choices such as auxiliary losses and input parameters along their own
compute-optimal frontiers. \cref{sec:physics} translates the loss into
background rejection and projects the performance at larger budgets, followed by a discussion and
conclusions in \cref{sec:conclusion}.
% .............................................................................
\section{Architecture and Framework}
\label{sec:tagging:gn3}
% .............................................................................

Except for the warm-up studies of \cref{sec:warmup}, all results in this paper use the tagger,
dataset, and training setup described in this section. The architecture used throughout this study is a transformer
flavor tagger, based on the ATLAS flavor-tagging
GN- family~\cite{Aad:2933089,ATL-PHYS-PUB-2026-001}, trained within the
$\salt$~\cite{Barr:2025djz} framework\footnote{\url{https://gitlab.cern.ch/aft/algorithms/salt}}. Throughout this study we use the same structure as the ATLAS Carpe Datum scaling study~\cite{ATL-SOFT-PUB-2026-002} (the $\mathrm{GN3EPCLV01}$ variant~\cite{ATL-PHYS-PUB-2026-005} without the charge prediction task), and vary architecture size, shape, tasks, optimizer, hyperparameters, training dataset size and input features. The model
ingests three complementary input modalities at the per-jet level:
charged-particle tracks (with their full set of impact-parameter and
quality features), particle-flow objects~\cite{ATLAS:2017ghe}, and
soft electron~\cite{ATLAS:2019qmc} and muon information. Jet-level
features are broadcast to each per-object input, and the resulting
per-object vectors are passed through dedicated initialization
networks before being processed by a shared transformer encoder.
The encoder produces contextualized embeddings for each input
object, which are summarized into a global jet representation
through attention pooling. The pooled jet embedding, together with
the per-object embeddings, is then consumed by a set of
task-specific heads for the primary jet-flavor classification and
a number of auxiliary objectives. The full input feature set is listed in \cref{tab:TaggerInputs}.

\paragraph{Dataset} The JetSet2 dataset\footnote{\url{https://gitlab.cern.ch/atlas/open-data/jetset2-open-data}}~\cite{ATLAS:JetSet2} contains $\sim 11$B jets, constructed by combining
$t\bar{t}$ and $Z' \to q\bar{q}$ samples with $VH \to b\bar
b, c\bar c, \tau^+ \tau^-$, $Z' \to t\bar t$, multi-jet, and
$\gamma^* \to \tau^+ \tau^-$ samples, reweighted to the mean $\pt$ and $\eta$ spectrum across flavors. To allow comparisons to the TN25-86M model trained in~\cite{ATL-SOFT-PUB-2026-002}, we use the same mixture of jets that make up the $\sim7.7$B-jet dataset used for the training of that model. The mean particle multiplicity per jet is $\bar n_p = 42.6 \pm 12.7$, dominated by particle-flow objects and tracks. Throughout this paper we quote dataset size as $D$, the number of \emph{unweighted jets}, and every figure axis uses those units. Where relevant, we also refer to the same budget in \emph{tokens} (per-jet constituents passed through the transformer), $T \approx 42.6\,D$. The two differ only by that constant factor. %The flat-$\pt$ reweighting is applied throughout training (\cref{sec:metrics}), and the per-jet weights $w_i$ lower the effective sample size to $\widehat{\mathrm{ESS}} = \bigl(\sum_i w_i\bigr)^2 \big/ \sum_i w_i^2 \approx 3.8\times10^9$ jets, about half the nominal count. 
The FLOP estimate~\cite{kaplan2020scalinglawsneurallanguage} for a forward$+$backward pass at parameter count $N$ on a budget of $D$ jets for transformers is\footnote{For $N$ parameters, the cost of the forward pass of a transformer is $C_\mathrm{forward} \approx 2N$ due to the combined multiplication and addition operations in matrix multiplications. 
Computing the gradient via backpropagation requires propagating
gradients with respect to both the weights and the intermediate
activations of the model, so that $C_\mathrm{backward} \approx 2 C_\mathrm{forward}$.}
\begin{equation}
\label{eq:compute}
C \;=\; 6\,N\,T \;\approx\; 256\,N\,D.
\end{equation}
Batch sizes $b$ are reported in jets throughout. \cref{fig:flops_per_jet} in Appendix~\ref{app:flops} shows that the approximation of \cref{eq:compute} holds against measured FLOPs across the family of models spanned in this work.  

\paragraph{Training}
The model is trained to optimize five loss functions. The jet classification loss, $\mathcal{L}_{\text{jet-class}}$, uses cross-entropy, where the per-class weights inside $\mathcal{L}_{\text{jet-class}}$ compensate
for the unequal flavor abundances in the training sample. They are set to
$w_i = N_b / N_i$, with $N_i$ the number of jets of flavor $i$, giving
$w_b{=}1.00$, $w_c{=}3.64$, $w_s{=}5.00$, $w_{ud}{=}1.82$, $w_g{=}1.54$,
$w_\tau{=}9.09$. The vertexing loss,  $\mathcal{L}_{\text{vertex}}$, applies binary cross-entropy to each pair of tracks to identify which tracks share a vertex. The track-origin loss, $\mathcal{L}_{\text{track-origin}}$, and track-type classification, $\mathcal{L}_{\text{track-type}}$, each use categorical cross-entropy. The jet \pt regression loss,  $\mathcal{L}_{\pt}$, is an $L1$ loss (mean absolute error) on the ratio of true to reconstructed jet \pt. The model is trained with the AdamW optimizer~\cite{loshchilov2019decoupledweightdecayregularization} to minimize the direct
sum of the five task losses,
\begin{equation}
\mathcal{L}_{\text{total}} = \sum_{i=1}^{5} \mathcal{L}_i.
\end{equation}
Throughout this paper we train and analyze multiple variants of this architecture: a \emph{multi-task} variant that activates all five heads and minimizes $\mathcal{L}_{\text{total}}$ above, a \emph{single-task} variant that keeps the jet-classification head only and minimizes $\mathcal{L}_{\text{jet-class}}$, and any pair of $\mathcal{L}_{\text{jet-class}}$ and $\mathcal{L}_i$. The variants share every other architectural choice. 

Auxiliary objectives lower the jet classification loss at a fixed number of training samples, but also increase the computational cost of training. Their usefulness therefore depends on whether the gain in classification loss outweighs this additional cost. The hyperparameter optimization of \cref{sec:completep} is carried out \emph{independently} for each variant to study how the optimal hyperparameters $(b^\star, \eta^\star)$ themselves scale with $(N, T)$ when the target changes. Comparing the resulting scaling laws lets us isolate whether the auxiliary tasks help or hurt compute-efficient scaling of the primary jet-classification objective, and whether the auxiliary objectives have a different optimal scaling path from the primary one.

The learning rate follows a linear warmup from $\eta_{\mathrm{peak}}/100\to\eta_{\mathrm{peak}}$ over $8\%$ of the steps $K=\frac{D}{b}$,
followed by cosine annealing\footnote{DeepSeek
\citep{deepseekai2024deepseekllmscalingopensource} adopts a multi-step schedule that, after 2000 warmup steps, holds $\eta$
constant at $\eta_\text{peak}$ for $80\%$ of training and decays only in the
final steps, similar to Warmup-Stable-Decay (WSD) \citep{wen2024understandingwarmupstabledecaylearningrates}. This makes
checkpoints from the constant phase reusable across token budgets, since the
trajectory up to the decay is independent of the target $D$, without
sacrificing the performance of cosine annealing. It
reduces the cost of sweeping $D$, at the price of one or two decay phases per budget.} from $\eta_{\mathrm{peak}}\to\eta_{\mathrm{peak}}/10$. From here on the peak learning rate is the quantity being scaled, and we refer to it simply as $\eta$. 

% =============================================================================
\section{Scaling laws}
\label{sec:scaling-laws}
% =============================================================================

A scaling strategy starts with a model class: an architecture family with a small set of tunable hyperparameters. Some of those hyperparameters control \emph{model complexity} (e.g., the parameter count $N$, set by the width $w$ and depth $l$ of our transformer model), others control the \emph{training dynamics} (e.g., the batch size $b$ and the peak learning rate $\eta$). A handful of further choices, such as the learning-rate schedule shape, the optimizer, and a fixed attention-head dimension of $d_\mathrm{head}=32$ (one head per $32$ units of width, so the head count grows with $w$), are held constant across the family. We thus treat $(w, l)$ and $(b, \eta)$ as the four axes to tune at every scale and keep rest of the hyperparameters fixed as architecture-family conventions. 

Learning curves were empirically observed to follow predictable power laws across translation, vision, speech, and language well before the LLM era~\cite{hestness2017deeplearningscalingpredictable}, including joint functional fits for model and dataset size~\cite{rosenfeld2019constructivepredictiongeneralizationerror}. The famous Refs.~\cite{kaplan2020scalinglawsneurallanguage,hoffmann2022trainingcomputeoptimallargelanguage} used such laws to dictate, given a target compute budget, how large a model to build and how much data to train it on. Ref.~\cite{weng2026scaling} surveys this line of work, including the competing functional forms, and the practicalities of what might cause a fit to fail.

\paragraph{Why scaling?} For a fixed model size $N$ trained on infinite, \emph{non-repeated} data, the validation loss decreases as data $D$ accumulates, until some critical budget $D^{\mathrm{crit}}_N$ past which further data lowers the loss only marginally and eventually plateaus when the model saturates its capacity\footnote{Scaling laws under data repetition show that repeated tokens retain the value of fresh ones only for the first few epochs~\cite{muennighoff2025scalingdataconstrainedlanguagemodels}, but ultimately decay in value and bottleneck the achievable performance due to overfitting~\cite{lovelace2026prescriptivescalinglawsdata,vigl2026neuralscalinglawsboosted}. We assume non-repeated data throughout and refer to training and validation loss interchangeably as ``the loss'', and unless stated otherwise we report the final validation loss. We treat compute-optimal scaling under data repetition in \cref{sec:repetition_main}.}. Increasing $N$ pushes $D^{\mathrm{crit}}_N$ higher and lowers the loss saturation value, until $N$ is large enough that the saturation value reaches an irreducible floor $L_\infty$. This floor is set jointly by the architecture family, the input features, and the
optimizer.

\begin{figure}[!htbp]
    \centering
    \includegraphics[width=0.8\textwidth]{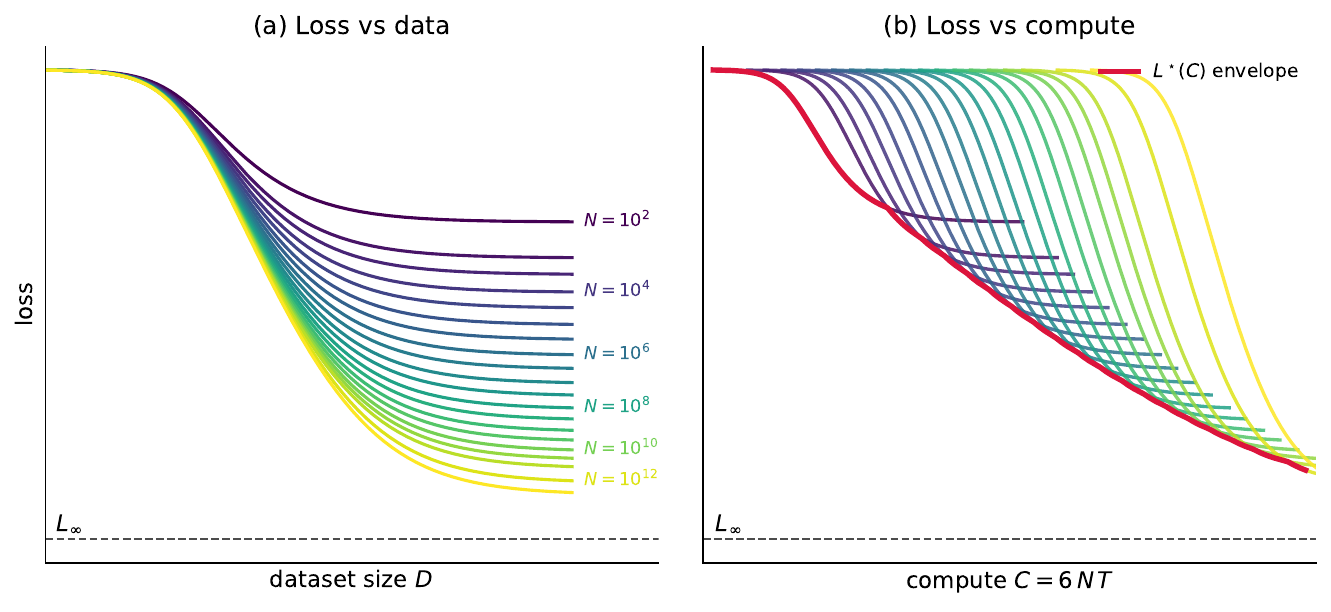}
    \caption{Sketch of how empirical scaling laws emerge. \textbf{(a)} loss against dataset size $D$ for a sequence of model sizes $N$: every curve follows a data-driven descent and saturates at a model-capacity plateau, lower for larger $N$, with all sufficiently large models approaching the irreducible floor $L_\infty$ (dashed). \textbf{(b)} the same curves against compute $C = 6NT \propto ND$: larger models shift rightward and the lower envelope (red) traces the compute-optimal frontier $L^\star(C)$.}
    \label{fig:sketch_scaling}
\end{figure}

A scaling law is a quantitative description of this behavior: the per-$N$ saturation generates the family of loss curves in \cref{fig:sketch_scaling}(a). Re-plotting the same curves against compute $C \approx 6\,N\,T$ shifts each one rightward by an amount proportional to $N$ (\cref{fig:sketch_scaling}(b)); the compute-optimal frontier $L^\star(C)$ is the lower envelope of the resulting family.

It is not surprising for a larger model to reach a lower loss than a smaller one at a fixed token budget; the question is which model wins at fixed compute. Since a larger model spends more compute per step, a smaller one can be trained on more data for the same compute, and the envelope records whichever choice wins. The picture also shows that scaling sets in only after an initial, mostly data-driven descent, before model size becomes a binding constraint. For the case studied here, that scale happens to be a few hundred million jets (\cref{fig:scaling_onset}).

The asymptotic floor $L_\infty$ characterizes the particular optimization problem in addition to the underlying physics. The information-theoretic limit is the Bayes risk $L_{\mathrm{Bayes}}$, the lowest loss any predictor could reach on the same inputs, which $L_\infty$ may approach but in general does not equal\footnote{Even with unlimited data, gradient descent on a structurally biased model family
is not guaranteed to recover the Bayes predictor.}. A tagger that ingests tracks and particle-flow objects operates on a hand-designed summary of the raw detector readout, so the Bayes risk on such summaries is itself bounded from below by the Bayes risk on the raw readout,
\begin{equation}
\label{eq:floor-hierarchy}
L_{\mathrm{Bayes}}^{\mathrm{raw}} \;\le\; L_{\mathrm{Bayes}} \;\le\; L_\infty.
\end{equation}
Lowering the floor therefore involves not only scaling but also moving toward less compressed inputs, which we study in \cref{sec:input_variants}. 

The achievable loss decreases as a power law in the model size $N$, the dataset size $D$ ($T$ in tokens), and the compute budget $C \approx 6\,N\,T$ (\cref{eq:compute}). The three axes are not independent: at fixed $C$, increasing $N$ leaves fewer tokens $T = C/(6\,N)$ in budget, so $N$ and $D$ trade off against each other. The substance of a scaling law is then twofold. First, at every $C$ there exists a unique allocation $(N^\star(C), D^\star(C))$ that minimizes the loss $L^\star(C)$; this allocation is the lower envelope just described. Second, both the location $(N^\star(C), D^\star(C))$ of that minimum and its value $L^\star(C)$ are themselves smooth, power-law functions of $C$, so that given a target compute budget $C$ too large for direct grid search\footnote{Since $L$ decreases with $C$, any fraction $f$ of the budget spent on tuning leaves
only $(1-f)\,C$ for the final model and costs performance accordingly. The aim is therefore to train the final model once, on nearly the entire budget.}, one can \emph{predict} both $(N^\star(C), D^\star(C))$ and $L^\star(C)$ from a much cheaper grid of small-scale runs.

\paragraph{Parametric forms.} Viewed as the envelope of the hand-off mechanism, the most direct object to fit is the frontier itself: the compute-optimal loss $L^\star(C)$ together with the allocation $(N^\star(C), D^\star(C))$ that achieves it. Each is parameterized as a power law in $C$~\cite{kaplan2020scalinglawsneurallanguage,hoffmann2022trainingcomputeoptimallargelanguage},
\begin{equation}
\label{eq:lstar-direct}
L^\star(C) \;=\; L_\infty \;+\; \Bigl(\tfrac{C_c}{C}\Bigr)^{\!\gamma},
\qquad
N^\star(C) \;\propto\; C^{\,a},
\qquad
D^\star(C) \;\propto\; C^{\,1-a},
\end{equation}
where the constraint $a + (1-a) = 1$ follows from $C = 6\,N\,T$ and $L_\infty$ is the irreducible-loss floor along the frontier. This form is the cheapest to fit and the most robust, but it makes a statement only about the optimal trajectory and not the suboptimal $(N, D)$ points.
% while saying nothing about the off-optimal $(N, D)$ points.

A more ambitious target is modeling the full loss landscape $L(N, D)$: with it in hand, both the compute-optimal frontier and the predicted loss at any $(N, D)$ come for free. The canonical parameterization is the additive ``Chinchilla'' form~\cite{hoffmann2022trainingcomputeoptimallargelanguage}, where the two power-law terms decouple cleanly:
\begin{equation}
\label{eq:chinchilla}
L(N, D) \;=\; L_\infty \;+\; \tfrac{A}{N^{\alpha}} \;+\; \tfrac{B}{D^{\beta}}.
\end{equation}
With $(\alpha, \beta, A, B, L_\infty)$ known, the compute-optimal frontier at fixed $C = 6\,N\,T$ is closed-form:
\begin{equation}
\label{eq:frontier}
N^{\star}(C) \;\propto\; C^{\,\beta/(\alpha+\beta)},
\qquad
D^{\star}(C) \;\propto\; C^{\,\alpha/(\alpha+\beta)},
\qquad
L^\star(C) - L_\infty \;\propto\; C^{-\alpha\beta/(\alpha+\beta)},
\end{equation}
which fixes the exponent $a = \beta/(\alpha+\beta)$ of \cref{eq:lstar-direct}.

The first LLM scaling study~\cite{kaplan2020scalinglawsneurallanguage} concluded $N^\star \propto C^{0.73}$, while the later Chinchilla analysis~\cite{hoffmann2022trainingcomputeoptimallargelanguage} found $N^\star \propto C^{0.5}$. The difference in the fitted coefficients has been traced to counting non-embedding rather than total parameters, mismatched warmup and schedule lengths at small scale, and uneven tuning across the grid~\cite{pearce2024reconcilingkaplanchinchilla,porian2025resolvingdiscrepanciescomputeoptimalscaling,weng2026scaling}. 

\paragraph{Fitting approaches.} The compute-optimal point can be extracted in three main ways originally proposed in~\cite{hoffmann2022trainingcomputeoptimallargelanguage}:
\begin{enumerate}
\item \emph{Fixed model sizes, vary $D$ (training-curve method).} Train a handful of $N$ values to different token budgets and read off $N^\star(C),D^\star(C)$ from the envelope of the per-$N,D$ loss curves.
\item \emph{IsoFLOP profiles.} For each compute budget $C_i$, sweep $N$ at $D = C_i/(6N)$ and locate the $N$ that minimizes the loss; connect the per-$C$ minima to recover $N^\star(C),D^\star(C)$ directly.
\item \emph{Parametric fit.} Fit \cref{eq:chinchilla} jointly to all $(N, D)$ training runs (e.g., with a Huber loss~\cite{10.1214/aoms/1177703732}). This recovers the full $L(N, D)$ surface rather than just the frontier, at the cost of being more sensitive to the assumed functional form.
\end{enumerate}
We apply all three in \cref{sec:warmup,sec:final} and compare the resulting allocations.

The fitted coefficients are not only a property of the model class, data, and task, but also depend on how the hyperparameters of each cell are chosen as the $(N,D)$ grid grows, and a mis-tuned schedule can introduce large biases. What we want from a scaling law prediction is then twofold:
\begin{enumerate}
\item \emph{Predict the loss.} Given a budget $C$, predict the best performance $L^\star(C)$ a model in this architecture family can achieve if trained optimally at $(N^{\star}(C),D^{\star}(C))$.
\item \emph{Predict the recipe.} Given the same budget, predict the architectural choices (size, shape) \emph{and} the training hyperparameters $(\eta, b, \dots)$ that realize $L^\star(C)$.
\end{enumerate}

\paragraph{The envelope as a two-level optimum.} When compute is the constraint\footnote{Compute is the binding constraint whenever a single pass over the dataset would exceed the training budget, as for internet-scale text in LLM training. In HEP, Monte Carlo simulation can in principle generate arbitrarily large training datasets, but in practice can be limited by CPU and storage resources. Most public datasets are modest in size and standard practice is to repeat passes over the dataset with early stopping. Overparameterized models can gain additional performance through \emph{double descent}~\cite{nakkiran2019deepdoubledescentbigger,vigl2025doubledescentoverparameterizationparticle}, but a fixed dataset ultimately limits how far the loss can be reduced (\cref{sec:repetition_main}).}, a model of size $N$ can be trained on at most $T = C/(6\,N)$ tokens, and minimizing the loss becomes a matter of convergence speed within a \emph{limited} number of steps through $T$. The levers are the architecture's compute efficiency together with $(b,\eta)$, which set the number of optimizer steps $T/b$ and the magnitude of each update.

Each point on the compute-optimal envelope $L^\star(C)$ of \cref{fig:sketch_scaling}(b) therefore corresponds to a different training configuration: a model of size $N^\star(C)$ trained for $T^\star(C) = C/(6\,N^\star(C))$ tokens, with $(b^\star, \eta^\star)$ tuned for that $(N^\star, T^\star)$ pair. Formally, the envelope is a two-level optimum,
\begin{equation}
\label{eq:l-star-envelope}
L^\star(C) \;=\; \min_{(N, T)\,:\,6\,N\,T \,=\, C}\, L^\star_{\mathrm{cell}}(N, T),
\qquad
L^\star_{\mathrm{cell}}(N, T) \;\equiv\; \min_{b,\,\eta}\, L(N, T, b, \eta),
\end{equation} where the inner minimization tunes $(b, \eta)$ in each $(N, T)$ cell and the outer minimization allocates compute between $N$ and $T$. 

\begin{figure}[!htbp]
%\begin{wrapfigure}{r}{0.5\textwidth}
    \centering
    \includegraphics[width=0.5\textwidth]{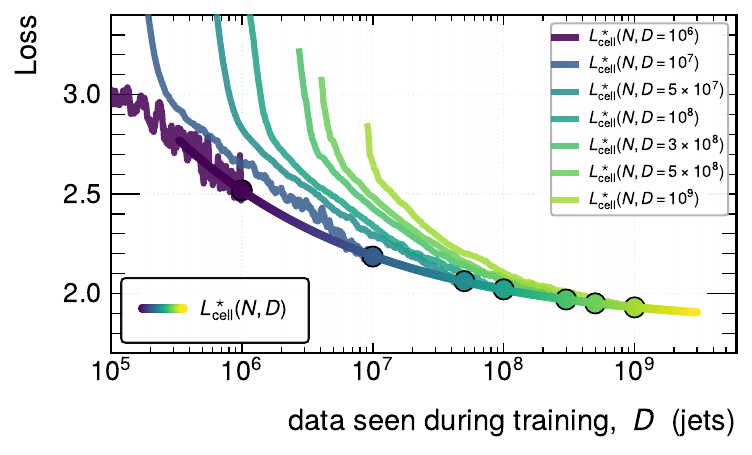}
    \caption{Optimal validation loss $L^\star_{\mathrm{cell}}(N,T)$ at fixed $N$ as a
  function of token budget $T = 42.6\,D$. Each loss curve is a separate run with
  $(b, \eta)$ tuned for the budget at which it ends, and the envelope passes
  through their endpoints. This is the inner minimization of
  \cref{eq:l-star-envelope}. The outer minimization over $N$ yields the envelope
  targeted by scaling laws.}
    \label{fig:cell_envelope}
%\end{wrapfigure}
\end{figure}

\Cref{fig:cell_envelope} shows the inner minimization at fixed $N$. For a sequence
of token budgets $T_1 < T_2 < \dots < T_n$, the $k$-th loss curve is a separate
run of $T_k$ tokens with $(b_k, \eta_k) = (b^\star(N, T_k), \eta^\star(N, T_k))$,
and its final loss is $L^\star_{\mathrm{cell}}(N, T_k)$. The envelope
$L^\star_{\mathrm{cell}}(N, T)$ is the curve through these $n$ endpoints. Run $k$
touches the envelope only at $T = T_k$ and lies above it for all $T < T_k$. Tracing $L^\star_{\mathrm{cell}}(N, T)$ over $T$ is therefore
\emph{not} the same as following one run to increasing $T$, but it requires a
separately tuned run for every $T_k$. Every curve in \cref{fig:sketch_scaling}
should be read as an envelope constructed this way.

\paragraph{Comparing architectures at scale} This is a nested constrained optimization over architecture families $\mathcal{A}$, i.e., which one reaches the lowest loss under a fixed budget of some resource $Q$,
\begin{equation}
\label{eq:aspect-opt}
\min_{\mathcal{A}}\;\;
\min_{\substack{N,\, D,\, w/l,\, b,\, \eta \,:\\ Q_{\mathcal{A}}(N, D, w/l, b)\, \le\, Q_{\max}}}\;
L_{\mathcal{A}}(N, D, w/l, b, \eta).
\end{equation}
Only $Q_{\max}$ is held fixed across families: the model size $N$, the dataset size $D$, the shape $\WL$, and the hyperparameters are all free in the inner minimization, so each family is compared at its own optimal allocation rather than at a common parameter count, training tokens or steps budget. $Q$ can be training FLOPs, GPU-hours, energy, or the number of unique training jets, and the ranking of two families might not be the same under each. 

Inductive biases that offer gains in data-limited regimes generally introduce additional computational cost. Their value at scale depends on whether the reduction in loss is sufficient to compensate for the extra compute that could have been spent on training more efficient architectures on more data. \citet{brehmer2025doesequivariancematterscale} showed that equivariant models can outperform non-equivariant ones at a fixed compute budget, and a recent systematic architecture comparison~\cite{Favaro:2026amj} on jet tagging found that equivariant transformers outperform standard transformers on fixed-size datasets when geometric information is relevant, on boosted jet tagging on JetClass~\cite{qu_2022_6619768} and ATLASTop~\cite{Aad_2024}. In contrast, little to no performance difference was observed for flavor tagging on the ATLAS JetSet~\cite{jetset} dataset.

We use $Q = 6NT = C$ in \cref{sec:ablations} to compare the multi-task tagger and the jet-classification-only variant, and the different input configurations, each trained at its own joint
$(N^\star, D^\star, \etastar, \bstar, \ARstar)$ at fixed compute.

% =============================================================================
\section{Warm-up: The Scaling Law Lifecycle}
\label{sec:warmup}
% =============================================================================

Before turning to the full tagger, we study systems where the full $L(N,D)$ surface, from the initial data-driven descent to the irreducible floor, is visible. We first consider a synthetic teacher--student problem with a known floor to assess whether the different fitting approaches can recover it. We then turn to a system that uses \emph{compressed} jets, where the onset and power-law descent can be observed at low compute budgets and dataset sizes.

\paragraph{Teacher--student toy.} 

\begin{wrapfigure}{r}{0.5\textwidth}
    \centering
    \includegraphics[width=0.5\textwidth]{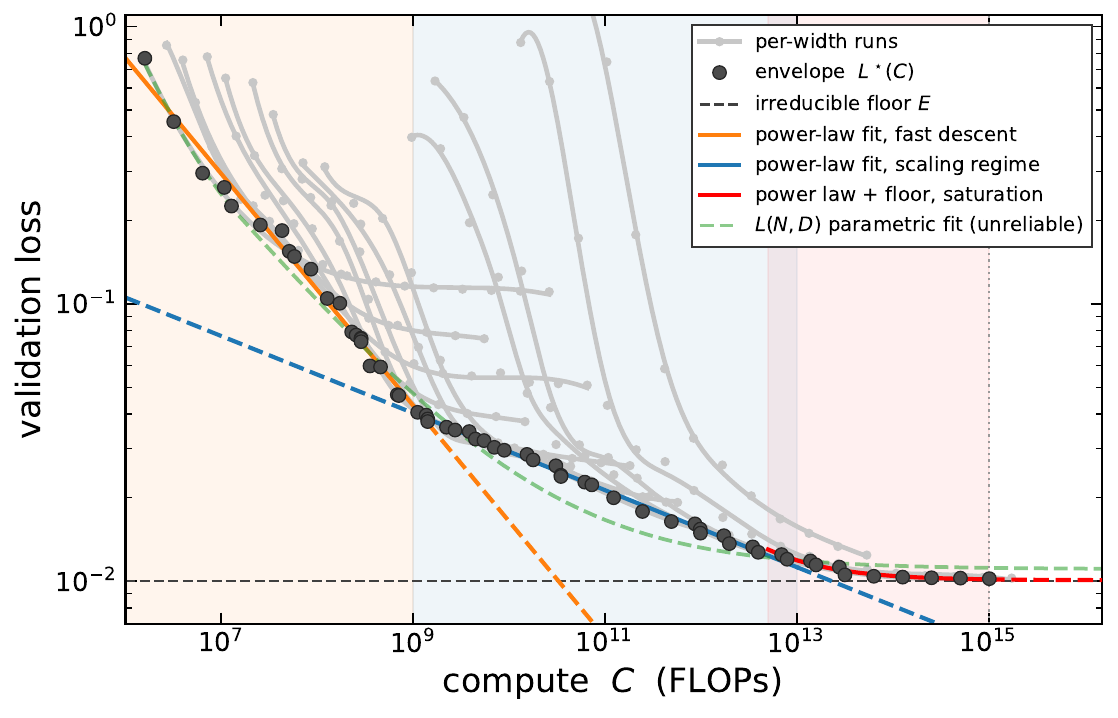}
    \caption{Compute-optimal envelope $L^\star(C)$ of the teacher--student problem (Appendix~\ref{app:tutorial}), with three power-law fits: an initial data-driven descent, a scaling regime with a pure power law, and a tail asymptoting to $L_\infty + A\,C^{\gamma}$ that recovers the known floor ($L_\infty = 0.0101$ against the true $\sigma^2 = 0.01$, dotted). The full-surface fit (Approach 3, green dashed) mis-fits the envelope across the regimes and flattens onto a biased floor. Inside the scaling regime nothing signals where the power law ends.}
    \label{fig:frontier_regimes}
\end{wrapfigure}

A scaling law is most useful when it predicts not only the descent but where the descent ends. We therefore consider a synthetic problem for which the true floor is known exactly. A fixed, randomly initialized MLP (the teacher) defines the target. MLP students of growing width are trained to imitate it on Gaussian inputs $x \sim \mathcal{N}(0, I)$ with label noise, $y = \mathrm{teacher}(x) + \sigma\varepsilon$. Data is drawn fresh at every step, so each run is single-pass and $D$ counts examples seen. The floor is known exactly, the label noise no model can predict $L_\infty = \sigma^2$, which lets any fitted floor be graded against the truth. A grid over $(N, D)$ is fitted with all three approaches of \cref{sec:scaling-laws}, and a step-by-step tutorial\footnote{Code and step-by-step instructions: \url{https://github.com/mvigl/scaling-laws-demo}} is also provided. All three approaches agree on how a compute budget should be split between model size and dataset size, with the optimal $N^\star$ and $D^\star$ each growing roughly as $\sqrt{C}$, so doubling the compute budget requires an increase by a factor of $\sqrt{2}$ in both. As can be seen in \cref{fig:frontier_regimes}, the compute-optimal envelope passes through three regimes: an initial, mostly data-driven descent with a steep effective slope; a scaling regime, where a pure power law $A\,C^{\gamma}$ fits the frontier with no visible curvature; and, only very close to the floor, a tail where the bend appears and a floored fit $L_\infty + A\,C^{\gamma}$ recovers the known $L_\infty$. From inside the scaling regime the envelope carries no information about where it ends and the floor stays invisible until the frontier is nearly on it. This is also the situation we will encounter in \cref{sec:final} where the measured jet-tagging frontier sits in the scaling regime, with no visible curvature and too far from the floor for $L_\infty$ to be estimated reliably. The full-surface form of \cref{eq:chinchilla}, fitted once to the whole grid with a single $(\alpha, \beta, L_\infty)$ cannot match every regime at once: it sits below the measured envelope through the scaling regime and then flattens onto $L_\infty = 0.011$, about $10\%$ above the truth. We therefore recommend the direct frontier fits of Approaches 1 and 2, which make no assumption about the shape of $L(N,D)$ away from the frontier.

\paragraph{Compressed jets.} A small set-transformer autoencoder is trained for one pass over $5$M JetClass jets~\cite{qu_2022_6619768} and then frozen. Each jet enters as a set of up to $40$ \pt-sorted constituents with $17$ features. The encoder compresses the set into a single latent vector of dimension $\ell$, and a transformer decoder expands it back into the constituents. The objective is a reconstruction loss on the constituent kinematics with auxiliary charge, particle-identification and multiplicity terms, so no flavor label enters the autoencoder. Its only role is to compress the variable-length constituent inputs into a fixed-size latent representation, on which the boosted-jet tagging problem is then solved by MLP probes. The information discarded in the latent sets the floor the probes can reach, so a poorer autoencoder makes the whole life of the scaling law cheaper to observe and reachable at small compute and dataset budgets.

\begin{figure}[!htbp]
    \centering
    \includegraphics[width=0.9\textwidth]{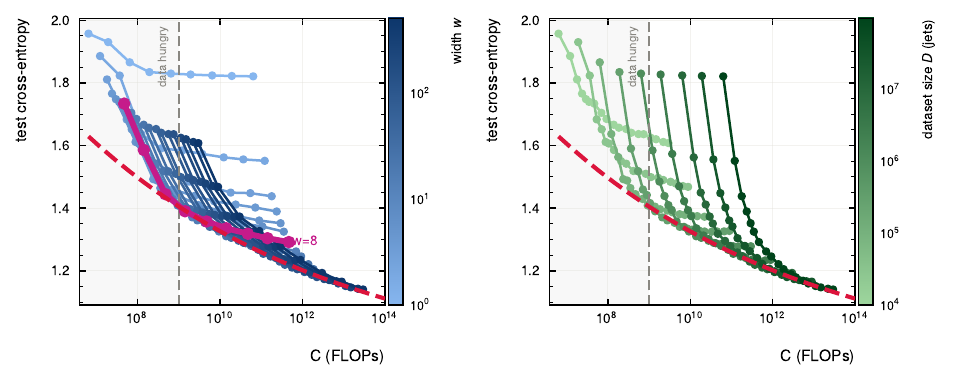}
    \caption{Test cross-entropy of MLP probes on a frozen jet latent space ($\ell = 128$) against training compute, colored by probe width $w$ (left) and by dataset size $D$ (right). The dashed red line is the compute-optimal envelope. In the shaded data-hungry region the envelope is traced by the training curve of a \emph{single} small model (the highlighted $w=8$ run): every larger model lies on or above it, and nothing yet distinguishes model sizes. Scaling in the compute-optimal sense only sets in past a task-dependent dataset budget ($\sim 10^5$ jets here), after which the allocation $(N^\star, D^\star)$ starts to matter.
    }
    \label{fig:simple_traj}
\end{figure}

The scaling study is run on one-hidden-layer MLP probes over the $128$-dimensional latents, for up to $10^8$ held-out jets, on a $10$-way boosted-jet classification target with a cross-entropy loss. \cref{fig:simple_traj} shows the onset condition of the scaling law on this system. Below a task- and dataset-dependent compute threshold, the envelope is dominated by a single small model rather than a sequence of optimal models. Only above this threshold does the characteristic scaling behavior emerge, which is not visible with datasets smaller than $\sim 10^5$ jets in this case.

\begin{figure}[!htbp]
    \centering
    \includegraphics[width=\textwidth]{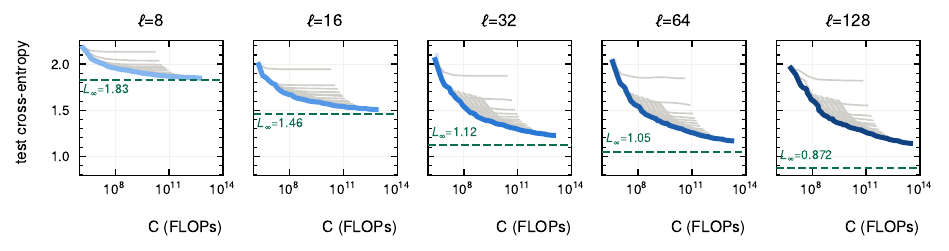}
    \caption{Compressed-representation scaling across latent dimensions $\ell = 8$--$128$: test cross-entropy on the $10$-way classification target against compute. Gray: test cross-entropy per probe width. Blue: latent's compute-optimal envelope. Larger latents lower the reachable loss while the envelope shape repeats.}
    \label{fig:simple_series_cloud}
\end{figure}

Where the onset and the irreducible floor $L_{\infty}$ sit depends on how much of the original information the latent keeps. \cref{fig:simple_series_cloud} repeats the study for $\ell = 8$ to $128$, on the same classification target. Richer latents lower the reachable loss and push the descent to higher compute, while the shape of the envelope, onset followed by a power-law descent, is the same at every $\ell$.

% =============================================================================
\section{A scaling recipe}
\label{sec:completep}
% =============================================================================

This section addresses \cref{eq:l-star-envelope} by predicting the joint optimum $(b^\star(N, T), \eta^\star(N, T))$ as a closed-form function of the grid axes, so that every run can be launched at its own optimum without a dedicated search. The strategy is to remove the dependencies one axis at a time. We first adopt a parameterization under which $\etastar$ becomes independent of the model-shape axes $(w, l)$, and thus size $N$, reducing the search to the $(b, T)$ plane at a fixed reference shape. We then sweep a small grid at that reference, fit the residual $(b, T)$ dependence to a multiplicative form, and inherit the result back across all $(w, l)$. We then characterize the full $\mathrm{loss}^\star(b, T;\etastar)$ surface around the joint locus to read off $b^\star(T;\etastar)$. Finally, we fix the remaining degree of freedom in $(w,l)_{|N}$ by imposing the optimal aspect ratio.

\subsection{Removing the model size dependence}
The first factor to remove is the model-size and shape dependence from $\etastar$. Naively, the optimal $\eta$ can shift by orders of magnitude as the model grows. The per-parameter update magnitude scales with width and depth, so the same $\eta$ in the optimizer produces a different effective step at different $(w, l)$. Sweeping $\eta$ at every $(w, l)$ for each $(b, T)$ is the cost we want to avoid. The $\mu$P framework~\cite{yang2022tensorprogramsvtuning} resolves the \emph{width} half of this problem. Per-parameter initialization, learning-rate, and Adam-$\epsilon$ multipliers keep activations $\Theta(1)$ as $w$ grows and make the optimal $\eta$ width-invariant. \emph{CompleteP}~\cite{dey2026dontlazycompletepenables} extends this with a depth-residual rescaling that absorbs the depth-axis growth into the network's forward dynamics. The \emph{Complete(d)P} refinement~\cite{mlodozeniec2025completedhyperparametertransfermodules} adds the treatment of the QK-norm gains under AdamW.

Under this parameterization the $\eta$ search shrinks from a four-dimensional problem in $(w, l, b, T)$ to a two-dimensional one in $(b, T)$, calibrated once at a reference $\Nref(w_\mathrm{ref}, \Lref)$. Every downstream cell at a different $(w, l)$ inherits $\etastar$ from the reference.
That is, $\arg\min_\eta L(w, l,b,T; \eta)$ coincides across all studied $(w, l)$ cells and the optimum is a function of $(b, T)$ alone\footnote{We will see that while $\etastar$ is not a function of model size $N$ with this prescription, there exists a specific aspect ratio $\ARstar(N)$ that yields an optimal loss at fixed $\etastar$.}:

\begin{equation}
\label{eq:completedp}
\arg\min_\eta L(w, l,b,T; \eta)=\etastar(b,T)
\end{equation}

The per-parameter-type multipliers used for every run in this work are those of \cref{tab:cmup-table} and the implementation is integrated into the $\salt$~\cite{Barr:2025djz} framework.

\begin{table}[!htb]
    \centering
    \caption{Per-parameter-type learning-rate (LR), AdamW-$\epsilon$ and weight-decay multipliers of the Complete(d)P recipe used for every run in this work, relative to the base values $\etabase$, $\epsilon_{\mathrm{base}}$, $\lambda_{\mathrm{base}}$; $\mN$ and $\mL$ are the width and depth ratios to the base model. The learning-rate and weight-decay columns follow Table~1 of \cite{dey2026dontlazycompletepenables} at $\alpha = 1$ and the QK-norm row in Table~1 of \cite{mlodozeniec2025completedhyperparametertransfermodules}; weight decay is the PyTorch AdamW convention, so $\lambda_{\mathrm{base}}\cdot\mN$ keeps $\eta\lambda$ width-independent.}
    \label{tab:cmup-table}
    \begin{tabular}{lccc}
        \toprule
        Parameter type & LR & AdamW $\epsilon$ & weight decay \\
        \midrule
        Embedding (input) & $\etabase$ & $\epsilon_{\mathrm{base}}$ & $\lambda_{\mathrm{base}}$ \\
        Hidden weights (attn, FFN) & $\etabase \cdot \mN^{-1}$ & $\epsilon_{\mathrm{base}} \cdot \mL^{-1}$ & $\lambda_{\mathrm{base}} \cdot \mN$ \\
        Unembedding / readout & $\etabase$ & $\epsilon_{\mathrm{base}}$ & $\lambda_{\mathrm{base}}$ \\
        Biases, LN/RMSNorm gains & $\etabase$ & $\epsilon_{\mathrm{base}}$ & $\lambda_{\mathrm{base}}$ \\
        QK-norm gains & $\etabase \cdot \mL^{-1}$ & $\epsilon_{\mathrm{base}} \cdot \mL^{-1}$ & $\lambda_{\mathrm{base}}$ \\
        \bottomrule
    \end{tabular}
\end{table}

\begin{figure}[!htb]
    \centering
\includegraphics[width=0.9\textwidth]{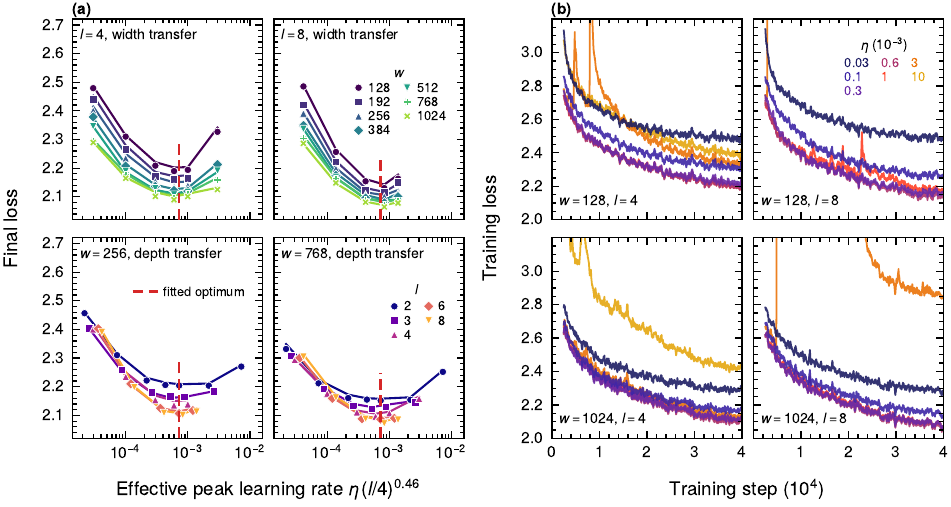}
    \caption{Empirical validation of the Complete(d)P recipe on the $(w, l, \eta)$ coordinate sweep at $(b, T) = (500, 2{\times}10^7)$. \textbf{(a)} Final validation loss against the effective peak learning rate $\eta\,(l/4)^{0.46}$ for seven widths at $l = 4$ and $l = 8$ (top) and five depths at $w = 256$ and $w = 768$ (bottom). The optimum lies at the same $\eta$ for all widths, with the residual drift in $l$ factored out. \textbf{(b)} Training loss against the training step for the $2{\times}2$ slice $w \in \{128, 1024\}$, $l \in \{4, 8\}$, one curve per peak learning rate $\eta$.}
    \label{fig:big_sweep_validation}
\end{figure}

\paragraph{Empirical validation}
We sweep a coordinate grid $(w, l, \eta)$ at fixed $(b, T)$ and compare $\arg\min_\eta L$ across cells. The value of $\arg\min_\eta = \etastar$ is extracted at every $(w, l)$ by fitting a parabola to the final validation loss as a function of $\log\eta$. \cref{fig:big_sweep_validation} shows how the $\etastar$ that minimizes the loss at fixed $(b,T)$ transfers perfectly across width $w$, while a mild residual dependence on depth $l$ remains. The $l$ dependence can be factored out by rescaling $\eta$ by a power of the depth ratio,
\begin{equation}
\label{eq:eta-L-rescale}
\etastar(b, T; l) \;=\; \left(\tfrac{\Lref}{l}\right)^{\gamma_l}\etastar(b, T; \Lref) ,
\end{equation}
so that the effective peak learning rate $\etastar\,(l/l_\mathrm{ref})^{\gamma_l}$ is constant across $(w, l)$. The fitted depth exponent is consistent with the value $\gamma_l=1/2$ used in the rest of this work. 

\begin{table*}[t]
  \centering
  \caption{Multiplicative calibration of the optimal learning rate
  $\etastar = \etabase\,(b/b_\mathrm{ref})^{\kappa}
  \,(T_\mathrm{ref}/T)^{\lambda}$
  (\cref{eq:eta-multiplicative}), with anchors $b_\mathrm{ref}=500$ and
  $T_\mathrm{ref}=2\times10^{7}$ jets.
  Shown for the total (multi-task) and jet-classification-only objectives. The
  per-head multi-task breakdown is given in \cref{tab:eta-fit-full}.}
  \label{tab:eta-fit}
  \begin{tabular}{l c c c}
    \toprule
    optimization target & $\etabase\,[10^{-4}]$ & $\kappa$ & $\lambda$ \\
    \midrule
    Total (multi-task)      & $6.05 \pm 0.18$ & $+0.429 \pm 0.014$ & $+0.098 \pm 0.009$ \\
    Jet-classification-only & $2.91 \pm 0.16$ & $+0.548 \pm 0.025$ & $+0.174 \pm 0.019$ \\
    \bottomrule
  \end{tabular}
\end{table*}

% =============================================================================
\subsection{Finding the joint $\etastar(\bstar, T)$}
\label{sec:multiplicative}
% =============================================================================

Having removed the $(w, l)$ dependence from $\etastar$, we now measure the residual $(b, T)$ dependence. In the large-step limit, $K = T/b \gg 1$, it is well described by a separable power law with $(\kappa, \lambda)$ treated as free parameters.
\begin{equation}
\label{eq:eta-multiplicative}
\etastar(b, T) \;=\; \etabase\,\Bigl(\tfrac{b}{\bref}\Bigr)^{\!\kappa}\Bigl(\tfrac{\Tref}{T}\Bigr)^{\!\lambda},
\end{equation}

\begin{wrapfigure}{r}{0.45\textwidth}
    \centering
    \includegraphics[width=0.95\linewidth]{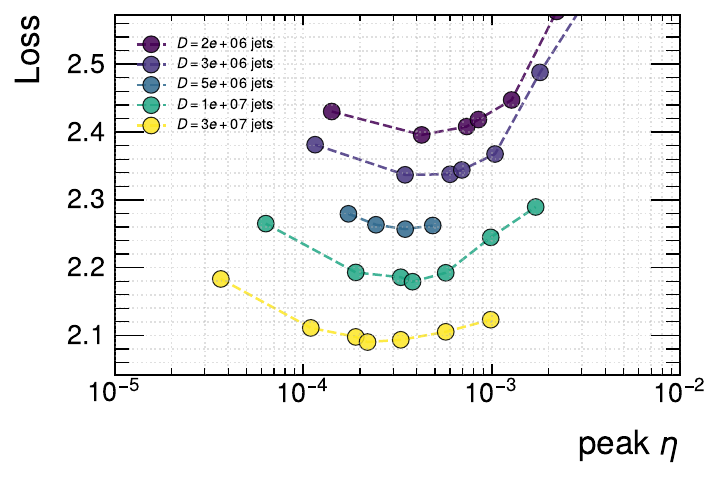}
    \caption{Final validation loss against peak learning rate at $N_{\mathrm{ref}}$ and $b = 100$, one
curve per dataset size $D$. The minimum of each curve is the extracted $\eta^\star(b, T)$.}
    \label{fig:eta_star_b}
\end{wrapfigure}

\begin{figure}[!htbp]
    \vspace{2mm}
    \centering
    \begin{subfigure}[b]{0.5\textwidth}
        \centering
        \includegraphics[width=\linewidth]{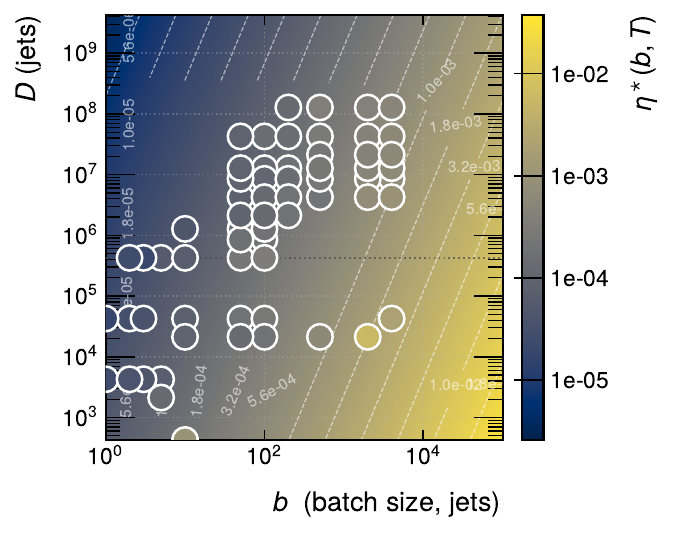}
        \caption{Optimal learning rate $\etastar(b, T)$.}
        \label{fig:eta_star_heatmap_Tcut:eta}
    \end{subfigure}\hfill
    \begin{subfigure}[b]{0.5\textwidth}
        \centering
        \includegraphics[width=\linewidth]{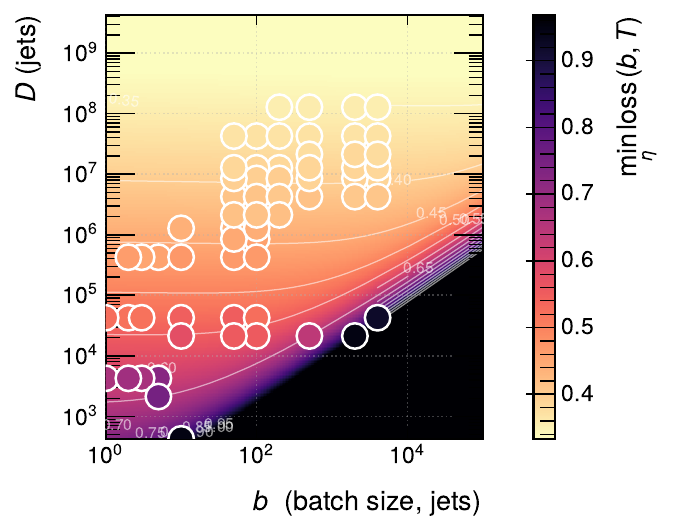}
        \caption{Loss at the optimal $\etastar$: $\mathrm{Loss}^\star(b, T) = \min_\eta L$.}
        \label{fig:eta_star_heatmap_Tcut:loss}
    \end{subfigure}
    \caption{The $(b, T)$ landscape at the reference architecture $\Nref$ for the jet-classification-only variant. \textbf{(a)} the measured optimal learning rate $\etastar(b, T)$, colored by $\log_{10}\etastar$, with iso-$\etastar$ contours from the multiplicative fit of \cref{eq:eta-multiplicative}. \textbf{(b)} the validation loss at that optimal learning rate, $\mathrm{loss}^\star(b, T) = \min_\eta L$, with iso-loss contours. The loss is essentially flat in $b$ and steepens only in the few-step corner (large $b$, small $T$), so $\bstar$ is left largely unconstrained. White circles mark the swept cells.}
    \label{fig:eta_star_heatmap_Tcut}
    \vspace{2mm}
\end{figure}

\begin{figure}[!htbp]
    \vspace{2mm}
    \centering
    \begin{subfigure}[b]{0.5\textwidth}
        \centering
        \includegraphics[width=\linewidth]{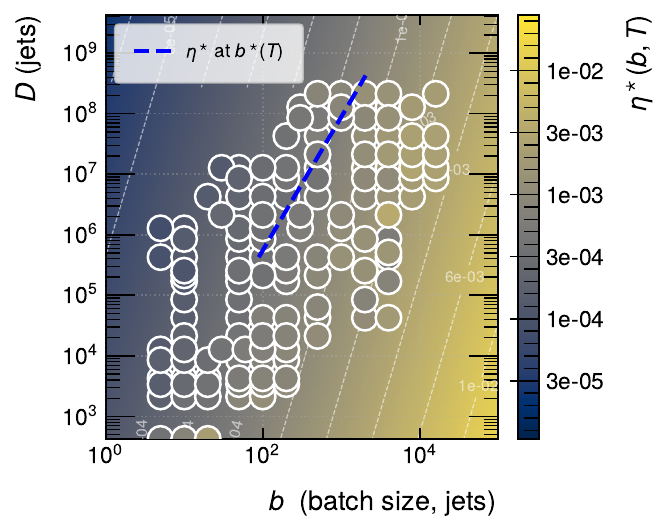}
        \caption{Optimal learning rate $\etastar(b, T)$.}
        \label{fig:eta_star_loss_star_heatmap:eta}
    \end{subfigure}\hfill
    \begin{subfigure}[b]{0.5\textwidth}
        \centering
        \includegraphics[width=\linewidth]{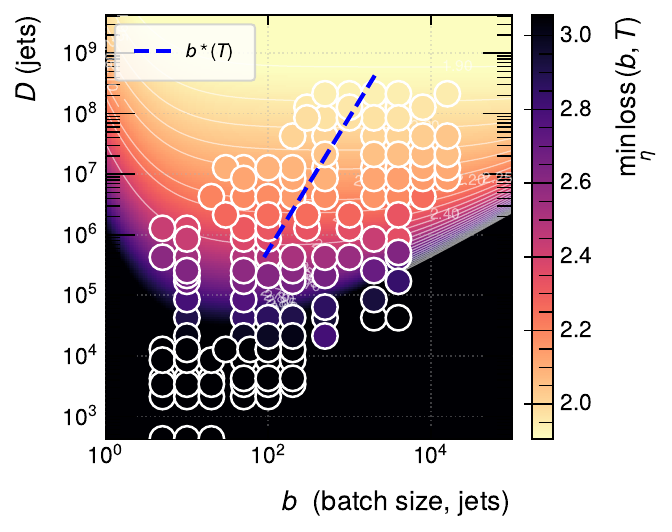}
        \caption{Loss at the optimal $\etastar$: $\mathrm{Loss}^\star(b, T) = \min_\eta L$.}
        \label{fig:eta_star_loss_star_heatmap:loss}
    \end{subfigure}
    \caption{The $(b, T)$ landscape at the reference architecture $\Nref$ for the multi-task variant. \textbf{(a)} the measured optimal learning rate $\etastar(b, T)$, colored by $\log_{10}\etastar$, with iso-$\etastar$ contours from the multiplicative fit of \cref{eq:eta-multiplicative}. \textbf{(b)} the validation loss at that optimal learning rate, $\mathrm{loss}^\star(b, T) = \min_\eta L$, with iso-loss contours. The loss now forms a shallow minimum in $b$ whose location grows as $\bstar \propto \sqrt{T}$. White circles mark the swept cells.}
    \label{fig:eta_star_loss_star_heatmap}
    \vspace{2mm}
\end{figure}

At a fixed reference size $\Nref$ we sweep $(b, T, \eta)$ and, for each $(b, T)$ cell, extract the empirical $\etastar$ as the per-cell argmin over $\eta$ (\cref{fig:eta_star_b}). The fitted exponents depend on the loss optimization target: the jet-classification-only and the summed multi-task objectives yield different $(\kappa,\lambda)$. \cref{fig:eta_star_heatmap_Tcut}(a) and \cref{fig:eta_star_loss_star_heatmap}(a) show the measured $\etastar(b,T)$ landscape for the jet-classification-only and multi-task variants respectively, each with the multiplicative fit of \cref{eq:eta-multiplicative} overlaid as iso-$\etastar$ contours. The total loss variant and jet-classification-only variant fits are compared in \cref{tab:eta-fit}, while the full per-head breakdown (\cref{tab:eta-fit-full} in Appendix~\ref{app:pertask_hp}) shows every individual multi-task head clustering at $\kappa_{\mathrm{eff}}\approx0.42$--$0.45$ and $\lambda_{\mathrm{eff}}\approx0.07$--$0.13$. The jet-classification-only objective is steeper on both axes, with $\kappa_{\mathrm{eff}}=0.548\pm0.025$ and $\lambda_{\mathrm{eff}}=0.174\pm0.019$, and a smaller absolute learning rate $\etabase=(2.91\pm0.16)\times10^{-4}$, indicating that single-task training requires a more aggressive batch-size scaling and a stronger learning-rate decay with the data budget than the multi-task objective.

With $\etastar(b, T)$ known, the remaining freedom is the batch size: we seek the $\bstar$ that, with $\eta$ already on its optimum, minimizes the loss,
\begin{equation}
\label{eq:bstar}
\bstar(T) \;=\; \arg\min_b\, \Lstar\bigl(b, \etastar(b, T), T\bigr).
\end{equation}
For jet classification alone, $\Lstar(b, \etastar, T)$ is essentially flat in $b$ across the whole sweep and steepens only in the few-step corner ($K = T/b$ small), so $\bstar$ is formally as small as one and any $b$ short of the few-step regime is equally good. For the summed multi-task loss the picture changes since at low $b$ the loss diverges and the optimum grows as $\bstar \propto \sqrt{T}$ (\cref{fig:eta_star_loss_star_heatmap}). One or more of the auxiliary tasks therefore drives a non-trivial optimal batch size. Optimizing for any single task inside the multi-task loss likewise produces its own optimal batch size, so the tasks reshape one another's hyperparameter optima (per-task plots in Appendix~\ref{app:pertask_hp}).

\begin{table}[t]
\centering
% \vspace{2mm}
\caption{Fitted power-law exponents for optimal batch size and learning rate
as a function of token budget $T$, as reported by studies with differing
optimizers and tuning protocols. Adapted from \cite{shulgin2026derivinghyperparameterscalinglaws}.}
\begin{tabular}{lccc}
\toprule
 & Batch Size & Learning Rate & Optimizer \\
\midrule
This work & $b^\star \propto T^{0.45\pm 0.05}$ & $\eta^\star \propto T^{+0.10\pm 0.02}$ & AdamW~\cite{loshchilov2019decoupledweightdecayregularization} \\
DeepSeek~\cite{deepseekai2024deepseekllmscalingopensource}        & $b^\star \propto T^{0.3271}$      & $\eta^\star \propto T^{-0.1250}$              & AdamW~\cite{loshchilov2019decoupledweightdecayregularization} \\
StepLaw~\cite{li2025predictablescaleistep}         & $b^\star \propto T^{0.571}$       & $\eta^\star \propto N^{-0.713}\, T^{0.307}$   & AdamW~\cite{loshchilov2019decoupledweightdecayregularization} \\
von R\"utte et al.~\cite{vonrütte2026scalingbehaviordiscretediffusion} & $b^\star \propto T^{0.8225}$ & $\eta^\star \propto T^{0.2806}$          & LaProp~\cite{ziyin2021lapropseparatingmomentumadaptivity} \\
Filatov et al.~\cite{filatov2026optimalscalingneedsoptimal} & $b^\star \propto T^{0.45 \pm 0.07}$ & $\eta^\star \propto T^{-0.28 \pm 0.07}$  & Scion~\cite{pethick2025trainingdeeplearningmodels} \\
\bottomrule
\end{tabular}
% \vspace{2mm}
\label{tab:empirical-scaling}
\end{table}
\newpage
% =============================================================================
\subsection{Optimal aspect ratio and considerations on GPU-hours}
\label{sec:aspect}
% =============================================================================

\begin{wrapfigure}{r}{0.35\textwidth}
    \centering
% \begin{figure}[!htbp]
    \centering
    \includegraphics[width=\linewidth]{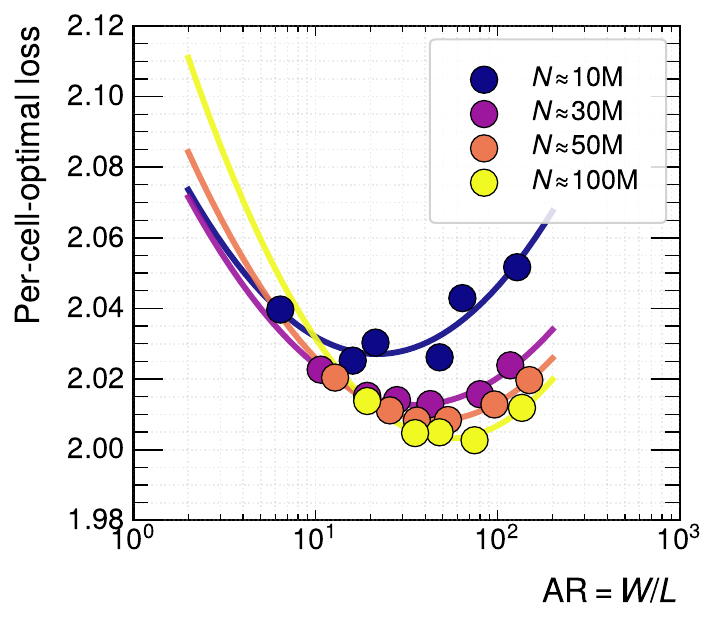}
    \caption{Final validation loss against aspect ratio $\mathrm{AR} = w/l$ for four
    model sizes, $N \approx 10$M, 30M, 50M and 100M, with each point trained at its
own $(\eta^\star, b^\star)$. Large AR is a wide, shallow network and small AR a narrow, deep one. The
minimum $\mathrm{AR}^\star(N)$ drifts to larger AR as $N$ grows, so width should grow faster than depth.}
    \label{fig:ar_validation}
% \end{figure}
\end{wrapfigure}

At any
shape $(w, l)$, setting $\bstar(T)$ and $\etastar(\bstar, T)$ minimizes the loss for that shape; it
does not follow, however, that all $(w, l)$ pairs with the same parameter count $N$ reach the same
loss. Instead, the loss may depend on the aspect ratio $\mathrm{AR} = w/l$. We define the
optimal aspect ratio as the one minimizing the per-cell-optimal loss at fixed $N$,
\begin{equation}
\label{eq:ar-opt}
\ARstar(N) \;=\; \arg\min_{(w, l)\,:\,N(w, l) = N}\, \Lstar(N, w, l),
\end{equation}
where each candidate $(w, l)$ is already trained at its own $(\etastar, \bstar)$. We find that
$\ARstar$ grows roughly as $\ARstar(N) \propto \sqrt{N}$: as the model grows, width should
increase faster than depth. This is consistent with earlier observations that model shape can affect scaling
behavior~\cite{alabdulmohsin2024gettingvitshapescaling}. \cref{fig:ar_validation} shows how the
per-cell-optimal loss traces a shallow parabola in $\log\mathrm{AR}$ at each model size, and its
minimum drifts to larger $\mathrm{AR}$ as $N$ grows\footnote{It is interesting to note that the optimal
growth $\ARstar(N) \propto \sqrt{N}$ is realized to good approximation by scaling width alone at
fixed depth: since $N \propto l\,w^2$, a fixed-$l$ scan gives $w \propto \sqrt{N}$ and
hence $\mathrm{AR} \propto \sqrt{N}$. This may explain why previous work observed smooth scaling
laws on jet tagging tasks when varying only the model
width~\cite{ATL-SOFT-PUB-2026-002,vigl2026neuralscalinglawsboosted}. Although such models sit away
from the optimal aspect ratio, they scale in the correct direction, so that the entire family
remains uniformly detuned.}. 

\begin{figure}[!htbp]
    \centering
    \includegraphics[width=0.85\textwidth]{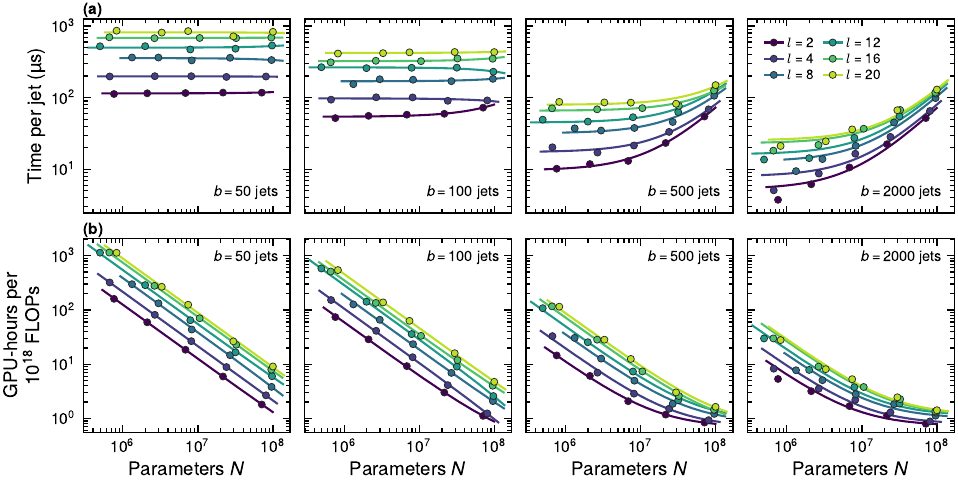}
    \caption{Measured throughput on an H200 against the parameter count $N$, one color per depth $l$ and one column per batch size $b$. Points are measurements of a full training step, and lines are linear fits in $N$ at fixed $b$ and $l$. \textbf{(a)} Time per jet. Throughput improves as $b$ grows and amortizes the per-step overhead, leveling off once the device is compute-bound. \textbf{(b)} The same runs as GPU-hours per $10^{18}$ FLOPs of theoretical compute. Equal theoretical compute costs more wall-clock time for deeper networks, and small models at small batch size reach only a fraction of the device peak.}
    \label{fig:throughput}
\end{figure}

While a shallower model than
$\ARstar$ provides a higher loss, it can require fewer GPU-hours at fixed theoretical compute
since deeper networks can be slower per FLOP. Likewise, when $\bstar$ is small, hardware considerations may favor a larger batch to increase
throughput. \citet{hoffmann2022trainingcomputeoptimallargelanguage} (page 4) adopt models shallower than the prescription of \citet{levine2021depthtowidthinterplayselfattention} because the
shallower configurations achieved better wall-clock performance on their hardware. \cref{fig:throughput} shows the fitted time per
jet as a function of model size at four batch sizes, and converts it into
GPU-hours per unit of theoretical compute.  

\begin{figure}[!htbp]
    \centering
    \includegraphics[width=0.6\textwidth]{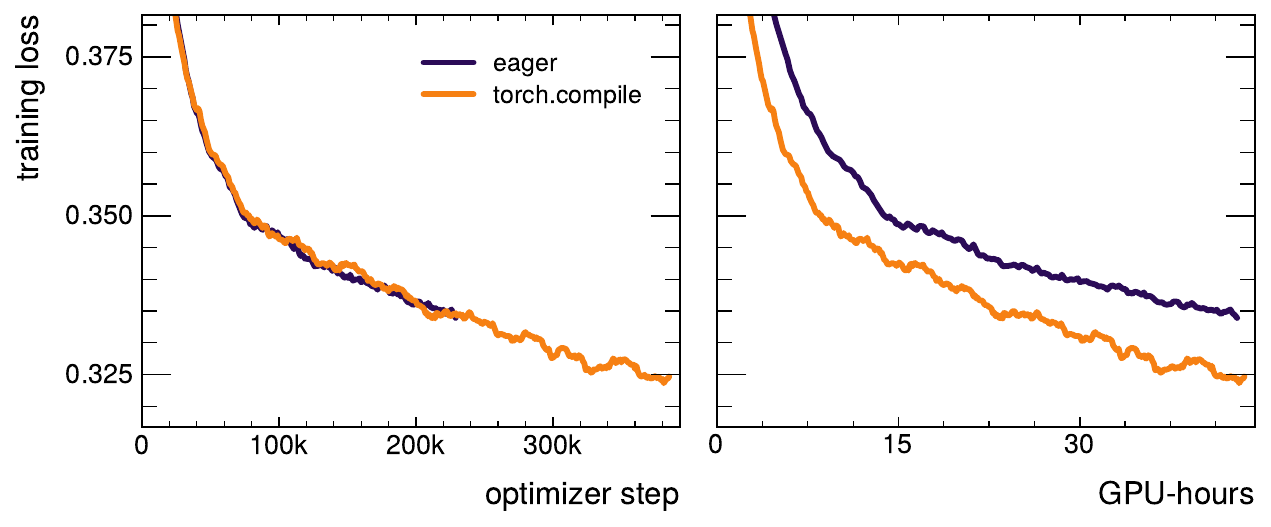}
    \caption{Training loss against optimizer steps (left) and A100 GPU-hours (right), for a model trained eagerly and with the training graph compiled by \texttt{torch.compile}. Both runs spent the same amount of compute. Model and hyperparameters are identical, so the loss per step coincides while
    the compiled run performs $1.7\times$ more optimizer steps per GPU-hour.}
    \label{fig:compile}
\end{figure}

Scaling laws are stated in FLOPs, but training time and energy consumption depend on GPU-hours. Lowering the GPU-hours per FLOP is therefore as much a gain as an architecture that lowers the loss at fixed FLOPs, and a large part of the work in improving scaling efficiency is on the hardware and software side, for example, fused attention kernels~\cite{dao2022flashattentionfastmemoryefficientexact}, low-precision arithmetic and compiler-generated kernels (\cref{fig:compile}).  

% =============================================================================

\section{Compute-optimal scaling laws}
\label{sec:final}
% =============================================================================

With the hyperparameter recipe in hand, we can solve the outer minimization of \cref{eq:l-star-envelope} that produces the scaling law itself. We train a handful of models on a grid of model sizes and dataset budgets $(N, D)$, following the prescription:
\begin{enumerate}
\item Pick a model size\footnote{Ideally one sweeps several aspect ratios per $N$ to locate $\ARstar(N)$. We found that scaling $N$ through width alone at fixed depth, so that $\mathrm{AR} \propto \sqrt{N}$, would give clean scaling laws in this study, since the offset from the optimum stays smooth along the frontier, but at the cost of suboptimality in loss at every $C$. For the results we present here we use the optimal aspect ratio $\ARstar$ from scans at fixed $N$.} $N$ and a dataset budget $D$ (equivalently a token budget $T \approx 42.6\,D$).
\item Read off the optimal batch size $\bstar(T;\etastar)$.
\item Insert $\bstar$ into the multiplicative law \cref{eq:eta-multiplicative} to obtain the matching $\etastar(\bstar, T)$.
\item Repeat over a grid in $(N, D)$. To also enable the IsoFLOP method (approach~2), include $(N, D)$ cells that share a common compute value: span a handful of compute budgets and, within each, sweep $(N, D)$ allocations with $6NT \approx \text{const}$.
\end{enumerate}

\subsection{Onset of the scaling regime in JetSet2}

\begin{figure}[!htbp]
    \begin{subfigure}[t]{\textwidth}
        \centering
        \includegraphics[width=0.9\linewidth]{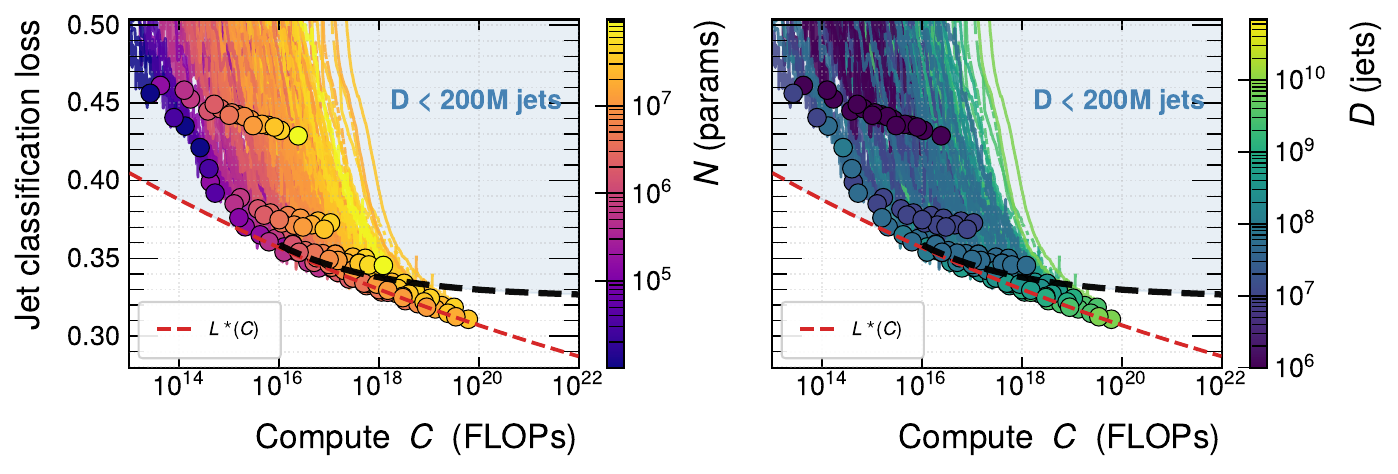}
    \end{subfigure}
    \caption{Onset of scaling on the multi-task loss variant. Jet classification loss against compute $C = 6NT$. The portion of the compute-optimal frontier a few-hundred-million-jet dataset can probe (up to $C \sim 10^{17}$ FLOP) is still the data-driven transient, and the asymptotic scaling slope sets in only beyond it. The same curves are colored by model size $N$ \textbf{(left)} and by dataset size $D$ \textbf{(right)}. Circles mark the final per-cell validation loss.}
    \label{fig:scaling_onset}
\end{figure}

\cref{fig:scaling_onset} shows the full set of runs that feeds the three approaches: every trained model's jet classification loss against cumulative compute, colored by model size in the left panel and by dataset size in the right. The common lower envelope is the compute-optimal frontier. Scaling only sets in once the data is large enough that model complexity is a binding constraint. Even the smallest architecture we can build inside our set of conventions ($\sim 10$k parameters) keeps improving steeply with data up to $D \sim$ a few hundred million jets, and only beyond that does model size start to bottleneck the loss. The portion of the $L^\star(C)$ envelope that a dataset of a few hundred million jets can probe, up to $C \sim 10^{17}$ FLOP, is barely inside the scaling regime.

\subsection{Fitting the scaling law}

We extract $L^\star(C)$ and the allocation $(N^\star(C), D^\star(C))$ with each of the three fitting approaches of \cref{sec:scaling-laws}, tracking the scaling of the total loss. The four auxiliary heads follow the same scaling form, each descending along its own envelope toward its own irreducible floor, and are collected in Appendix~\ref{app:aux_scaling}.

\begin{figure}[!htbp]
    \centering
    \includegraphics[width=\textwidth]{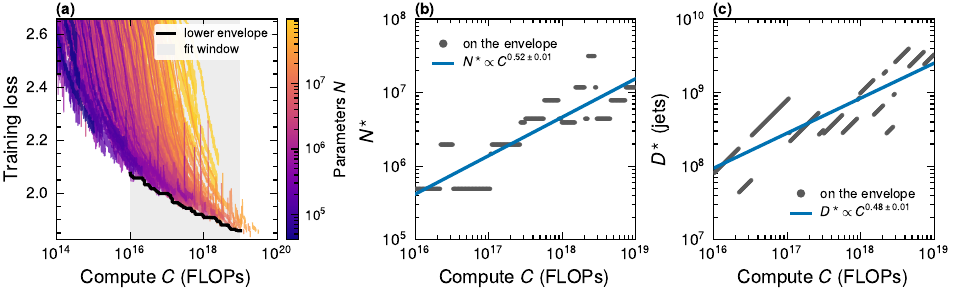}
    \caption{Approach 1 for the multi-task variant. \textbf{(a)} Total training loss against cumulative compute $C = 6N\,T$, one lightly smoothed curve per run colored by $N$. The black line is the lower envelope, the empirical compute-optimal frontier $L^\star(C)$. \textbf{(b,c)} The model size $N^\star$ and dataset size $D^\star$ on the envelope at each compute budget, with the fitted power laws of \cref{eq:approach1-exponents}.}
    \label{fig:approach1}
\end{figure}

\paragraph{Approach 1: training-curve envelope.} The first approach reads the compute-optimal frontier directly off the lower envelope of the per-run loss-versus-compute curves, with no assumed surface form. \cref{fig:approach1} shows the multi-task runs against cumulative compute $C = 6N\,T$, colored by model size. The envelope is the lower boundary of the architecture family and reading the model size and dataset size that lie on the envelope at each compute level gives the compute-optimal allocation directly. Both follow clean power laws,
\begin{equation}
\label{eq:approach1-exponents}
N^\star(C) \propto C^{\,0.52}, \qquad D^\star(C) \propto C^{\,0.48},
\end{equation}
matching the Chinchilla finding that model size and dataset size should scale together~\cite{hoffmann2022trainingcomputeoptimallargelanguage}.

\begin{figure}[!htbp]
    \centering
    \includegraphics[width=0.9\textwidth]{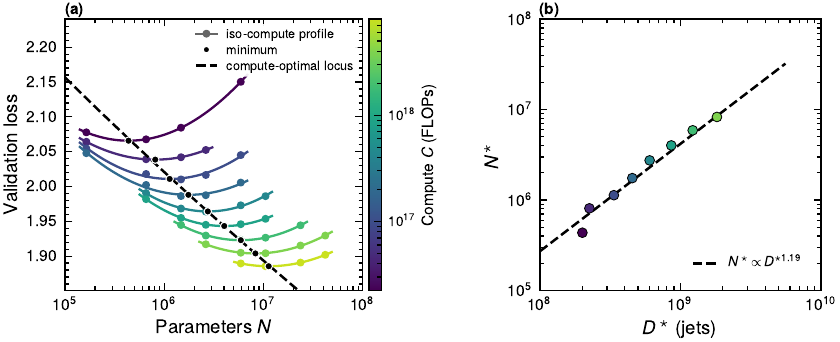}
    \caption{Approach 2 (IsoFLOP profiles) for the multi-task variant. \textbf{a)} Total final validation loss against model size $N$ along iso-compute slices. The dashed line is the compute-optimal trajectory. \textbf{b)} the resulting model-size--dataset allocation.}
    \label{fig:approach2}
\end{figure}

\paragraph{Approach 2: IsoFLOP profiles.} The second approach reads the loss-minimizing allocation directly off each iso-compute slice. \cref{fig:approach2} shows the final validation loss versus $N$ along iso-compute slices, each a shallow parabola whose minimum is the compute-optimal model size at that budget. The exponents are $N^\star \propto C^{0.55}$ and $D^\star \propto C^{0.45}$.

\begin{figure}[!htbp]
    \centering
    \includegraphics[width=0.45\textwidth]{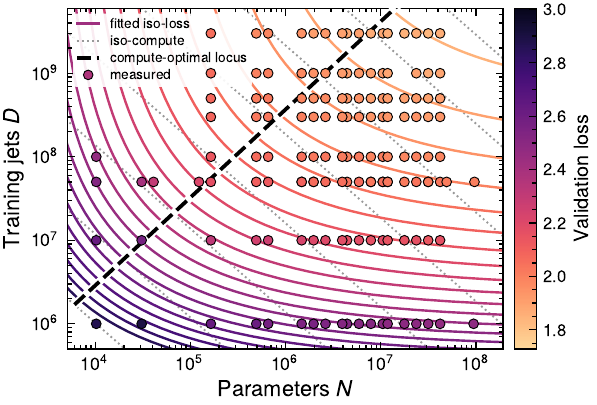}
    \caption{Approach 3 for the multi-task variant: iso-loss contours of the fitted total validation loss surface $L(N, D)$, with the measured cells on the same color scale. Dotted lines are iso-compute slices. The dashed line is the compute-optimal locus implied by the fit, the $(N^\star, D^\star)$ that minimize the loss at each compute budget.}
    \label{fig:isoflop_surface}
\end{figure}

\paragraph{Approach 3: full-surface fit.} The third approach fits the full $L(N, D)$ surface of \cref{eq:chinchilla} jointly to all cells. \cref{fig:isoflop_surface} shows the fitted surface in the $(N, D)$ plane with the measured cells and the implied compute-optimal locus, which gives $N^\star \propto C^{0.49}$ and $D^\star \propto C^{0.51}$.

\paragraph{All three approaches agree} $N^\star(C)$ and $D^\star(C)$ from the three approaches agree on the near-equal $\sqrt{C}$ dependence. Coefficients are reported in \cref{tab:nd_allocation} and compared to previous scaling studies.

\begin{table}[!htbp]
  \centering
  \small
  \begin{tabular}{lcc}
    \toprule
    Method & $a$ & $b$ \\
    \midrule
    This work (Approach 1) & $0.52 \pm 0.01$ & $0.48 \pm 0.01$ \\
    This work (Approach 2) & $0.55 \pm 0.02$ & $0.45 \pm 0.02$ \\
    This work (Approach 3) & $0.49 \pm 0.03$ & $0.51 \pm 0.03$ \\
    \midrule
    Carpe Datum~\citep{ATL-SOFT-PUB-2026-002}        & $0.10$   & $0.90$   \\
    Kaplan et al.~\citep{kaplan2020scalinglawsneurallanguage}   & $0.73$            & $0.27$            \\
    Hoffmann et al.~\citep{hoffmann2022trainingcomputeoptimallargelanguage} & $0.49$            & $0.51$            \\
    \bottomrule
  \end{tabular}
  \vspace{2mm}
  \caption{Compute-optimal allocation exponents, where
$N^\star(C) \propto C^{a}$ and $D^\star(C) \propto C^{b}$. All three approaches agree on $a \simeq b \simeq 0.5$, matching the
Hoffmann et al.\ result, where the tension with the earlier Kaplan et al.\
coefficients was attributed to non-embedding parameter count and hyperparameter tuning at small scale~\cite{pearce2024reconcilingkaplanchinchilla,porian2025resolvingdiscrepanciescomputeoptimalscaling,weng2026scaling}.
We observe a similar pattern relative to the Carpe Datum results: with tuned
hyperparameter scaling and higher dataset sizes, the compute-optimal allocation recovers the equal
$\sqrt{C}$ dependence of model and dataset size.}
\label{tab:nd_allocation}
\end{table}

% =============================================================================
\section{Compute-optimal scaling under data repetition}
\label{sec:repetition_main}
% =============================================================================

\begin{figure}[!htbp]
    \centering
    \includegraphics[width=0.9\textwidth]{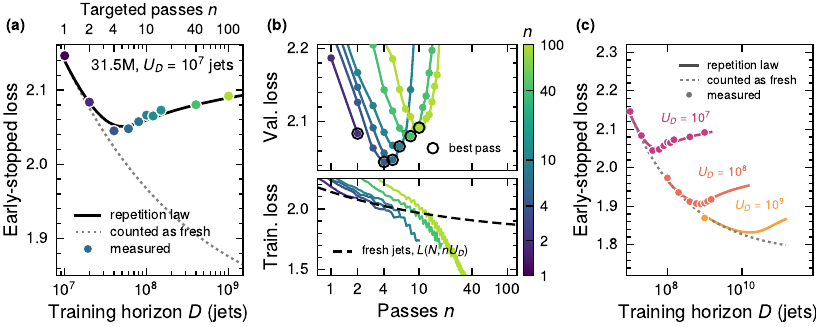}
    \caption{\textbf{(a)} Best early-stopped validation loss of runs targeting $n$ passes for a $31.5$M model on $U_D = 10^7$ unique jets, with \cref{eq:repetition-law} and the single-pass law counting repeated jets as fresh. \textbf{(b)} Validation (top) and training loss (bottom) of six of these runs against the number of passes $n$, and the single-pass law $L(N, nU_D)$ on fresh jets (dashed). \textbf{(c)} The same model on $10^7$, $10^8$ and $10^9$ unique jets against the training horizon.}
    \label{fig:rep_return}
\end{figure}

In practice, simulated data is finite, and the cost of generating more data can outweigh the return in performance, or it might simply not be possible to acquire more data in non-simulated settings. The available compute can then exceed what a single optimal pass over the data requires, and it can be spent on more parameters, on several passes over the \emph{unique dataset} $U_D$, or on both. \citet{muennighoff2025scalingdataconstrainedlanguagemodels} extend \cref{eq:chinchilla} with a decaying value of repeated passes and excess parameters, assuming that regularization keeps the validation loss from increasing, and \citet{lovelace2026prescriptivescalinglawsdata} model the final validation loss including overfitting, which they find to be set by the ratio of unique data to parameters, and also study the effect of weight decay on scaling. We model instead the best validation loss of early-stopped runs that target $n$ passes, trained with the hyperparameter recipe of \cref{sec:completep} for the \emph{training horizon} $D = nU_D$. The recipe minimizes the training loss at any horizon with unrepeated data. Under repetition, the validation loss starts to increase once the training loss crosses this lower bound for unrepeated data (\cref{fig:rep_return}b), consistent with the fact that a training loss below the optimum can only be reached by memorizing the training data. For any $U_D$, the best early-stopped loss is reached by the run whose training loss meets this bound at about the end of its schedule, so that early stopping is barely needed at the optimal training horizon. The intuition is that at this horizon, training and validation loss stay close through the annealed schedule, and with this negligible overfitting the recipe applies as for fresh jets\footnote{The hyperparameters at the optimal training horizon match the lowest loss at the lowest number of passes, set by a scan of $b$ and $\eta$ (Appendix \ref{sec:repetition}). The scaling recipe is thus optimal within the set of scalable hyperparameters we consider, although parameters we hold fixed, such as weight decay, may nevertheless affect the scaling behavior under data repetition~\cite{lovelace2026prescriptivescalinglawsdata}.}. Runs that target more passes cross the bound before their learning rate has annealed: early stopping cuts them short, and their loss pays for hyperparameters set for a horizon they do not complete. In such cases the compute cost of the run is considered only up to the early stopped passes $C = 6NT_Un^\star$.

\begin{figure}[htbp]
    \centering
    \includegraphics[width=0.82\textwidth]{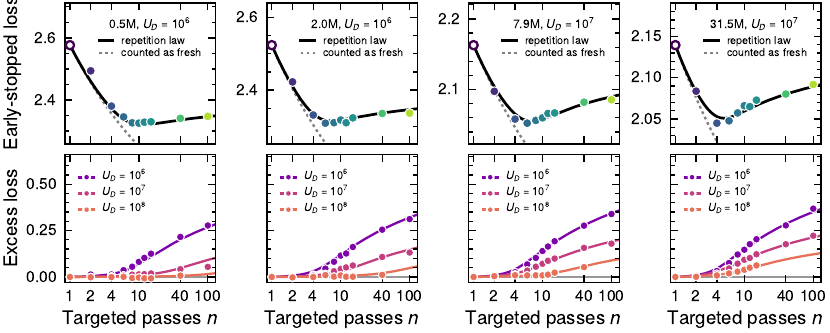}
    \caption{Top: early-stopped validation loss against the targeted number of passes for four model sizes, with \cref{eq:repetition-law} (solid) and the single-pass law with repeated jets counted as fresh (dotted). The open circles at $n = 1$ show the single-pass scaling law. Bottom: excess of the early-stopped loss over the single pass of the same model on as many fresh jets, for three unique sets.}
    \label{fig:rep_laws}
\end{figure}

We account for these effects by adding an empirical penalty term to the single-pass surface of \cref{eq:chinchilla}, evaluated at $n_{\mathrm{eff}}$ fresh-equivalent passes:
\begin{equation}
\label{eq:repetition-law}
\begin{gathered}
L(N, U_D, n) \;=\; \underbrace{L\bigl(N,\, U_D\,n_{\mathrm{eff}}\bigr)}_{\text{single pass over } U_D\,n_{\mathrm{eff}} \text{ jets}} \;+\; \underbrace{\delta\,\Bigl(\ln\frac{n}{n_{\mathrm{eff}}}\Bigr)^{p}}_{\text{cost of repeating}},\\[3pt]
n_{\mathrm{eff}} \;=\; \underbrace{1 + \frac{(n - 1)\,(n_{\mathrm{crit}} - 1)}{\bigl[(n - 1)^{s} + (n_{\mathrm{crit}} - 1)^{s}\bigr]^{1/s}}}_{\text{fresh-equivalent passes, saturating at } n_{\mathrm{crit}}},
\qquad
n_{\mathrm{crit}} \;=\; n_0\biggl[1 + \Bigl(\frac{1}{t_0}\,\underbrace{\frac{T_U}{N}}_{\mathclap{\text{unique tokens per parameter}}}\Bigr)^{\gamma}\,\biggr],
\end{gathered}
\end{equation}
where $T_U \approx 42.6\,U_D$ is the number of unique tokens. $n_{\mathrm{eff}}$ follows $n$ and saturates at $n_{\mathrm{crit}}$, which we find to be $n_0 \approx 4$ for models much larger than $T_U/t_0 \approx 0.5$M parameters per million unique jets and grows as $(T_U/t_0 N)^{\gamma}$ for smaller ones, so that models that are small compared to the data profit from many passes and large ones overfit in a few: the best training horizon of a $31.5$M model grows from 5 passes on $10^7$ unique jets to 17 on $10^9$ (\cref{fig:rep_return}c). The cost of repeating grows only logarithmically, and for $n = 1$ the law reduces to the single-pass surface. The excess over fresh data grows with $n$, more for larger models and smaller unique sets (\cref{fig:rep_laws}, bottom).

\begin{figure}[!htbp]
    \centering
    \includegraphics[width=0.82\textwidth]{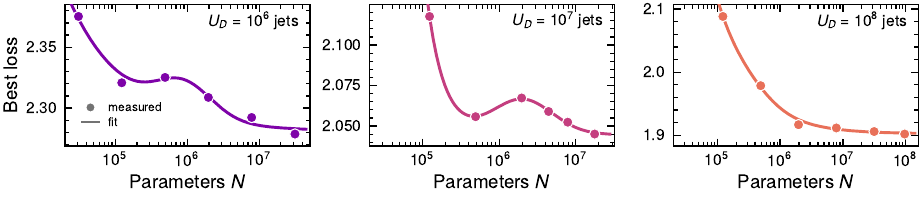}
    \caption{Best early-stopped validation loss over all targeted numbers of passes against model size on $10^6$, $10^7$ and $10^8$ unique jets, with an independent fit per unique set: a power-law descent plus a bump in $\ln N$ that is only as large as the data require.}
    \label{fig:rep_dd}
\end{figure}

Model-wise double descent surviving early stopping has been observed before~\cite{nakkiran2019deepdoubledescentbigger,vigl2025doubledescentoverparameterizationparticle}, including in jet tagging~\cite{ATL-SOFT-PUB-2026-002}, and we observe the same phenomenon here. At fixed $U_D$, the best early-stopped loss rises and falls again with model size, with the peak moving to larger models as $U_D$ grows and flattening past $U_D\sim10^8$ jets (\cref{fig:rep_dd}). 

\begin{figure}[htbp]
    \centering
    \includegraphics[width=0.9\textwidth]{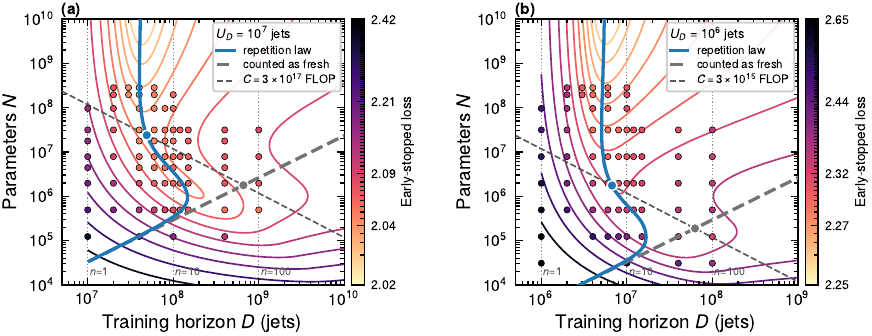}
    \caption{Loss of \cref{eq:repetition-law} in the plane of training horizon and model size (contours) at \textbf{(a)} $U_D = 10^7$ and \textbf{(b)} $U_D = 10^6$ unique jets, with the compute-optimal allocations of \cref{eq:repetition-law} (solid blue) and counting repeated jets as fresh (dashed), with their optima at $C = 3\times10^{17}$ and $3\times10^{15}$~FLOP (gray dashed lines). Dotted lines mark the number of passes. Plot style inspired by~\cite{lovelace2026prescriptivescalinglawsdata}.}
    \label{fig:rep_alloc_planes}
\end{figure}

When each run is charged its early-stopped schedule $C = 6NT_Un^\star$, repetition initially favors compute spent on model sizes and training horizons close to the single-pass compute-optimal allocation. Once the passes reach $n_{\mathrm{crit}}$, additional compute goes into parameters and the allocation reverts to larger models, which target fewer passes (\cref{fig:rep_alloc_planes}), in agreement with the overfitting form in~\cite{lovelace2026prescriptivescalinglawsdata}, with the difference that early stopping causes at most a plateau at large compute. The exponent $\gamma$ in \cref{eq:repetition-law} controls whether the law
develops a local minimum at small surplus compute. When it does, the
compute-optimal allocation stays at that minimum over a range of budgets until
the budget is large enough to jump to the large-model valley, an
\emph{optimal-compute-wise} double descent. At our fitted $\gamma \approx 0.89$ the effect is too weak to cause this, but it is not negligible: over a range of $C$ spanning roughly a decade, the frontier loss is essentially flat, before it resumes descending toward its final plateau. We describe the phenomenon in Appendix \ref{sec:repetition}.

% =============================================================================
\section{What to train at scale: tasks and inputs}
\label{sec:ablations}
% =============================================================================

We ask here whether adding auxiliary training tasks or extending the input set lowers the jet-classification loss at a given compute budget. We derive the hyperparameter scaling separately for each configuration.

\subsection{Auxiliary tasks}
\label{sec:why-aux}

\begin{figure}[!htbp]
    \centering
    \includegraphics[width=0.75\textwidth]{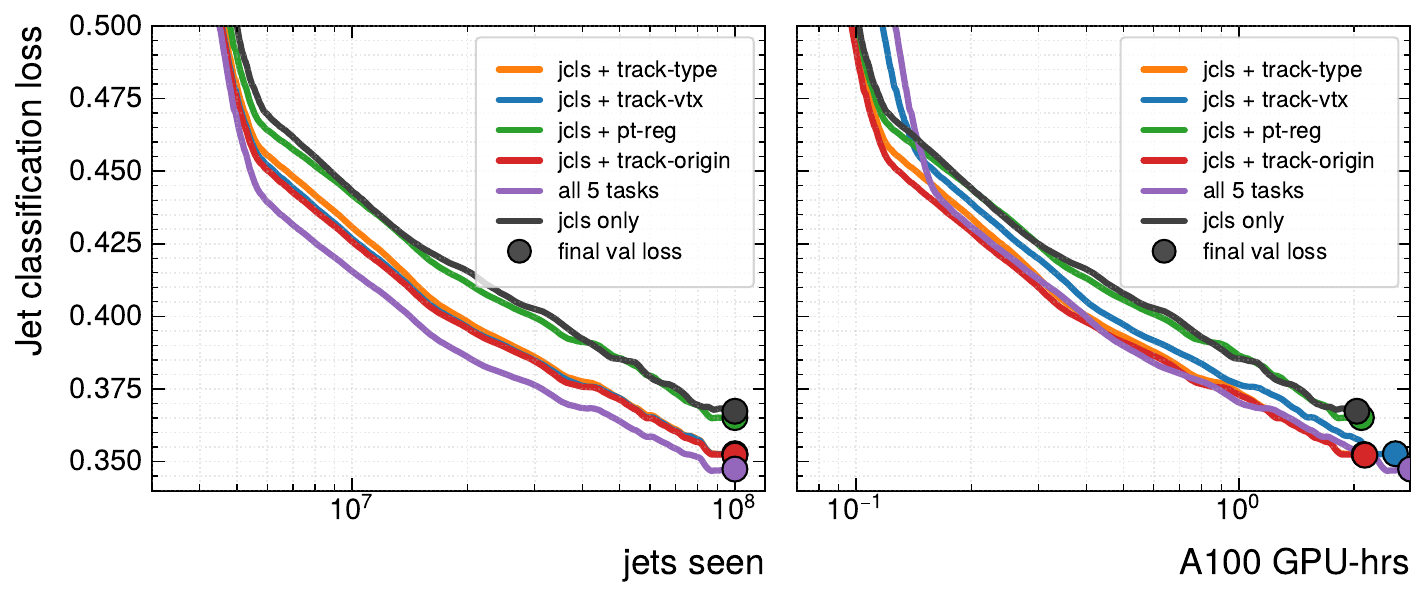}
    \caption{Jet-classification loss (jcls) for the classification-only baseline, for classification plus each single auxiliary task, and for all five tasks, against jets seen \textbf{(left)} and A100 GPU-hours \textbf{(right)}, at fixed model size $N\sim6.7$M. Per jet seen, every auxiliary task lowers the jet-classification loss. Per GPU-hour the gain is partly offset by the higher per-jet cost of the extra heads.}
    \label{fig:task_ablation}
\end{figure}

The multi-task variant carries four auxiliary heads, which are useful for reconstructing the $b$-hadron decay, but can also help the primary objective learn with fewer labeled examples. \cref{fig:task_ablation} (left) shows, for a fixed model size, the jet-classification loss against jets seen for the classification-only baseline, for classification plus each single auxiliary task, and for all five tasks together, with every run trained at its hyperparameter optimum (\cref{sec:multiplicative}). At a fixed number of jets seen, each auxiliary task lowers the jet-classification loss, and all five together reach the lowest. Wall-clock however increases from $73$~$\mu$s/jet for classification alone to $101$~$\mu$s/jet for all five. Re-expressed against GPU-hours (\cref{fig:task_ablation}, right) the per-jet gain is partly offset by this cost, and whether the multi-task setup wins depends on whether the loss it buys outweighs the throughput it spends. Both panels are taken at a fixed model size, while the relevant question is whether the auxiliary tasks still help once each variant is scaled compute-optimally, or whether they eventually become a bottleneck. On the other hand, auxiliary tasks can be useful inference targets in their own right, and a single model can amortize the computational cost across multiple tasks, rather than requiring a separate model for each.

\begin{figure}[!htbp]
    \centering
    \includegraphics[width=0.8\textwidth]{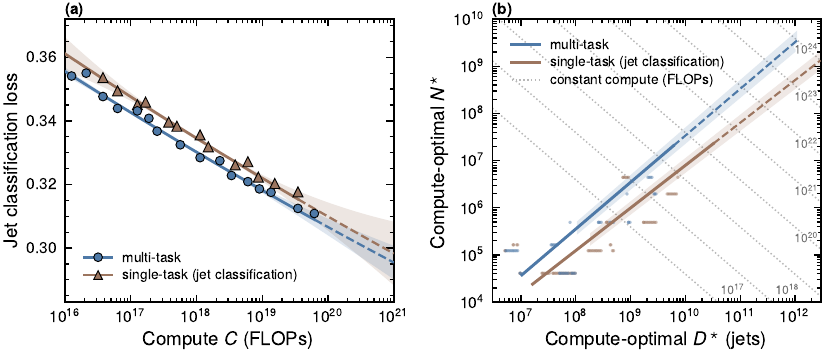}
    \caption{Multi-task (blue) versus single-task (brown) compute-optimal scaling, both with the full input set. \textbf{(a)} Final jet-classification validation loss on the compute-optimal frontier $L^\star(C)$ with $1\sigma$ bands. The multi-task frontier lies below the single-task one across the studied range. \textbf{(b)} The corresponding compute-optimal allocation $N^\star(D^\star)$ with $1\sigma$ bands. Dotted diagonals are lines of constant compute at decades from $C = 10^{17}$ to $10^{24}$~FLOP. At a given compute budget the multi-task objective prefers a larger model and a smaller dataset, so the auxiliary heads shift the compute-optimal split toward model size.}
    \label{fig:mt_vs_st}
\end{figure}

At fixed compute, the multi-task variant consistently achieves a lower jet-classification loss than the classification-only variant (\cref{fig:mt_vs_st}), so the auxiliary tasks continue to improve the primary objective at scale. Notably, the multi-task objective favors a larger model and a smaller dataset than the classification-only objective at fixed compute. The variants must therefore be compared along their respective compute-optimal frontiers rather than at fixed $N$ or $D$.

\subsection{Input variants}
\label{sec:input_variants}
\begin{figure}[!htbp]
    \centering
    \includegraphics[width=0.4\textwidth]{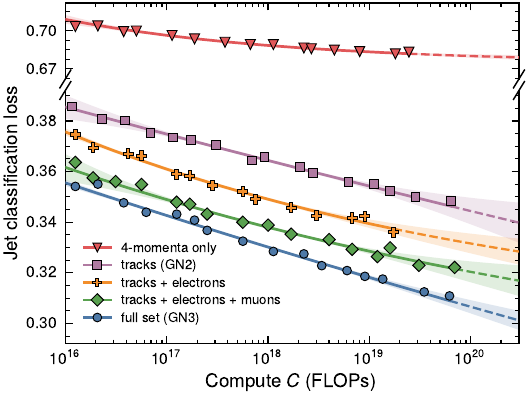}
    \caption{Compute-optimal jet-classification loss for each input level: the runs on the compute-optimal frontier and the fitted frontier $L^\star(C)$ with its $1\sigma$ band, dashed beyond the measured range. Each added modality lowers the loss at fixed compute, and the four-momenta-only model (upper axis range) lies far above the rest. Apart from the 4-momenta-only variant, the frontiers are close to parallel, so the input level mainly shifts the loss rather than changing the scaling exponent.}
    \label{fig:variant_frontier:loss}
\end{figure}

The same scaling study is repeated for the following input progression, adding one modality at a time after a kinematics-only baseline: using all track features (the GN2 set); adding soft electrons; adding muons; and finally adding particle-flow objects, which recovers the full GN3 input set. The exact variables entering each variant are listed in \cref{tab:TaggerInputs}. \cref{fig:variant_frontier:loss} fits a compute-optimal frontier $L^\star(C)$ to each input level.

Every added modality lowers the compute-optimal loss, with the largest single step from adding the full set of track features to bare four-momenta and persisting returns from electrons, muons, and particle-flow. Aside from the four-momenta-only variant, which is close to its floor, the fitted frontiers are close to parallel, so richer inputs shift the loss at fixed compute while leaving the scaling exponent nearly unchanged, favoring inputs as close to the raw detector signals as practical. This implies that performance should be improved in both directions: by scaling the model and dataset, and by moving to richer, lower-level inputs.

\subsection{Comparison with TN25-86M}

\begin{figure}[!htbp]
    \centering
    \includegraphics[width=0.65\textwidth]{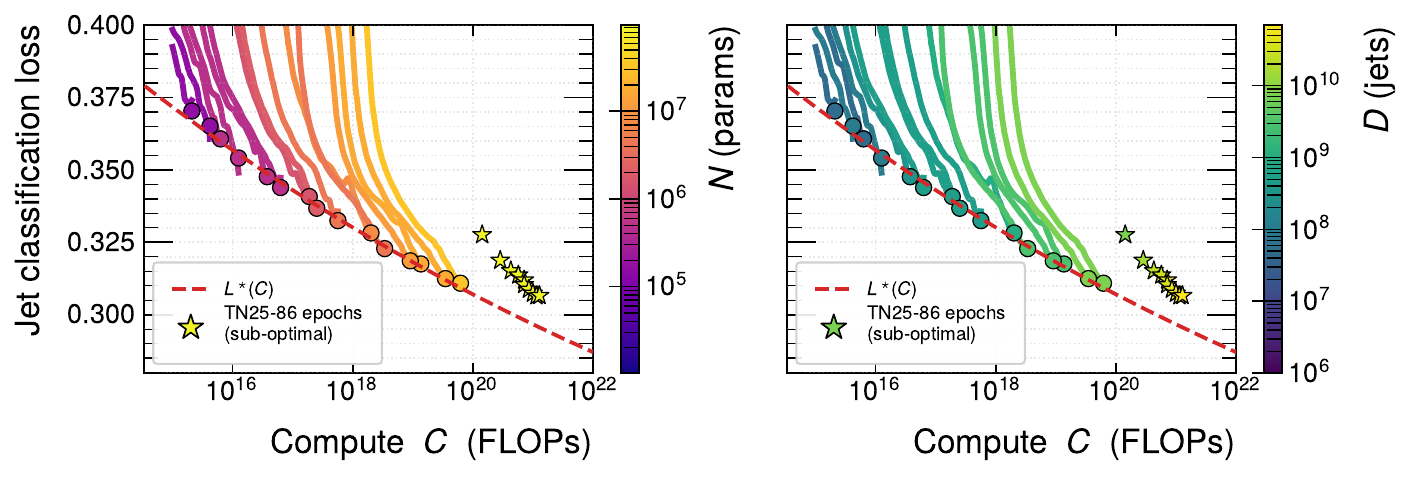}
    \caption{Jet-classification loss against compute, colored by model size \textbf{(left)} and dataset size \textbf{(right)}. The dashed line is the compute optimal loss $L^\star(C)$. The data-repeated TN25-86M run lies above the single-pass frontier.}
    \label{fig:task_jets_classification}
\end{figure}

Previous scaling-law studies on jet tagging~\cite{ATL-SOFT-PUB-2026-002,vigl2026neuralscalinglawsboosted} probe the high-compute region by repeating data on oversized networks, effectively reducing $L^\star(C)$ to a pure $L(D)$ law. The TN25-86M model of~\cite{ATL-SOFT-PUB-2026-002} therefore lies off the compute-optimal frontier by design, since it spends extra FLOPs sub-optimally to maximize performance on a fixed set of $\sim7.7$B jets, repeated over
multiple epochs, until overfitting eventually limits further gains. \cref{fig:task_jets_classification} shows how TN25-86M (stars indicate epochs) lands above the single-pass compute-optimal frontier $L^\star(C)$, so an optimally sized model trained on a larger set of fresh data, if available, would achieve the same loss at a lower compute cost.

% =============================================================================
\section{Physics performance}
\label{sec:physics}
% =============================================================================

The jet classification head has multiple classes, and a background rejection value is obtained by projecting the per-flavor scores $(p_b, p_c, p_\tau, p_{\text{light}})$ into a scalar flavor discriminant
\begin{equation}
\label{eq:discriminant}
D_b \;=\; \ln \frac{p_b}{f_c\,p_c + f_\tau\,p_\tau + (1 - f_c - f_\tau)\,p_{\text{light}}},
\end{equation}
with the background fractions $(f_c, f_\tau, f_{\mathrm{light}}=1-f_c- f_\tau)$ set by hand. A working point is a cut $D_b > t$ on the discriminant: the threshold $t$ sets the $b$-jet efficiency $\epsilon_b$, the fraction of true $b$-jets kept, and for each background flavor $f$ a mis-tag efficiency $\epsilon_f$ whose inverse $1/\epsilon_f$ is the background rejection. The fractions are fixed by hand because a deployed tagger is a frozen predictor that must be calibrated against data at a few working points. By the Neyman--Pearson lemma \cref{eq:discriminant} separates $b$-jets from the background optimally when the $f_x$ equal the prior probabilities of each background flavor in the sample being discriminated against, i.e., the background composition of the downstream analysis. Since that composition varies with kinematics, the optimal fractions are in general functions of $(\pt, \eta)$ and differ from one analysis to the next, so no single fixed choice is universally optimal.

\begin{figure}[!htbp]
    \centering
    \includegraphics[width=0.65\textwidth]{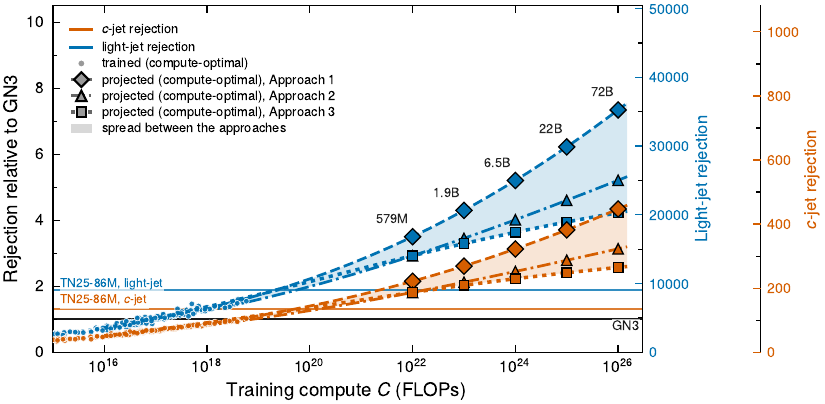}
    \caption{Light-jet (blue) and $c$-jet (orange) rejection of GN3~\cite{ATL-PHYS-PUB-2026-001}, TN25-86M~\cite{ATL-SOFT-PUB-2026-002}, and the compute-optimal models (model sizes are annotated for a few decades in compute). For the scaling laws projections, all three fitting approaches are shown. Evaluated at $70\%$ global $b$-jet efficiency, $20 < \pt < 250$~GeV, $|\eta| < 2.5$ on a $t\bar{t}$ sample.}
    \label{fig:atlas_evolution}
\end{figure}

\cref{fig:atlas_evolution} translates loss into background rejection on a $t\bar{t}$ sample at a fixed \emph{global} $b$-jet efficiency of $70\%$, using the GN3 discriminant fractions from~\cite{ATL-PHYS-PUB-2026-001} $(f_c, f_\tau) = (0.2, 0.1)$. It reports light-jet (blue) and $c$-jet (orange) rejection of GN3~\cite{ATL-PHYS-PUB-2026-001}, the TN25-86M model of~\cite{ATL-SOFT-PUB-2026-002}, and the projections for compute-optimal models trained on the full set of inputs and tasks, obtained with the three fitting approaches. The projected rejections follow from $R(L) = A e^{BL} + R_0$ (Appendix~\ref{app:bkg_rej_fits}). The three models share the same architecture and therefore belong to the same model class, and differ mainly in model size, amount of unique training data, and compute. \cref{fig:atlas_evolution} is thus not a comparison of tagger designs.
 The compute-optimal models are trained in a single pass over the data at the optimal model size, and the scaling law projections assume the same, with the corresponding model sizes annotated. GN3 has about 12M parameters and is trained for several passes over 350M unique jets, while TN25-86M has 86M parameters and is trained for several passes over about 7.7B unique jets. Both are trained until the validation loss stops improving.

\section{Conclusions}
\label{sec:conclusion}

We have presented a systematic procedure for deriving robust compute-optimal scaling laws, and demonstrated it for multi-task jet-tagging transformers on the $\sim11$ billion-jet JetSet2 dataset. The methodological result is that reliable scaling laws require the training dynamics to be scaled together with the model and dataset. Predictive hyperparameter transfer removes a source of bias that would otherwise be absorbed into the fitted scaling exponents and floor, allowing meaningful comparisons at scale, where each configuration must be given its own compute-optimal allocation.

Applied to the GN family, the scaling laws show a near-equal $\sqrt{C}$ dependence of model and dataset size. Auxiliary objectives consistently improve the primary jet-classification loss and shift the compute-optimal split toward larger models and smaller datasets, while richer input representations lower the loss but leave the scaling exponent approximately unchanged. The onset of the power-law regime occurs only after a substantial amount of data has been seen, and fits based on smaller datasets can return biased scaling exponents even when the resulting curves appear well described by a power law. Large full-simulation datasets such as JetSet2 are therefore a prerequisite for establishing how HEP models behave at scale, and the ability of LHC experiments to produce and retain such samples is a major asset for machine-learning research in HEP. 

This suggests a natural progression from small-scale studies to large-scale, full-simulation benchmarks. A method should demonstrate that its advantage persists as model size, dataset size, and compute increase, that it remains competitive when given the full input information available to the experiment, and ultimately that it holds up on real collision data. The framework presented here can serve as a practical blueprint for future
scaling studies.

The full study required approximately 97k A100 GPU-hours, substantial by HEP standards but modest compared with industrial-scale models. The observed scaling behavior suggests that considerably larger budgets could be productively invested, particularly if the field moves toward generic foundation models. Rather than training separate models for individual applications, one could pretrain a single model at scale, and then adapt the resulting representation to different experiments, detector configurations, and physics tasks through fine-tuning.

\begin{table}[H]
\small
\centering
\caption{The number of runs, jets seen, training FLOPs, and A100 GPU-hours per study, with the total.}
\label{tab:compute_budget}
\resizebox{0.8\textwidth}{!}{%
\begin{tabular}{lrrrr}
\toprule
Study & Runs & Jets seen & Training FLOPs & GPU-hours (A100) \\
\midrule
$\mu$P transfer                       &  324 & $6.5\times10^{9}$  & $2.8\times10^{19}$ & 0.34k \\
$\eta^\star$/$b^\star$ + aspect ratio & 1395 & $1.1\times10^{11}$ & $1.8\times10^{20}$ & 4.5k \\
Compute-optimal grid (all variants)                 & 1219 & $1.5\times10^{12}$ & $8.4\times10^{21}$ & 92k \\
\midrule
\textbf{Total} & \textbf{2938} & $\mathbf{1.6\times10^{12}}$ & $\mathbf{8.6\times10^{21}}$ & \textbf{97k} \\
\bottomrule
\end{tabular}}
\end{table}

\section*{Acknowledgments}

\noindent We thank Samuel Klein for his feedback on the manuscript. L.H. is supported by BMFTR Project SciFM 05D25WO2. M.K. is supported by the US Department of Energy (DOE) under Grant No. DE-AC02-76SF00515. N.H. is supported by the NSF AI Institute for Artificial Intelligence and Fundamental Interactions under Cooperative Agreement PHY-2525568. A.F. is supported by the Alexander von Humboldt Foundation through a Feodor Lynen Research Fellowship.

% =============================================================================
\bibliographystyle{unsrtnat}
\bibliography{main}

\newpage
\appendix

\section{Input variables}

\begin{table*}[h!bt]
    \caption{
    Input variables considered in this study.
    Muon variables are treated as additional track features, while particle-flow and electron inputs are separate input modalities.
    The five right-most columns indicate, with a colored cell, which variables enter each training variant:
    {\color{colFull}\rule{2.2mm}{2.2mm}}~full set (GN3),
    {\color{colNoflow}\rule{2.2mm}{2.2mm}}~Tracks$+$electrons$+$muons (GN3 minus the pflow modality),
    {\color{colGN2e}\rule{2.2mm}{2.2mm}}~Tracks$+$electrons (GN3 minus the the pflow and muon modality / GN2 plus the electron modality),
    {\color{colGN2}\rule{2.2mm}{2.2mm}}~tracks (GN2),
    {\color{colFourmom}\rule{2.2mm}{2.2mm}}~4-momenta only (all modalities).
    }
   \label{tab:TaggerInputs}
\begin{center}
\tiny
\setlength{\tabcolsep}{4pt}
\renewcommand{\arraystretch}{1.05}
\begin{tabular}{ @{} >{\raggedright\arraybackslash}p{1.6cm} >{\raggedright\arraybackslash}p{2.7cm} >{\raggedright\arraybackslash}p{0.45\textwidth} *{5}{>{\centering\arraybackslash}m{0.5cm}} @{} }
% \hline
\textbf{Input Modality} & \textbf{Variable} & \textbf{Description}
 & \rotatebox{90}{Full set (GN3)} & \rotatebox{90}{Tracks$+$electrons$+$muons}
 & \rotatebox{90}{Tracks$+$electrons} & \rotatebox{90}{Tracks (GN2)} & \rotatebox{90}{4-momenta} \\
\hline
\multirow{2}{*}{ Jet }
& \pt & Transverse momentum & \yF & \yN& \yE & \yG & \yM  \\
& $\eta$ & Pseudo-rapidity & \yF & \yN& \yE & \yG & \yM  \\
\hline
        \multirow{19}{*}{ Track }
        & $q/p$ & Track's charge over momentum (curvature) & \yF & \yN& \yE & \yG & \yM  \\
        & $\Delta\eta$ & Pseudorapidity relative to the jet $\eta$ & \yF & \yN& \yE & \yG & \yM  \\
        & $\Delta\phi$  & Azimuthal angle relative to the jet $\phi$ & \yF & \yN& \yE & \yG & \yM  \\
        & $d_0$  & Closest distance from the track to PV in the transverse plane & \yF & \yN& \yE & \yG &  \\
        & $z_0 \sin\theta$  & Distance to PV in the longitudinal plane, measured at point of closest approach in transverse plane & \yF & \yN& \yE & \yG &  \\
        & $\sigma(q/p)$ & Uncertainty on $q/p$ & \yF & \yN& \yE & \yG &  \\
        & $\sigma(\theta)$ & Uncertainty on track polar angle $\theta$ & \yF & \yN& \yE & \yG &  \\
        & $\sigma(\phi)$  & Uncertainty on track azimuthal angle $\phi$ & \yF & \yN& \yE & \yG &  \\
        & $s(d_{0})$  & Lifetime signed transverse IP significance & \yF & \yN& \yE & \yG &  \\
        & $s(z_{0}\sin\theta)$  & Lifetime signed longitudinal IP significance & \yF & \yN& \yE & \yG &  \\
        & nPixHits   & Number of pixel hits & \yF & \yN& \yE & \yG &  \\
        & nSCTHits   & Number of SCT hits & \yF & \yN& \yE & \yG &  \\
        & nIBLHits   & Number of IBL hits & \yF & \yN& \yE & \yG &  \\
        & nBLHits    & Number of B-layer hits & \yF & \yN& \yE & \yG &  \\
        & nIBLShared & Number of shared IBL hits & \yF & \yN& \yE & \yG &  \\
        & nIBLSplit  & Number of split IBL hits & \yF & \yN& \yE & \yG &  \\
        & nPixShared & Number of shared pixel hits & \yF & \yN& \yE & \yG &  \\
        & nPixSplit  & Number of split pixel hits & \yF & \yN& \yE & \yG &  \\
        & nSCTShared & Number of shared SCT hits & \yF & \yN& \yE & \yG &  \\
	\hline
	\multirow{5}{*}{Soft leptons }
        & leptonID & Electron and muon reconstruction label & \yF & \yN& & &  \\
        & muonQuality & Muon reconstruction quality (Loose, Medium, Tight) & \yF & \yN& & &  \\
        & $\mathcal{R}$ & Ratio of $q/p$ measured independently in the MS and the ID & \yF & \yN& & &  \\
        & $\mathcal{M}$ & Imbalance of the muon tri-momentum between the MS and the ID & \yF & \yN& & &  \\
        & $\mathcal{S}$ & Azimuthal Angular deviation between scattering centers belonging to a muon track & \yF & \yN& & &  \\
\hline
\multirow{28}{*}{Electron}
    & \pt                             & Electron transverse momentum & \yF & \yN& \yE & & \yM  \\
    & $p_\mathrm{T}^\mathrm{frac}$ & Electron \pt divided by the jet \pt & \yF & \yN& \yE & &  \\
    & $p_\mathrm{T}^\mathrm{rel}$                & Electron momenta transverse to the jet axis & \yF & \yN& \yE & &  \\
    & $\Delta R$  & Opening angle $\Delta R$ between the electron and the jet axis & \yF & \yN& \yE & &  \\
    & $|\eta|$   & Absolute pseudorapidity of the electron & \yF & \yN& \yE & &  \\
    & $\eta$                          & Electron pseudorapidity & \yF & \yN& \yE & & \yM  \\
    & $\phi$                          & Electron azimuthal angle & \yF & \yN& \yE & & \yM  \\
    & $E_\mathrm{T}$ & Transverse energy of the electron EM cluster & \yF & \yN& \yE & &  \\
    & $q/p$   & Electron track's charge over momentum & \yF & \yN& \yE & &  \\
    & $d_0^{\mathrm{BS}}$             & Transverse impact parameter relative to the beam spot & \yF & \yN& \yE & &  \\
    & $s(d_0^{\mathrm{BS}})$          & Significance of $d_0$ relative to the beam spot, $|d_0/\sigma(d_0)|$ & \yF & \yN& \yE & &  \\
    & \texttt{ptVarCone30}/\pt        & Track-based isolation with a variable cone ($R_\mathrm{max}=0.3$), divided by \pt & \yF & \yN& \yE & &  \\
    & nPixHits                        & Number of pixel hits on the electron track & \yF & \yN& \yE & &  \\
    & nSCTHitsInclDead                & Number of SCT hits on the electron track, including crossings of dead sensors & \yF & \yN& \yE & &  \\
    & $\Delta p / p$                  & Momentum lost by the track between the perigee and the last measurement point, divided by the momentum at perigee & \yF & \yN& \yE & &  \\
    & eProbabilityHT  & Likelihood probability based on transition radiation in the transition radiation tracker & \yF & \yN& \yE & &  \\
    & $\Delta\eta_1$                  & $\Delta\eta$ between the cluster position in the first EM layer and the extrapolated track & \yF & \yN& \yE & &  \\
    & $\Delta\phi_{\mathrm{res}}$     & $\Delta\phi$ between the cluster position in the second EM layer and the momentum-rescaled track, times the track charge & \yF & \yN& \yE & &  \\
    & $E/p$                           & Ratio of the EM cluster energy to the track momentum & \yF & \yN& \yE & &  \\
    & $R_{\mathrm{had}}$              & Ratio of $E_\mathrm{T}$ in the hadronic calorimeter to $E_\mathrm{T}$ of the EM cluster (used for $0.8<|\eta|<1.37$) & \yF & \yN& \yE & &  \\
    & $R_{\mathrm{had1}}$             & Ratio of $E_\mathrm{T}$ in the first hadronic-calorimeter layer to $E_\mathrm{T}$ of the EM cluster (used for $|\eta|<0.8$ or $|\eta|>1.37$) & \yF & \yN& \yE & &  \\
    & $E_{\mathrm{ratio}}$            & Ratio of the energy difference between the largest and second-largest energy deposits in the first EM layer to their sum & \yF & \yN& \yE & &  \\
    & $w_{\eta 2}$                    & Lateral shower width in the second EM layer, computed over $3\times 5$ cells & \yF & \yN& \yE & &  \\
    & $R_\phi$                        & Ratio of the energy in $3\times 3$ cells to the energy in $3\times 7$ cells centered on the electron cluster, in the second EM layer & \yF & \yN& \yE & &  \\
    & $R_\eta$                        & Ratio of the energy in $3\times 7$ cells to the energy in $7\times 7$ cells centered on the electron cluster, in the second EM layer & \yF & \yN& \yE & &  \\
    & $w_{s,\mathrm{tot}}$            & Total shower width in the first EM layer, computed over a window of $\Delta\eta\times\Delta\phi \approx 0.0625\times 0.2$ & \yF & \yN& \yE & &  \\
    & $f_1$                           & Fraction of cluster energy deposited in the first EM layer & \yF & \yN& \yE & &  \\
    & $f_3$                           & Fraction of cluster energy deposited in the third EM layer & \yF & \yN& \yE & &  \\
\hline
\multirow{5}{*}{PFlow}
        & \pt & PFlow transverse momentum & \yF && & & \yM  \\
        & energy & PFlow energy & \yF && & & \yM  \\
        & $\Delta\eta$ & Pseudorapidity relative to the jet $\eta$ & \yF && & & \yM  \\
        & $\Delta\phi$ & Azimuthal angle relative to the jet $\phi$ & \yF && & & \yM  \\
        & isCharged & Boolean indicating if charged PFlow & \yF && & &  \\ [0.1cm]
\end{tabular}
\end{center}
\end{table*}

\section{Data repetition: hyperparameter scaling and optimal-compute-wise double descent}
\label{sec:repetition}

\begin{figure}[!htbp]
    \centering
    \includegraphics[width=0.8\textwidth]{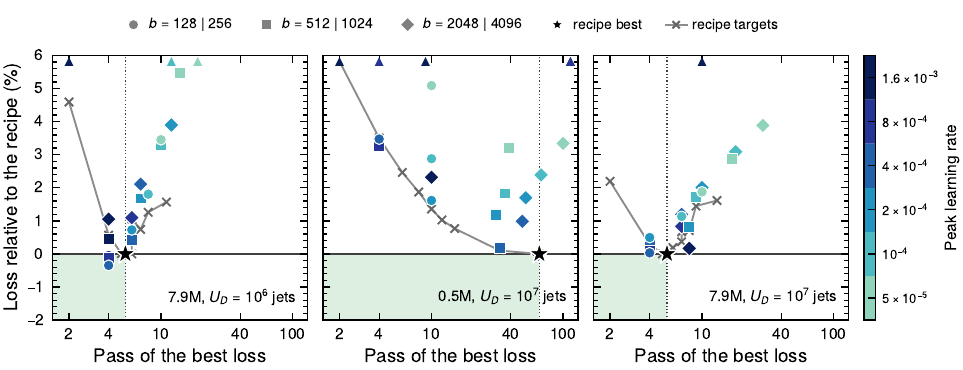}
    \caption{Best early-stopped loss of each point of a grid of $b$ (markers) and $\eta$ (colors), over the targeted pass counts run for it, relative to the best loss of the recipe (star), against the pass at which it is reached. The gray line connects the recipe's own runs, one per targeted pass count. A setting better than the recipe on both counts would sit in the shaded corner. Triangles: points above the range shown. Of the two batch sizes of each marker, the first is used on $10^6$ and the second on $10^7$ unique jets.}
    \label{fig:rep_hp_scan}
\end{figure}

At the optimal training horizon, the hyperparameters of the scaling recipe of \cref{sec:completep} yield a loss comparable with the lowest from independent scans of $b$ and $\eta$, and at the same number of epochs (thus compute), as shown in \cref{fig:rep_hp_scan}.

\begin{figure}[!htbp]
    \centering
    \includegraphics[width=0.75\textwidth]{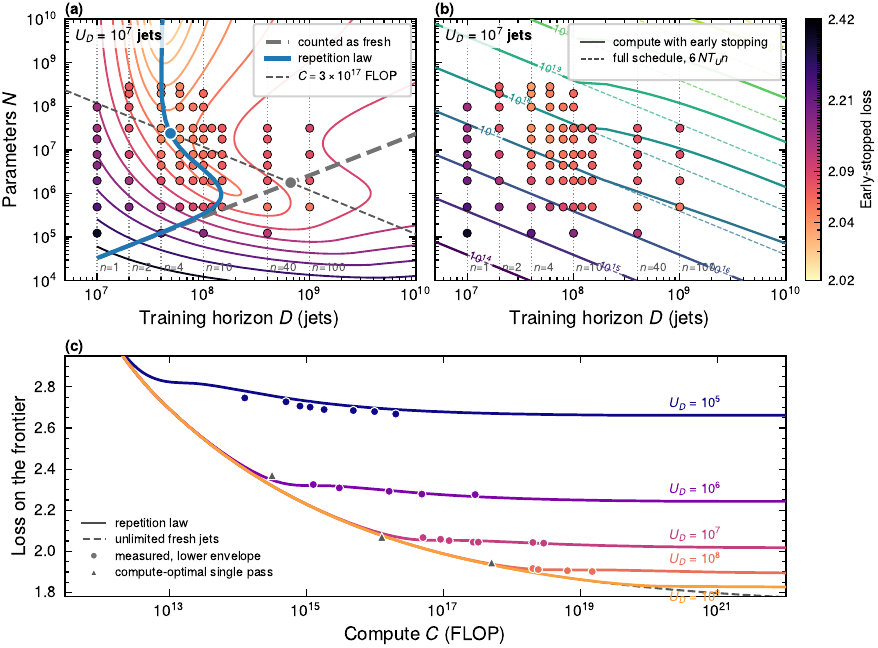}
    \caption{Allocation under repetition. \textbf{(a)} Contours of the loss of \cref{eq:repetition-law} at $U_D = 10^7$ in the plane of training horizon and parameters, the measured early-stopped losses on the same color scale, the compute-optimal allocations of \cref{eq:repetition-law} and of the single-pass law with repeated jets counted as fresh, and their optima at $C = 3\times10^{17}$ FLOP; dotted lines mark the number of passes. \textbf{(b)} Constant compute when runs stop early (solid, labeled) and when the full schedule is charged (dashed), with the runs colored by their loss. \textbf{(c)} Loss on the compute-optimal frontier for $U_D = 10^5$ to $10^9$, the frontier with unlimited fresh jets (dashed), the measured runs on their lower envelope (circles) and compute-optimal single passes (triangles).}
    \label{fig:rep_alloc}
\end{figure}

Each run is charged its early stopped schedule, $C = 6NT_Un^\star$, and the optimal allocation minimizes \cref{eq:repetition-law} at each compute budget. Early stopping shortens only schedules longer than the best one, which the allocation never selects (\cref{fig:rep_alloc}b). Within a few passes, repeated jets count nearly as fresh jets, which favors small models close to the single-pass compute-optimal allocation, trained for several passes.  Once the passes reach $n_{\mathrm{crit}}$, \cref{eq:repetition-law} puts additional compute into parameters, and the allocation reverts to larger models that target fewer passes (\cref{fig:rep_alloc}a), in agreement with Ref.~\cite{lovelace2026prescriptivescalinglawsdata}, while the loss on the frontier eventually flattens.

\begin{figure}[!htbp]
    \centering
    \includegraphics[width=0.6\textwidth]{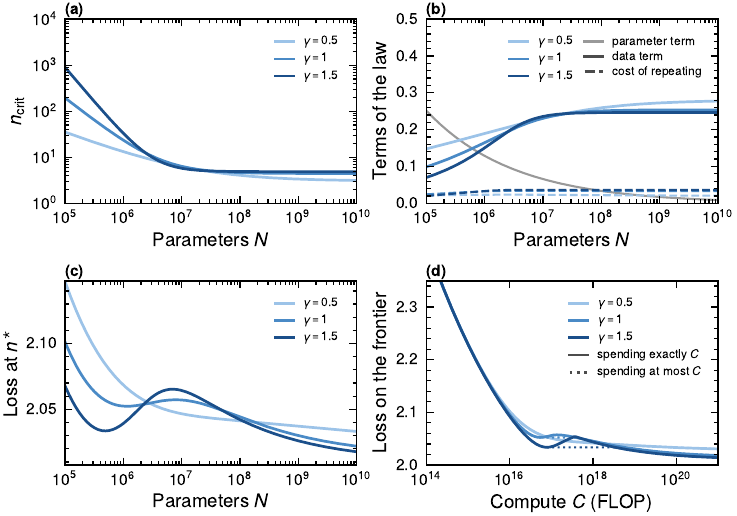}
    \caption{\cref{eq:repetition-law} at $U_D = 10^7$ unique jets with the onset exponent held at 0.5, 1 and 1.5 and the other parameters refitted. \textbf{(a)} $n_{\mathrm{crit}}$; \textbf{(b)} the terms of the law at its best target (the parameter term does not depend on $\gamma$); \textbf{(c)} the law at its best target against model size; \textbf{(d)} the loss on the compute-optimal frontier, spending at most (dashed) and exactly (solid) the budget.}
    \label{fig:gamma_anatomy}
\end{figure}

\begin{figure}[!htbp]
    \centering
    \includegraphics[width=0.6\textwidth]{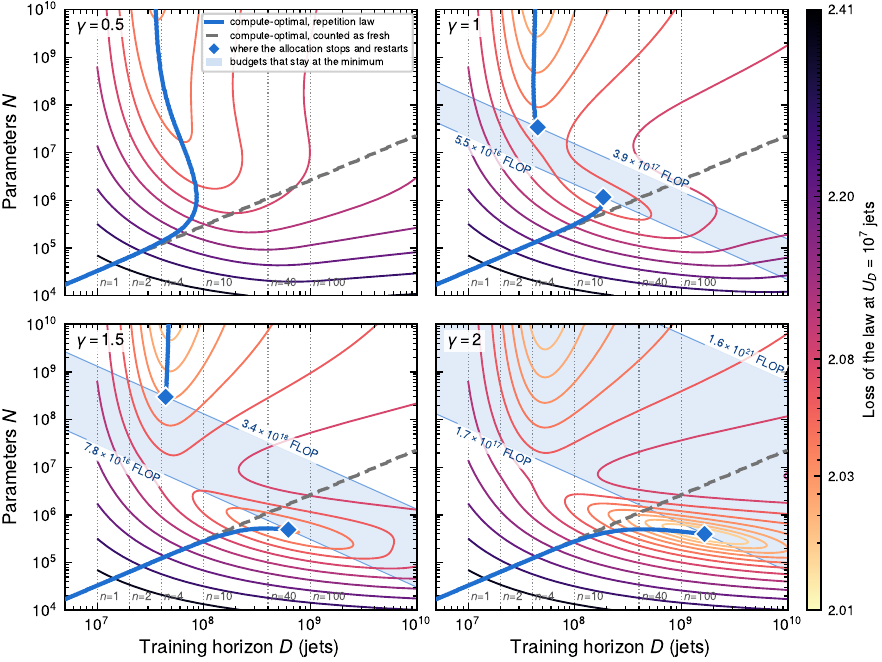}
    \caption{Allocation implied by \cref{eq:repetition-law} at $U_D = 10^7$ unique jets with the onset exponent $\gamma$ held at 0.5, 1, 1.5 and 2: contours of the loss in the plane of training horizon and parameters, and the compute-optimal allocation of \cref{eq:repetition-law} (solid) and of the single-pass law with repeated jets counted as fresh (dashed). Where the law has a minimum at small models, the allocation stops there (lower diamond), and then restarts once the budget exceeds the band. Dotted lines mark the number of target passes.}
    \label{fig:gamma_frontiers}
\end{figure}

\paragraph{Optimal-compute-wise double descent?} The exponent $\gamma$ of \cref{eq:repetition-law} controls how fast the number of useful passes grows as a model gets smaller relative to its unique set, $n_{\mathrm{crit}} \propto (T_U/N)^{\gamma}$ below $N = T_U/t_0$. It is fitted to $\gamma = 0.89 \pm 0.06$ (rms $0.35\%$), but holding $\gamma$ fixed at $0.8$ or $1.2$ still gives a reasonable fit (rms $0.36\%$ and $0.40\%$), and different values of $\gamma$ give rise to different behaviors of the compute-optimal allocation and loss. We thus repeat the fit to \cref{eq:repetition-law} with $\gamma$ held at 0.5, 1, 1.5 and 2, to qualitatively explore its effects. Its value sets how far the law rises in model size at fixed $U_D$ (\cref{fig:gamma_anatomy}), providing a possible quantitative explanation for the mild model-wise double descent we observed with early stopping: our fitted $\gamma<1$ implies only a mild residual peak that flattens at high $U_D$. On the compute axis, a larger value of $\gamma$ appears as an intermediate peak followed by a second descent in loss, meaning that for a brief range more compute, however spent, can hurt performance. Consequently there is a range of budgets over which the best allocation stays at the smaller model and leaves the rest of the budget unused, until the budget is high enough to ``jump" past the peak toward larger models. Given the empirical nature of \cref{eq:repetition-law}, we leave the interpretation of this term as driving early stopped model-wise double descent, and the apparent ``optimal-compute-wise" double descent, to future work, and qualitatively draw four scenarios in \cref{fig:gamma_frontiers}.

% =============================================================================
\section{Background Rejection fits}
\label{app:bkg_rej_fits}
\begin{figure}[!htbp]
    \centering
    \includegraphics[width=0.9\textwidth]{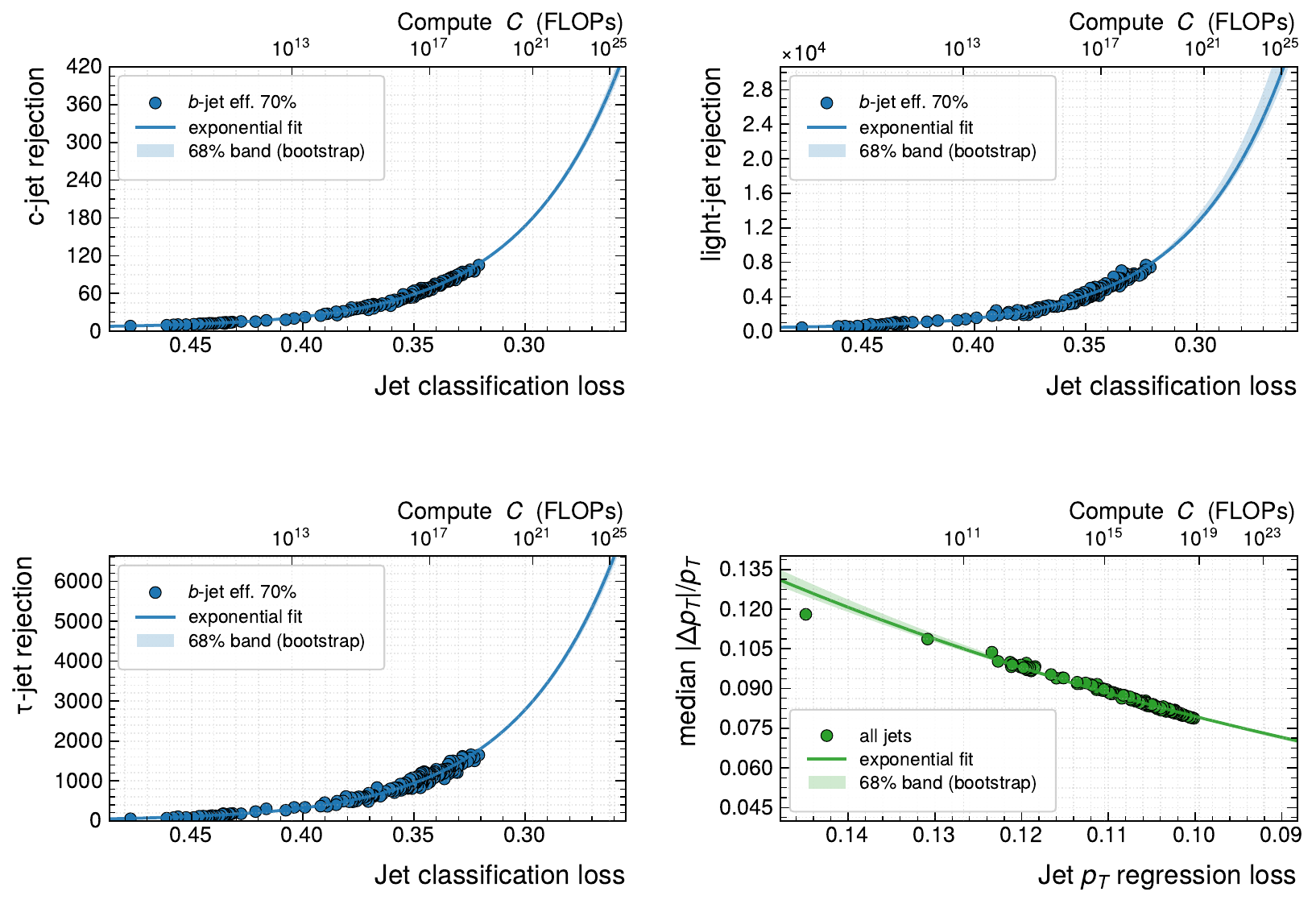}
    \caption{Downstream performance against training loss for the multi-task
  models. The first three panels
  show the $c$-, light- and $\tau$-jet rejection at the $70\%$ $b$-jet
  efficiency working point against the jet-classification loss. The fourth
  shows the median relative $\pt$ resolution against the $\pt$-regression
  loss. Curves are fits of $R(L) = A\,e^{BL} + C$, with the shaded $68\%$ interval from bootstrapping. The top axis converts loss into training
  compute through the compute-optimal envelope
  $L(C) = E + A\,C^{-\gamma}$.}
    \label{fig:metric_vs_loss}
\end{figure}

% =============================================================================
Figure~\ref{fig:metric_vs_loss} maps loss onto rejection with an exponential fit $R(L) = A e^{BL} + R_0$. The compute axis is taken from approach 1.

\begin{figure}[!htbp]
    \centering
    \includegraphics[width=0.9\textwidth]{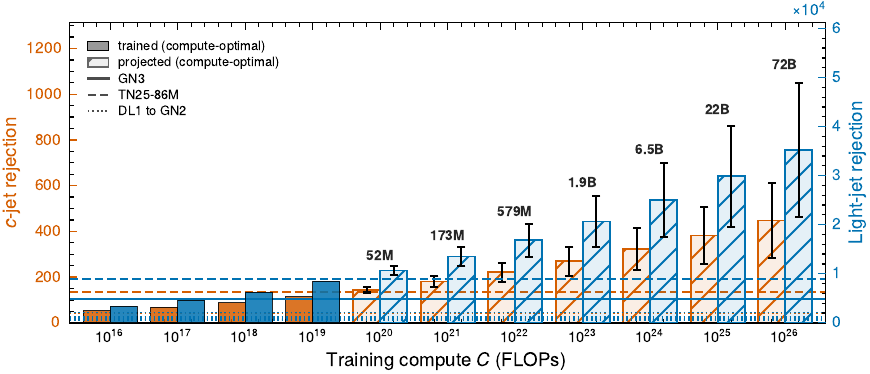}
    \caption{Light-jet rejection (blue) and $c$-jet rejection (orange) of the deployed ATLAS flavor taggers (DL1 to GN3), the TN25-86M model, and the compute optimal models. Error bars are the rejection-fit and $L^\star(C)$ uncertainties. Evaluated at $70\%$ global $b$-jet efficiency, $20 < \pt < 250$~GeV, $|\eta| < 2.5$ on a $t\bar{t}$ sample.}
    \label{fig:atlas_evolution_backup}
\end{figure}

\cref{fig:atlas_evolution_backup} shows the background rejections for the deployed ATLAS models up to GN3, TN25-86M, and the scaling-law projections for a compute-optimal model trained on the full set of inputs and tasks.

\section{Auxiliary-tasks scaling}
\label{app:aux_scaling}

The per-task scaling laws of the four auxiliary heads, together with the total loss, are collected here. Each descends along its own envelope toward its own irreducible floor, with the direct envelope fit annotated on the panels.

\begin{figure}[!htbp]
    \centering
    \includegraphics[width=0.9\textwidth]{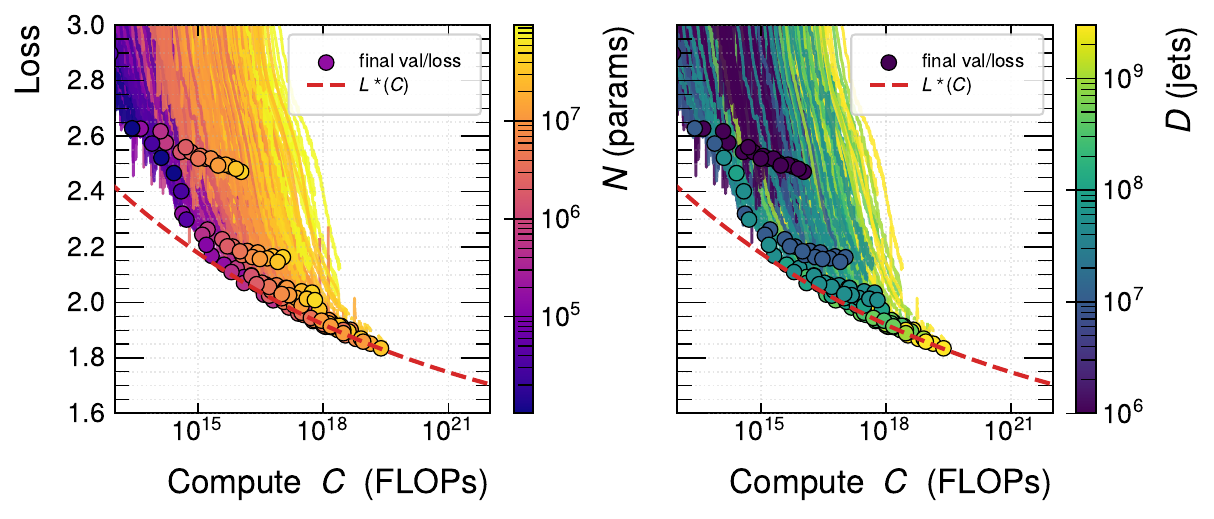}
    \caption{All trained multi-task runs: training loss against cumulative compute $C$. The same curves are colored by model size $N$ \textbf{(left)} and by dataset size $D$ \textbf{(right)}; circles mark the final per-cell validation loss. The shared lower envelope is the empirical compute-optimal frontier $L^\star(C)$ that Approaches 1--3 estimate.}
    \label{fig:loss_vs_compute_all}
\end{figure}

\begin{figure}[!htbp]
    \centering
    \includegraphics[width=0.9\textwidth]{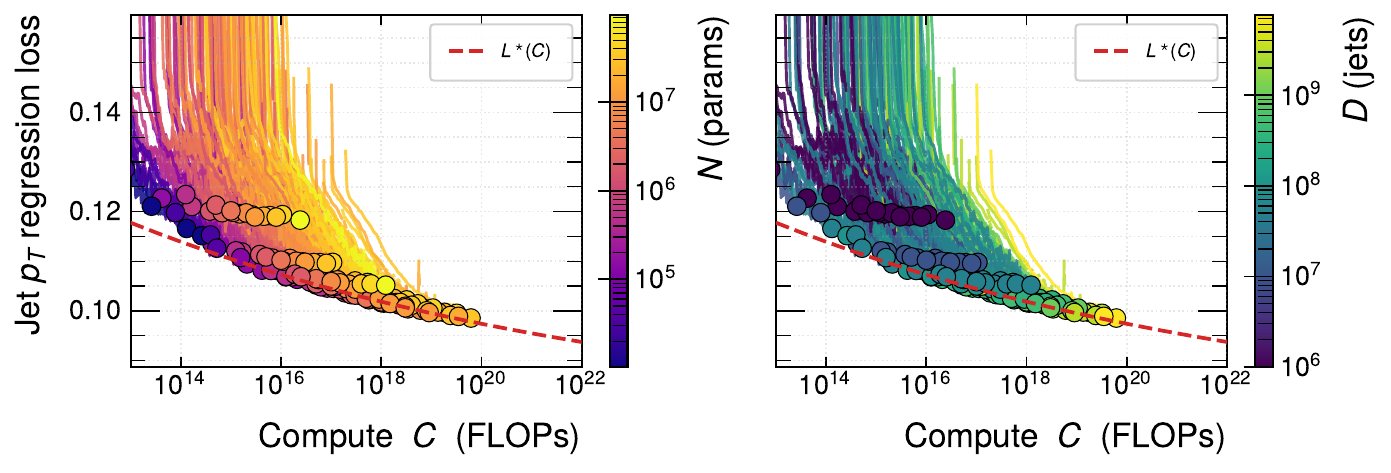}
    \caption{Jet-$\pt$ regression loss against compute. The same curves are colored by model size $N$ \textbf{(left)} and by dataset size $D$ \textbf{(right)}; circles mark the final per-cell validation loss.}
    \label{fig:task_jet_pt_regression}
\end{figure}

\begin{figure}[!htbp]
    \centering
    \includegraphics[width=0.9\textwidth]{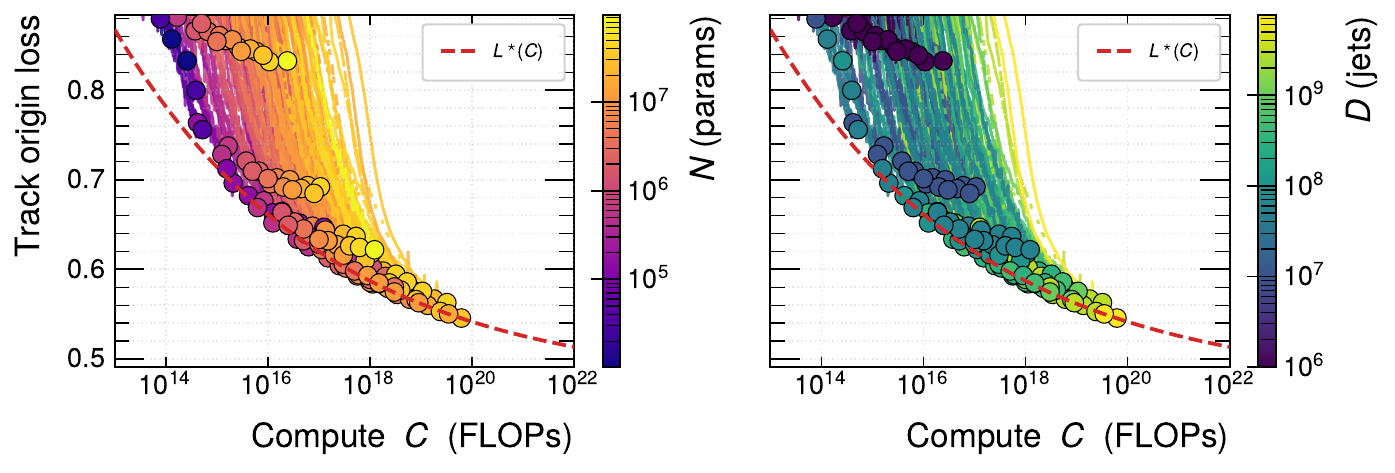}
    \caption{Track-origin loss against compute. The same curves are colored by model size $N$ \textbf{(left)} and by dataset size $D$ \textbf{(right)}; circles mark the final per-cell validation loss.}
    \label{fig:task_track_origin}
\end{figure}

\begin{figure}[!htbp]
    \centering
    \includegraphics[width=0.9\textwidth]{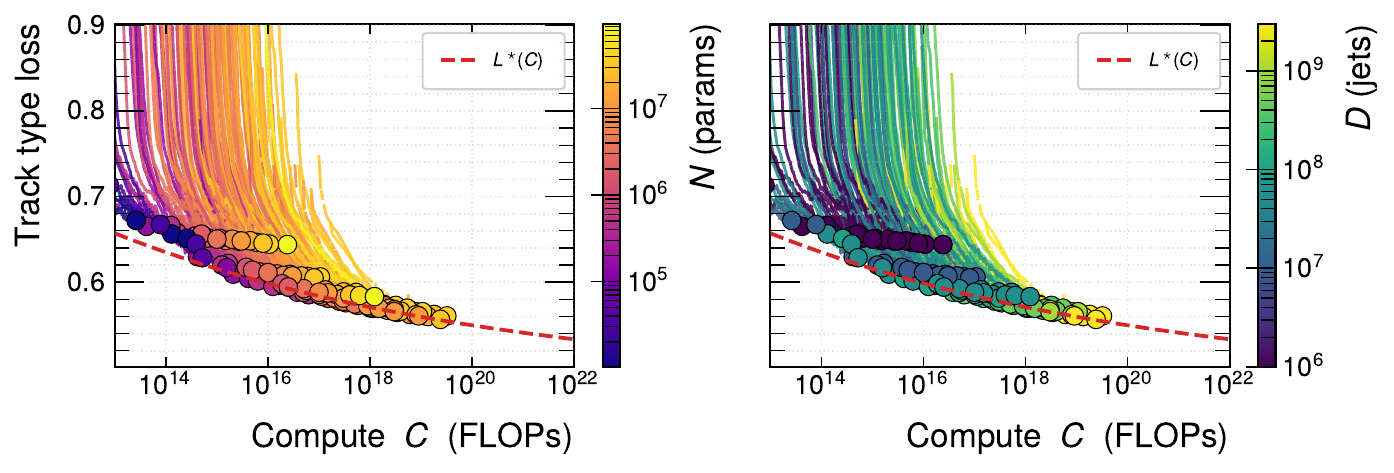}
    \caption{Track-type loss against compute. The same curves are colored by model size $N$ \textbf{(left)} and by dataset size $D$ \textbf{(right)}; circles mark the final per-cell validation loss.}
    \label{fig:task_track_type}
\end{figure}

\begin{figure}[!htbp]
    \centering
    \includegraphics[width=0.9\textwidth]{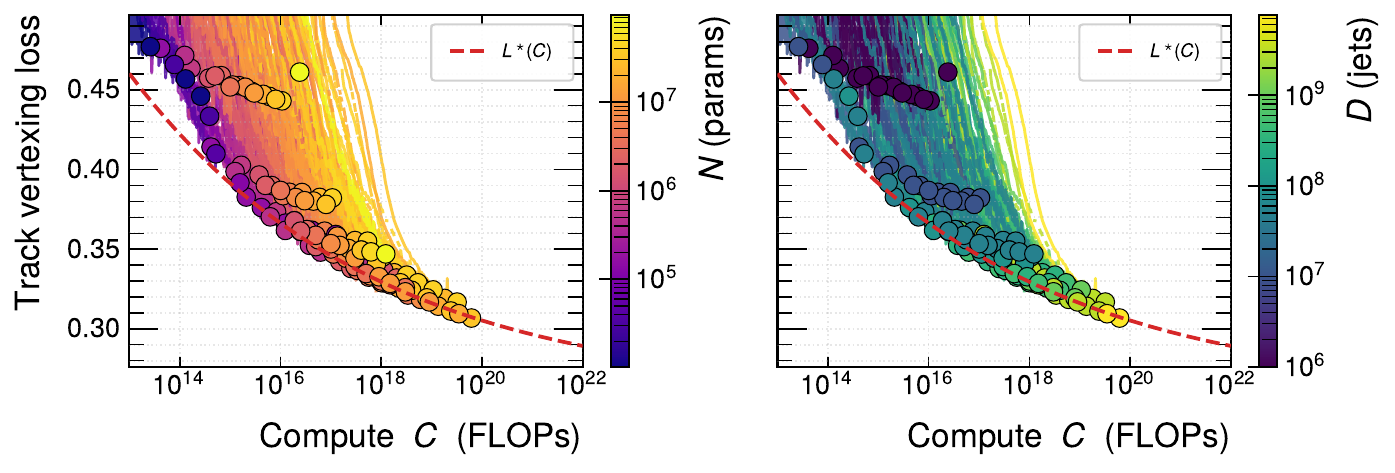}
    \caption{Track-vertexing loss against compute. The same curves are colored by model size $N$ \textbf{(left)} and by dataset size $D$ \textbf{(right)}; circles mark the final per-cell validation loss.}
    \label{fig:task_track_vertexing}
\end{figure}

\section{Single-auxiliary-task scaling}
\label{app:singleaux}

\begin{figure}[!htbp]
    \centering
    \includegraphics[width=0.75\textwidth]{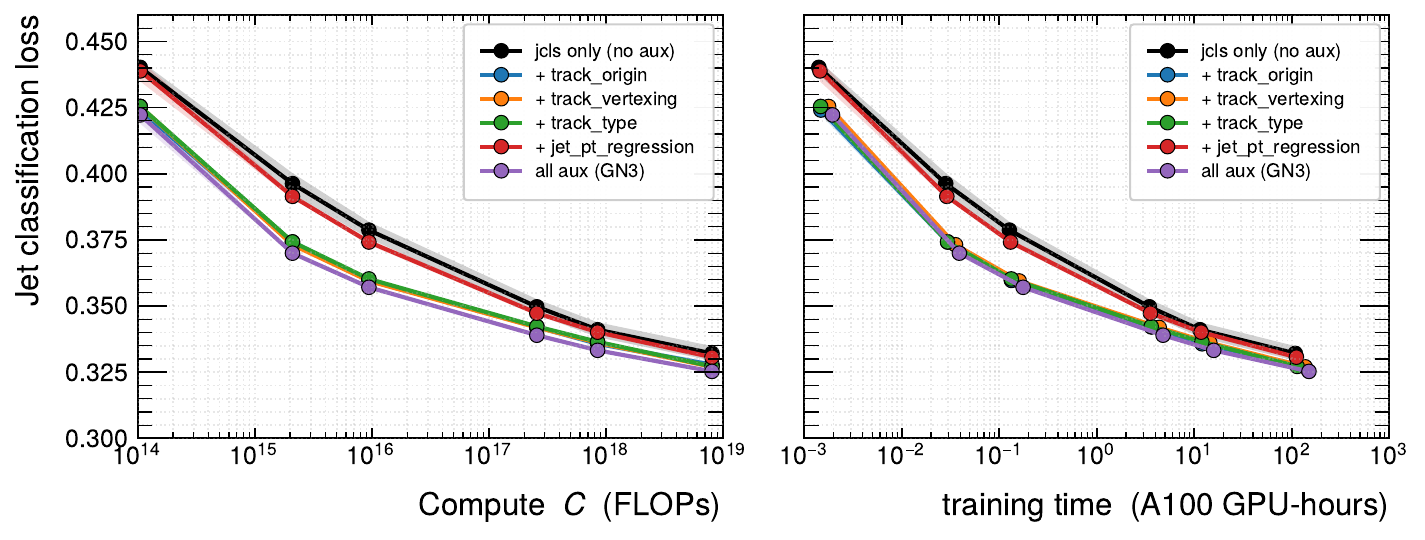}
    \caption{Single-auxiliary-task ablation: compute-optimal jet-classification loss for the classification-only baseline, for classification plus each single auxiliary head, and for all five together, against compute $C = 6NT$ \textbf{(left)} and A100 GPU-hours \textbf{(right)}. Bands are the seed-to-seed spread. The three track-level heads (origin, type, vertexing) each recover most of the full multi-task gain on their own, whereas jet-$\pt$ regression contributes least; per GPU-hour the ordering tightens as the throughput cost of the extra head is folded in.}
    \label{fig:singleaux}
\end{figure}

To identify which auxiliary heads drive the multi-task gain, we repeat the comparison with jet classification plus exactly one auxiliary task at a time, alongside the classification-only and full multi-task baselines. \cref{fig:singleaux} shows the resulting jet-classification loss both against compute and against A100 GPU-hours. Each auxiliary head improves on the classification-only baseline, although none individually recovers the full multi-task gain, indicating that the benefits are partly complementary. The three track-level heads provide the largest individual gains, while jet-$\pt$ regression is the weakest contributor. Accounting for wall-clock cost narrows these differences, but the full multi-task model remains competitive with or better than every single-auxiliary variant across the studied range.

\section{Full $\mu$P sweep}
\label{app:full_sweep}

\begin{figure}[!htbp]
    \centering
    \includegraphics[width=0.8\textwidth]{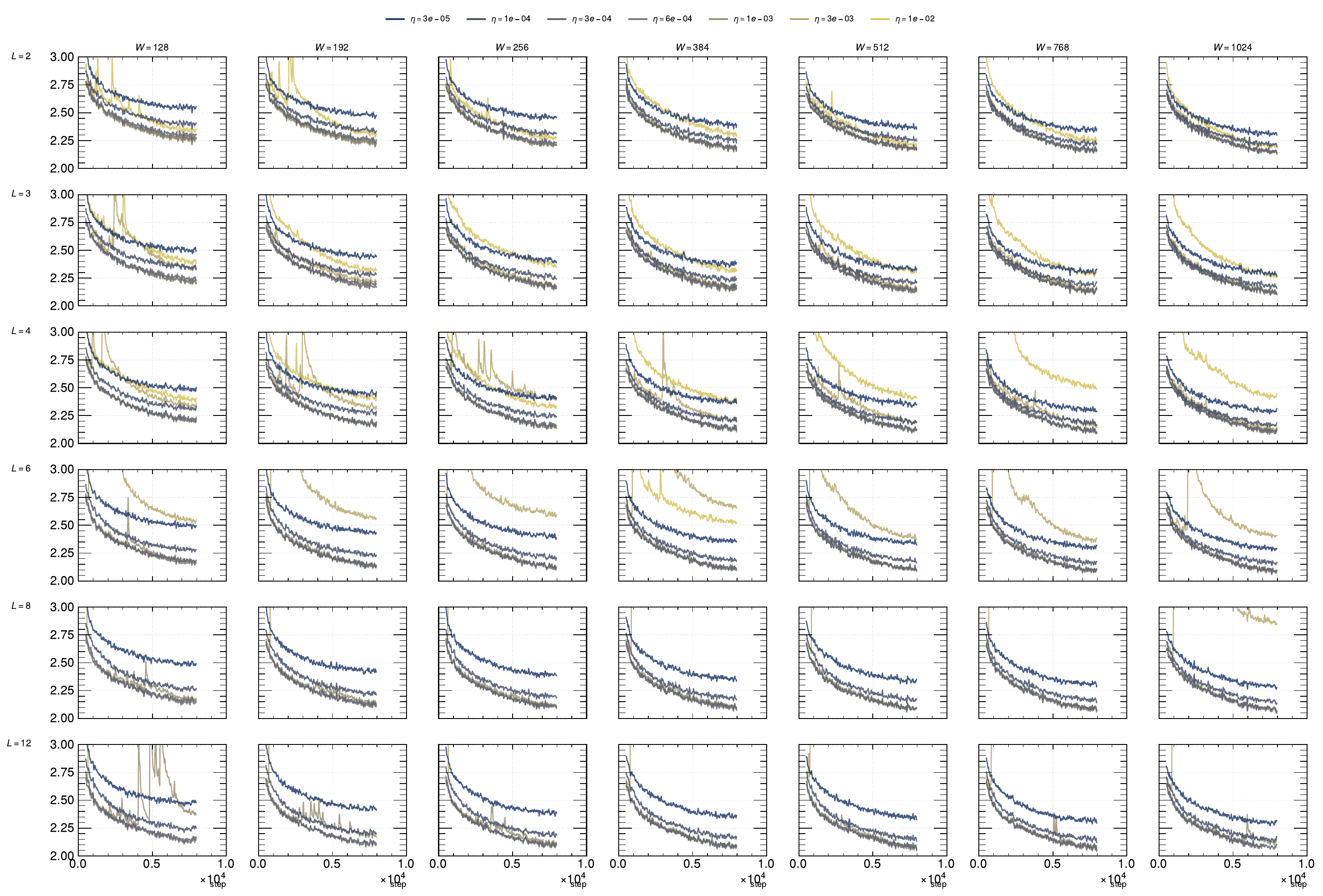}
    \caption{Training loss against optimizer step for every $(w, l)$ cell of the $\mu$P sweep (rows: depth $l$; columns: width $w$), with one curve per peak learning rate $\eta$ (color). Anchors $(b, T) = (500, 2{\times}10^7)$.}
    \label{fig:big_sweep_curves_combined}
\end{figure}

\begin{figure}[!htbp]
    \centering
    \includegraphics[width=0.7\textwidth]{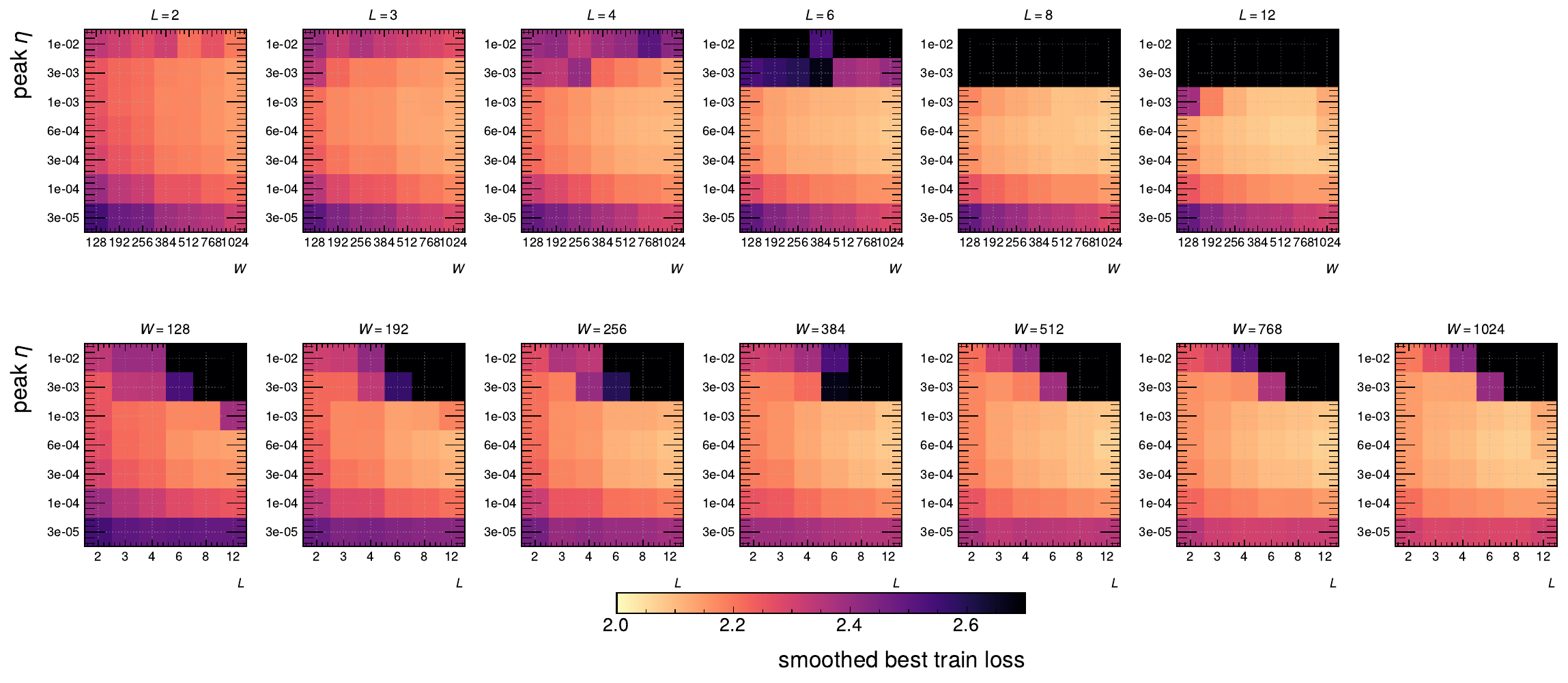}
    \caption{Smoothed best training loss across the $\mu$P sweep. \textbf{Top:} over peak $\eta$ and width $w$ at each depth $l$; \textbf{bottom:} over peak $\eta$ and depth $l$ at each width $w$. Anchors $(b, T) = (500, 2{\times}10^7)$.}
    \label{fig:big_sweep_heatmaps_combined}
\end{figure}

\cref{fig:big_sweep_curves_combined} shows the training loss against step for every learning rate at each $(w, l)$ cell, and \cref{fig:big_sweep_heatmaps_combined} the corresponding best-loss heatmaps over learning rate and model shape.

\begin{figure}[!htbp]
    \centering
    \includegraphics[width=0.3\textwidth]{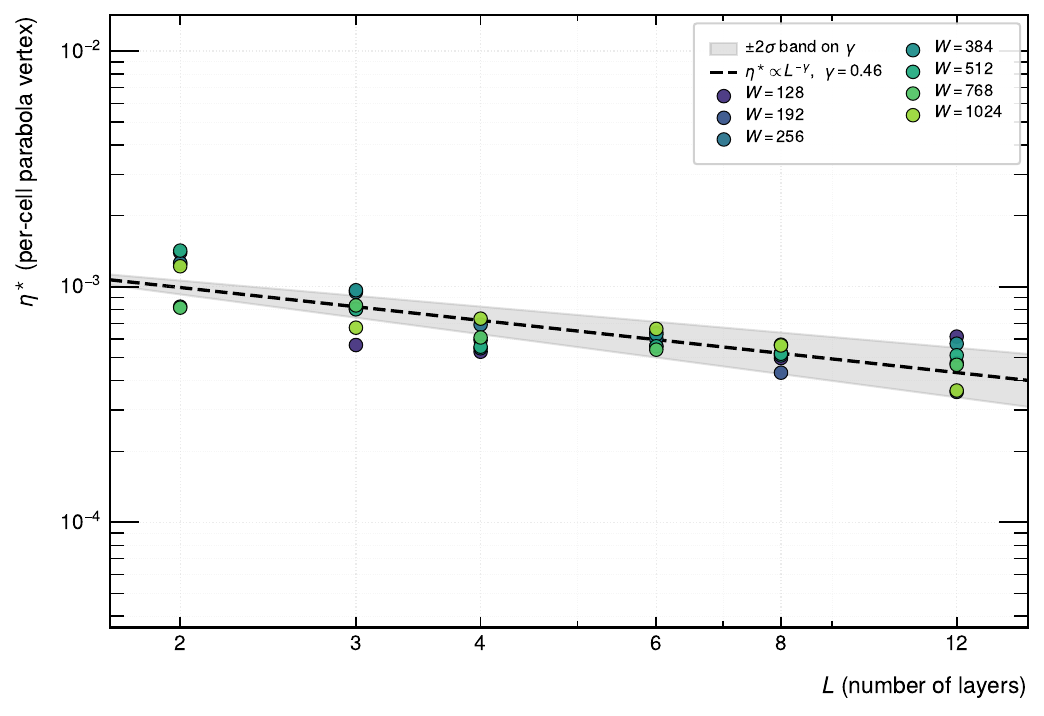}
    \caption{Fit of the depth exponent $\gamma_l = 0.464 \pm 0.049$ in $\etastar(l) = (\Lref/l)^{\gamma_l}\,\etastar(\Lref)$; after applying this multiplier the effective $\etastar$ reduces to a single value across the studied $l$ range.}
    \label{fig:big_sweep:gamma}
\end{figure}

\section{Per-task hyperparameter optima}
\label{app:pertask_hp}

\begin{figure}[!htbp]
    \centering
    \includegraphics[width=0.95\textwidth]{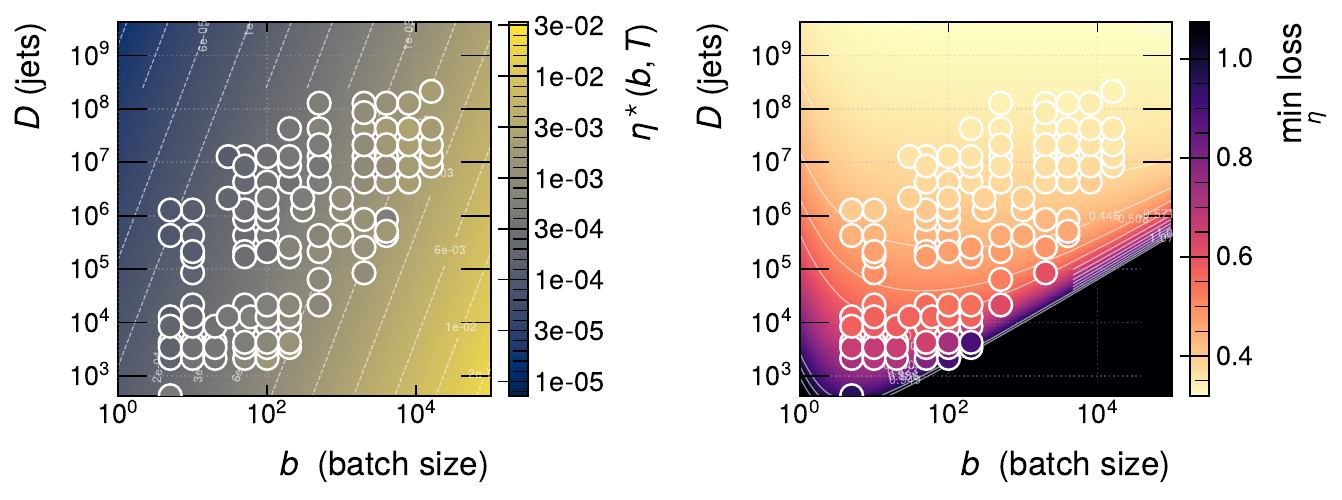}
    \caption{Per-task $(b, T)$ landscape for the jet-classification head: \textbf{(left)} the optimal learning rate $\etastar(b, T)$ with iso-$\etastar$ contours from the multiplicative fit, and \textbf{(right)} the loss at that optimum, $\min_\eta L$, with iso-loss contours; white circles mark the swept cells.}
    \label{fig:eta_loss_jcls}
\end{figure}

\begin{figure}[!htbp]
    \centering
    \includegraphics[width=0.95\textwidth]{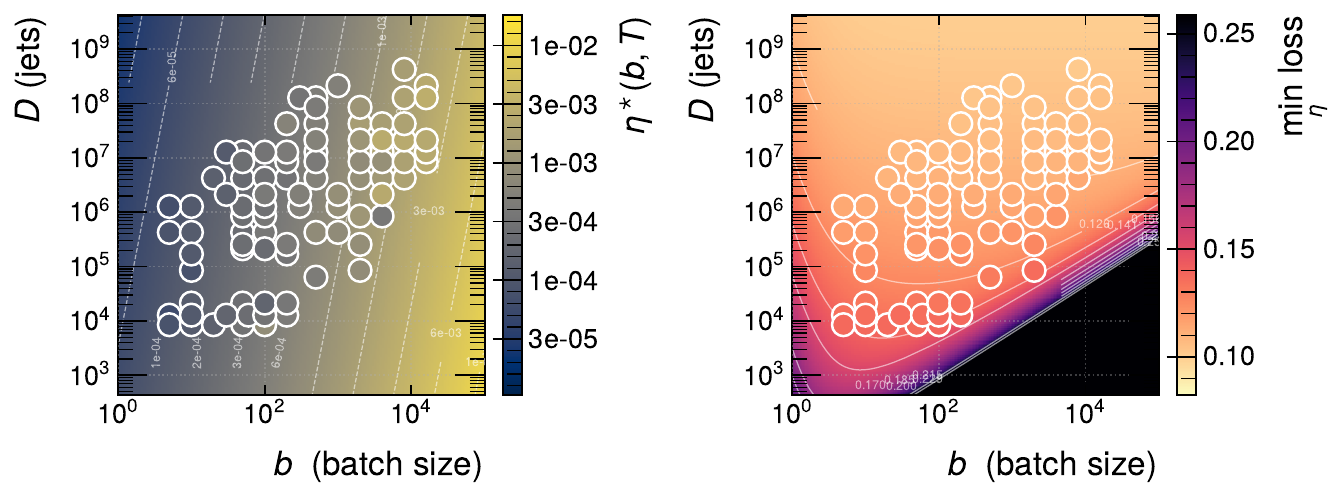}
    \caption{Per-task $(b, T)$ landscape for the $\pt$ regression head: \textbf{(left)} the optimal learning rate $\etastar(b, T)$ with iso-$\etastar$ contours from the multiplicative fit, and \textbf{(right)} the loss at that optimum, $\min_\eta L$, with iso-loss contours; white circles mark the swept cells.}
    \label{fig:eta_loss_ptreg}
\end{figure}

\begin{figure}[!htbp]
    \centering
    \includegraphics[width=0.95\textwidth]{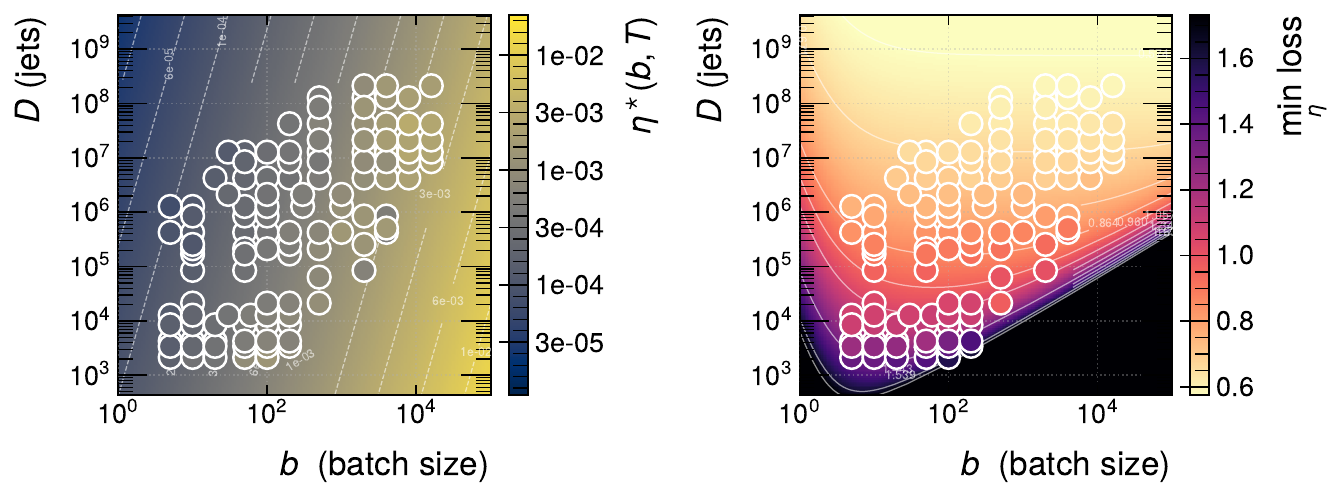}
    \caption{Per-task $(b, T)$ landscape for the track-origin head: \textbf{(left)} the optimal learning rate $\etastar(b, T)$ with iso-$\etastar$ contours from the multiplicative fit, and \textbf{(right)} the loss at that optimum, $\min_\eta L$, with iso-loss contours; white circles mark the swept cells.}
    \label{fig:eta_loss_torigin}
\end{figure}

\begin{figure}[!htbp]
    \centering
    \includegraphics[width=0.95\textwidth]{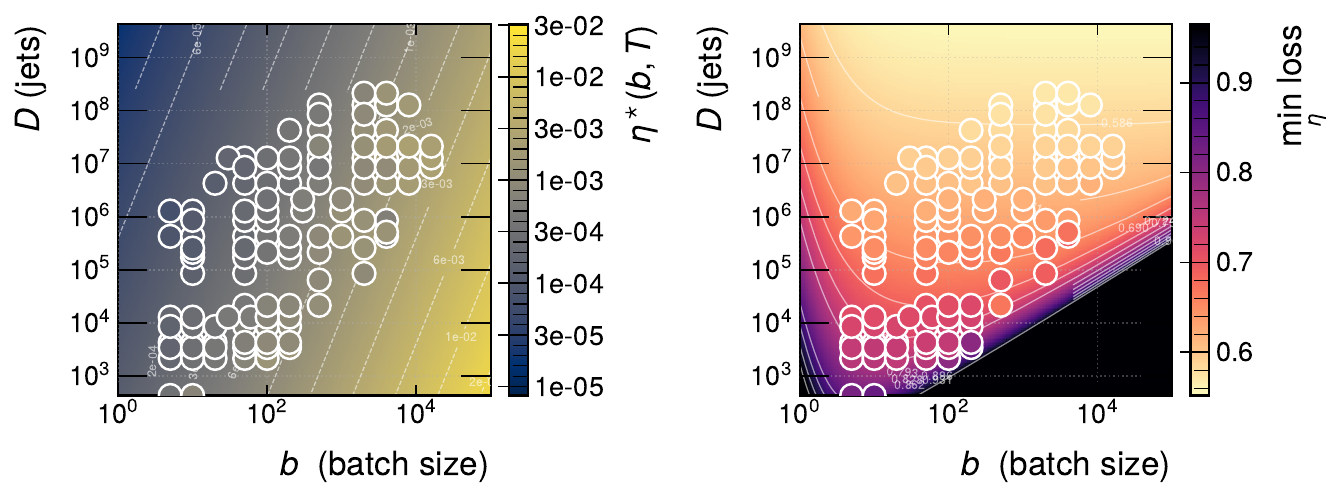}
    \caption{Per-task $(b, T)$ landscape for the track-type head: \textbf{(left)} the optimal learning rate $\etastar(b, T)$ with iso-$\etastar$ contours from the multiplicative fit, and \textbf{(right)} the loss at that optimum, $\min_\eta L$, with iso-loss contours; white circles mark the swept cells.}
    \label{fig:eta_loss_ttype}
\end{figure}

\begin{figure}[!htbp]
    \centering
    \includegraphics[width=0.95\textwidth]{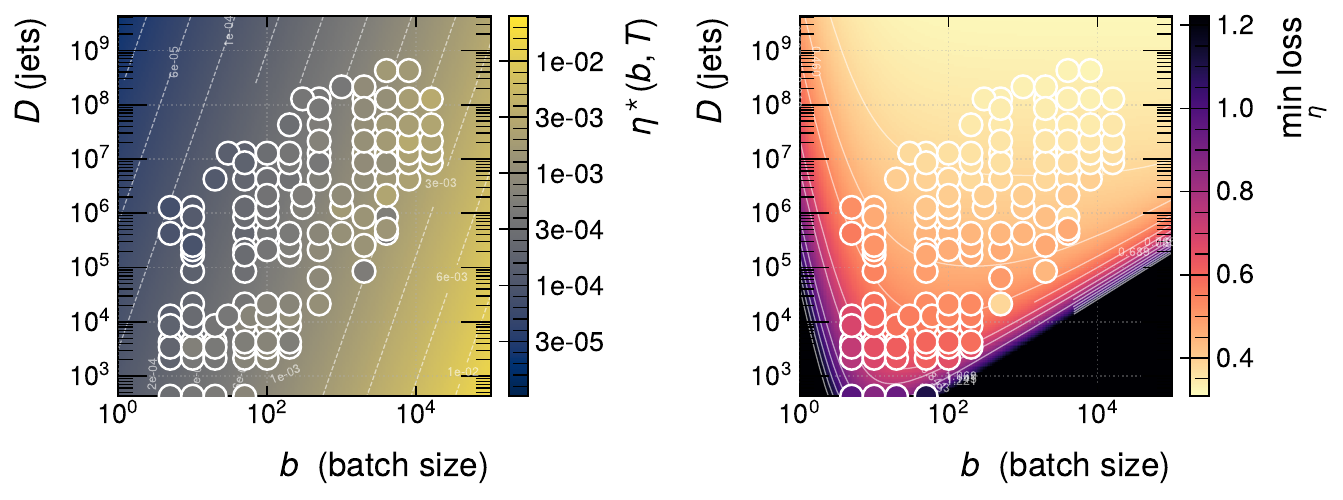}
    \caption{Per-task $(b, T)$ landscape for the track-vertexing head: \textbf{(left)} the optimal learning rate $\etastar(b, T)$ with iso-$\etastar$ contours from the multiplicative fit, and \textbf{(right)} the loss at that optimum, $\min_\eta L$, with iso-loss contours; white circles mark the swept cells.}
    \label{fig:eta_loss_tvtx}
\end{figure}

The multiplicative recipe of \cref{sec:multiplicative} is calibrated independently for each objective. \cref{tab:eta-fit-full} gives the full per-head $\etastar$ calibration, extending the main-text \cref{tab:eta-fit} with the five individual multi-task heads. \cref{fig:eta_loss_jcls,fig:eta_loss_ptreg,fig:eta_loss_torigin,fig:eta_loss_ttype,fig:eta_loss_tvtx} show the per-task $(b, T)$ landscapes. The fitted $\bstar(T)$ exponent differs across tasks, so each head has its own $\bstar$ growth, which is why optimizing for a single component differs from optimizing the summed loss.

\begin{table*}[]
  \centering
  \caption{Full per-head multiplicative $\etastar$ calibration, extending \cref{tab:eta-fit} with the five individual multi-task heads. Same functional form and anchors ($b_\mathrm{ref}=500$, $T_\mathrm{ref}=2\times10^{7}$ jets) as \cref{tab:eta-fit}.}
  \label{tab:eta-fit-full}
  \begin{tabular}{l c c c}
    \toprule
    optimization target & $\etabase\,[10^{-4}]$ & $\kappa_\mathrm{eff}$ & $\lambda_\mathrm{eff}$ \\
    \midrule
    Total (multi-task)            & $6.05 \pm 0.18$ & $+0.429 \pm 0.014$ & $+0.098 \pm 0.009$ \\
    \midrule
    \quad jet classification      & $5.68 \pm 0.20$ & $+0.453 \pm 0.014$ & $+0.125 \pm 0.010$ \\
    \quad jet $\pt$ regression    & $5.54 \pm 0.26$ & $+0.450 \pm 0.021$ & $+0.065 \pm 0.014$ \\
    \quad track origin            & $5.91 \pm 0.21$ & $+0.430 \pm 0.014$ & $+0.090 \pm 0.010$ \\
    \quad track type              & $5.81 \pm 0.20$ & $+0.438 \pm 0.014$ & $+0.125 \pm 0.009$ \\
    \quad track vertexing         & $5.85 \pm 0.23$ & $+0.422 \pm 0.017$ & $+0.108 \pm 0.011$ \\
    \midrule
    Jet-classification-only       & $2.91 \pm 0.16$ & $+0.548 \pm 0.025$ & $+0.174 \pm 0.019$ \\
    \bottomrule
  \end{tabular}
\end{table*}

% =============================================================================
\section{FLOP estimate validation}
\label{app:flops}

\begin{figure}[!htbp]
    \centering
    \includegraphics[width=0.75\textwidth]{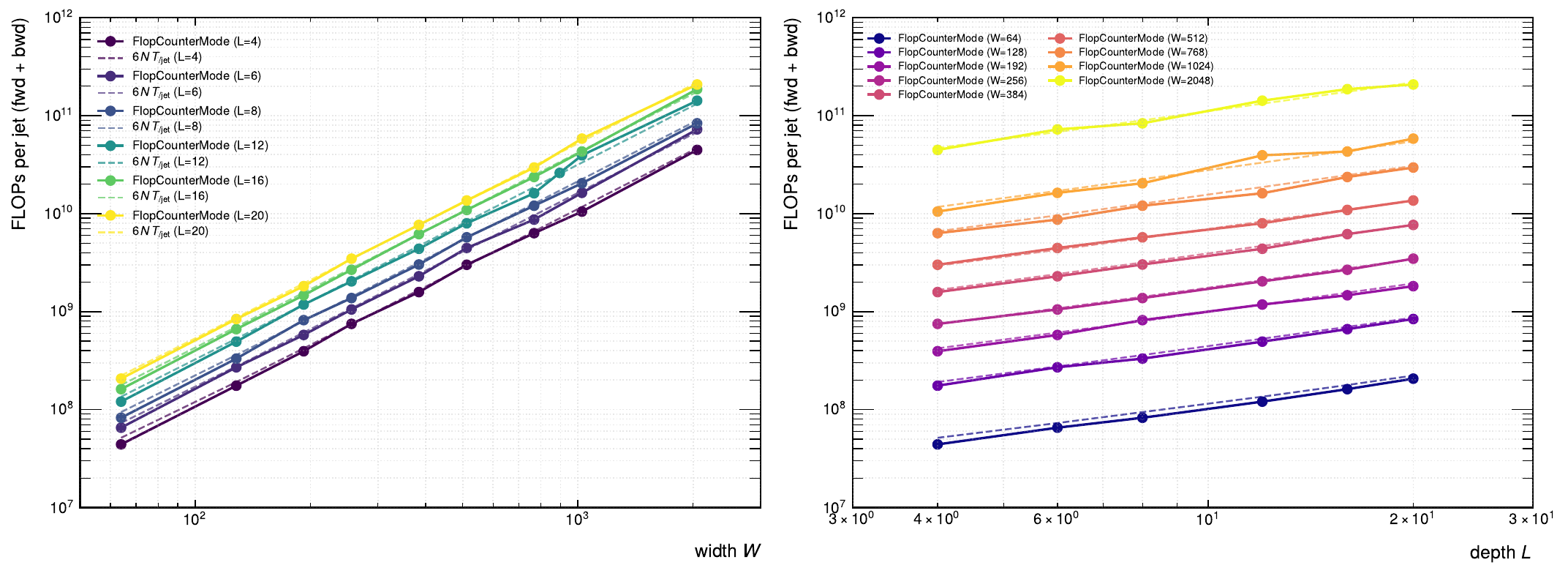}
    \caption{Forward-plus-backward FLOPs per jet measured with \texttt{FlopCounterMode} (solid) against width $w$ at fixed depth \textbf{(left)} and against depth $l$ at fixed width \textbf{(right)}, compared with the per-jet estimate $6\,N\,T_{\mathrm{jet}} \approx 256\,N$ of \cref{eq:compute} (dashed). The measurement follows the estimate across the model family, confirming that $C \approx 6\,N\,T$ is an adequate compute proxy for the architectures studied here.}
    \label{fig:flops_per_jet}
\end{figure}

The estimate $C \approx 6\,N\,T$ of \cref{eq:compute} assumes that the per-token cost is dominated by the parameter-count term and is otherwise independent of model shape. Per jet, it predicts $6\,N\,T_{\mathrm{jet}} \approx 256\,N$ FLOPs, where $T_{\mathrm{jet}} \approx 42.6$ is the mean number of constituents per jet. \cref{fig:flops_per_jet} compares this prediction with the forward-plus-backward FLOPs per jet measured with PyTorch's \texttt{FlopCounterMode} across the $(w, l)$ grid. The measurement tracks the estimate over more than three orders of magnitude in model size.

% =============================================================================
\section{Evaluation metrics and reweighting}
\label{sec:metrics}
% =============================================================================

This appendix gives the full argument behind treating the loss, rather than any single rejection number, as the performance metric of the trained tagger and shows how strongly a quoted rejection depends on the discriminant fractions (\cref{eq:discriminant}) and working point.

\paragraph{The flavor discriminant as a projection.} The scalar flavor discriminant $D_b$ of \cref{eq:discriminant} is built from the per-flavor scores $(p_b, p_c, p_\tau, p_{\text{light}})$ with the background fractions $(f_c, f_\tau, f_{\mathrm{light}}=1-f_c- f_\tau)$ set by hand. The per-flavor scores are themselves a projection of the encoder's latent representation, and \cref{eq:discriminant} projects them further down onto a single axis. A deployed tagger is a frozen predictor used by many downstream analyses, so its output must be well calibrated against data, which is in practice done for a fixed low-dimensional discriminant evaluated at a few \emph{working points}. A working point is a cut $D_b > t$ on the discriminant: the threshold $t$ sets the $b$-jet efficiency $\epsilon_b$, the fraction of true $b$-jets kept, and for each background flavor $f$ a mis-tag efficiency $\epsilon_f$, whose inverse $1/\epsilon_f$ is the background rejection. The price is that this scalar bottleneck fixes an arbitrary projection, biased toward whichever background the chosen $(f_c, f_\tau)$ favor.

\begin{figure}[!htbp]
    \centering
    \includegraphics[width=0.6\textwidth]{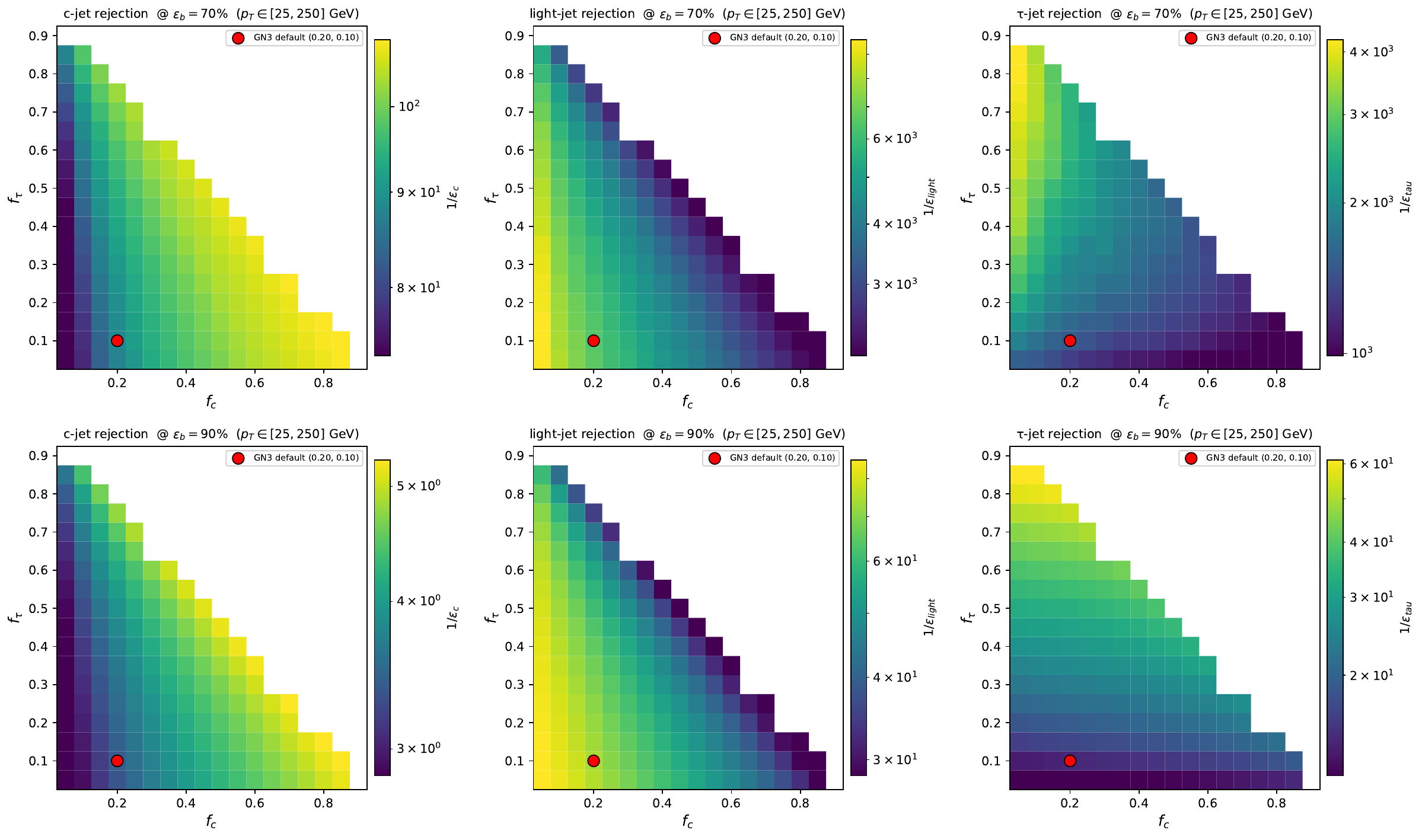}
    \caption{Background rejection at fixed $b$-jet efficiency as a function of the discriminant fractions $(f_c, f_\tau)$ of \cref{eq:discriminant}, for $c$-jets \textbf{(left)}, light-jets \textbf{(middle)}, and $\tau$-jets \textbf{(right)}, at \emph{global} $\epsilon_b = 70\%$ \textbf{(top)} and $90\%$ \textbf{(bottom)}, in $\pt \in [25, 250]$~GeV. The rejection varies by an order of magnitude or more across the plane and the three flavors trade off against one another. GN3 uses $(f_c, f_\tau) = (0.20, 0.10)$ (red circle); the fractions maximizing a given rejection differ per flavor, and the optimal choice for an analysis is set by its background composition through the Neyman--Pearson lemma. The reported rejection is thus a property of the chosen projection.}
    \label{fig:discriminant_scan}
\end{figure}

\paragraph{Dependence of the rejection on the fractions.} Because $(f_c, f_\tau)$ are free, the rejection quoted at a given $b$-jet efficiency varies by large factors across reasonable choices. \cref{fig:discriminant_scan} scans $(f_c, f_\tau)$ at fixed global efficiency: $c$-, light-, and $\tau$-jet rejection move in opposite directions and by an order of magnitude or more across the plane. GN3 adopts $(f_c, f_\tau) = (0.20, 0.10)$. No single choice is optimal for all three flavors at once, and the genuinely optimal fractions for a given analysis follow from its background composition through the Neyman--Pearson lemma. The same point trivially explains part of the gains seen when a downstream analysis is given the full latent space rather than a fixed discriminant threshold~\cite{Vigl_2024}: in that sense the tagger acts as a foundation model whose representation can be projected, or fine-tuned, optimally for the downstream task at hand, provided that representation is itself calibrated against data~\cite{Algren:2025mindthegap}.

\paragraph{$\pt,\eta$-dependence.} Calibration requirements also fix the working points, a discrete set of global $b$-jet efficiencies: for a fixed discriminant the threshold that yields a given efficiency drifts with $\pt$ and $\eta$, so a global $70\%$ maps to local, in-bin efficiencies that vary across $\pt$ and $\eta$.

\begin{figure}[!htbp]
    \centering
    \includegraphics[width=0.6\textwidth]{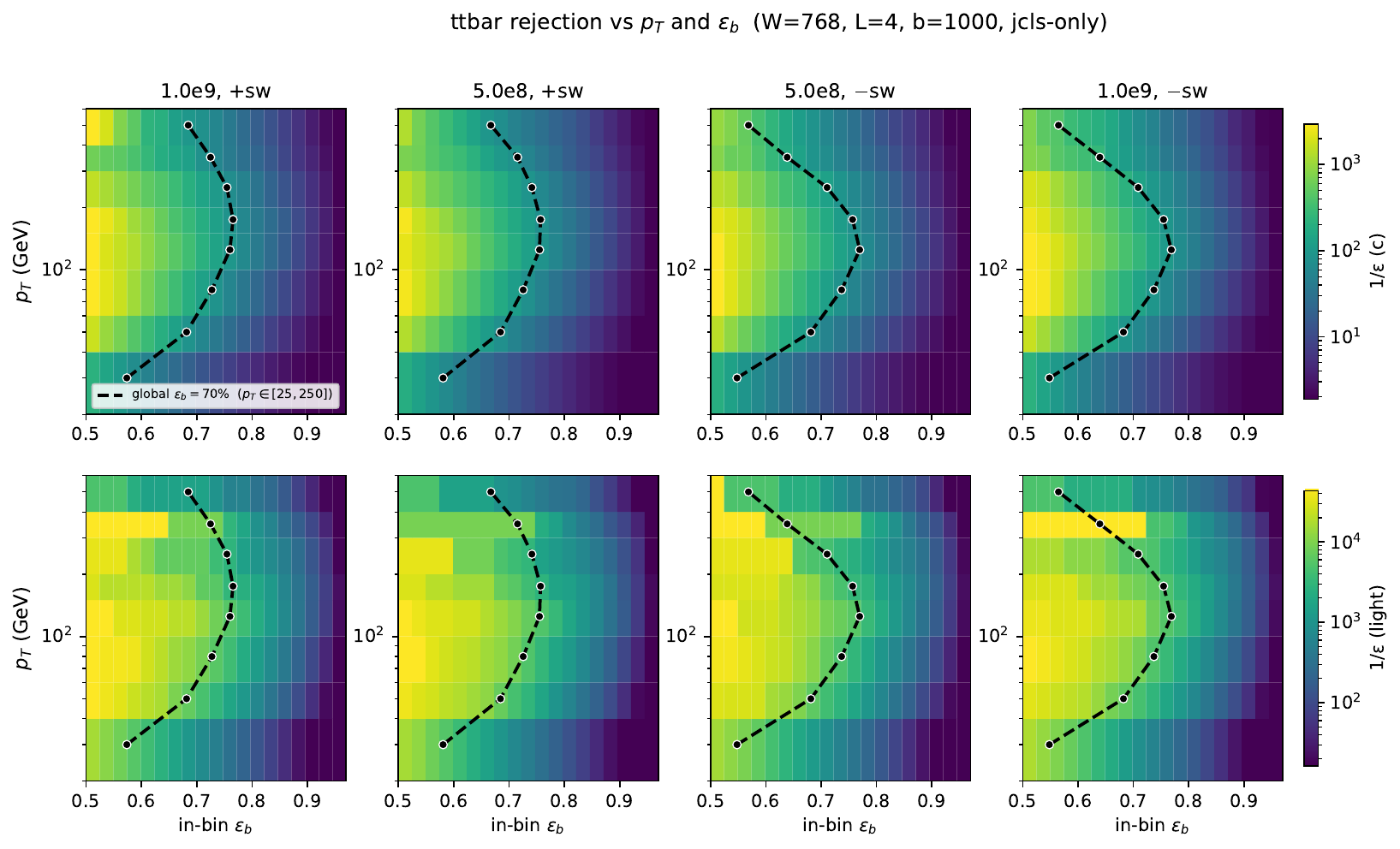}
    \caption{Background rejection $1/\epsilon$ against jet $\pt$ and in-bin $b$-jet efficiency $\epsilon_b$, for $c$-jets \textbf{(top)} and light-jets \textbf{(bottom)}, for a $w{=}768$, $l{=}4$ classification-only model trained at four settings: $10^9$ and $5\times10^8$ jets with ($+$sw) and without ($-$sw) sample weights. The discriminant uses the default GN3 fractions $(f_c, f_\tau) = (0.20, 0.10)$. At fixed in-bin efficiency the rejection landscape is nearly identical with and without sample weights; the dashed line, a global $\epsilon_b = 70\%$ over $\pt \in [25, 250]$~GeV, cuts across very different in-bin efficiencies and rejections.}
    \label{fig:sw_rej}
\end{figure}

\begin{figure}[!htbp]
    \centering
    \includegraphics[width=0.75\textwidth]{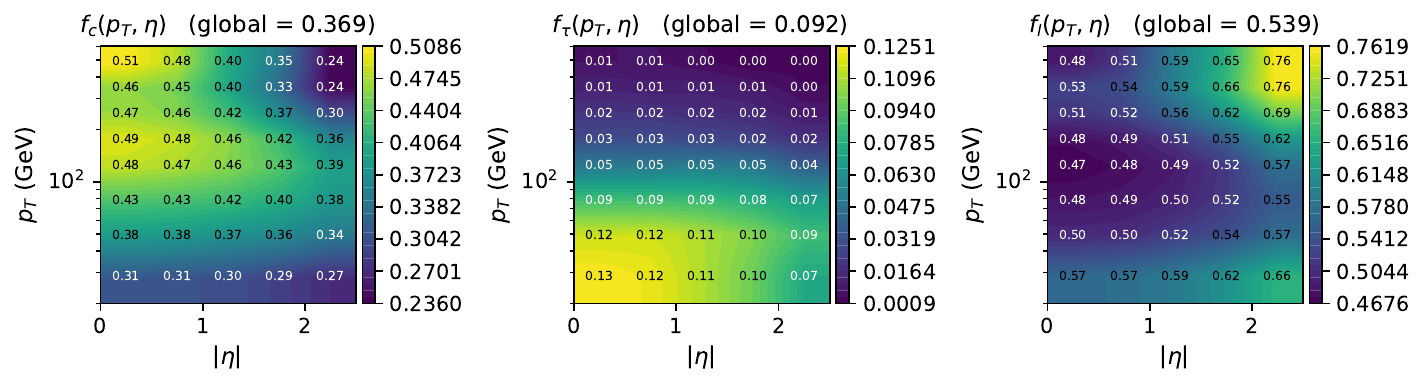}
    \caption{Background composition among non-$b$ jets of the $t\bar{t}$ test sample on the $(\pt, |\eta|)$ grid: $f_c$ \textbf{(left)}, $f_\tau$ \textbf{(middle)}, and $f_l$ \textbf{(right)}, with per-bin values annotated and the sample-global values in the titles. The composition is far from the fixed $(0.20, 0.10, 0.70)$ convention and varies strongly with kinematics, so the Neyman--Pearson-optimal fractions are per-jet, per-analysis quantities.}
    \label{fig:truthfrac_fractions}
\end{figure}

\paragraph{What a working point fixes.} Putting the two choices together, a working point is one location in the $(f_c, f_\tau)$ plane of \cref{fig:discriminant_scan}, which fixes the discriminant, together with one threshold on it, the dashed line in \cref{fig:sw_rej}, which fixes the efficiency. The rejection a deployed tagger reports is just the value read off at that one dot-and-line pair, i.e., the global efficiency and the rejection integrated along the dashed line in \cref{fig:sw_rej}. 
By the Neyman--Pearson lemma, the discriminant of \cref{eq:discriminant} separates $b$-jets from the background optimally when the $f_x$ are the prior probabilities of each background flavor in the sample being discriminated against. The right fractions are therefore the background composition of the analysis sample and, since that composition varies with kinematics, they are in general functions of $(\pt, \eta)$. \cref{fig:truthfrac_fractions} measures this composition on a dedicated $t\bar{t}$ test sample of $4.9$M jets. Every analysis, with its own background composition and kinematic selection, would prefer its own projection, which no fixed set of working points can provide.

In foundation-model terms, the right way to exploit such a network is to calibrate the latent space and give analyses full access to it, or to fine-tune the backbone, especially given the scale at which taggers are now being trained, the variety of inputs they ingest, and the multi-task objective they minimize.

\section{Input-variants scaling}
\label{app:input_variant_scaling}

\begin{figure}[!htbp]
    \centering
    \includegraphics[width=0.5\textwidth]{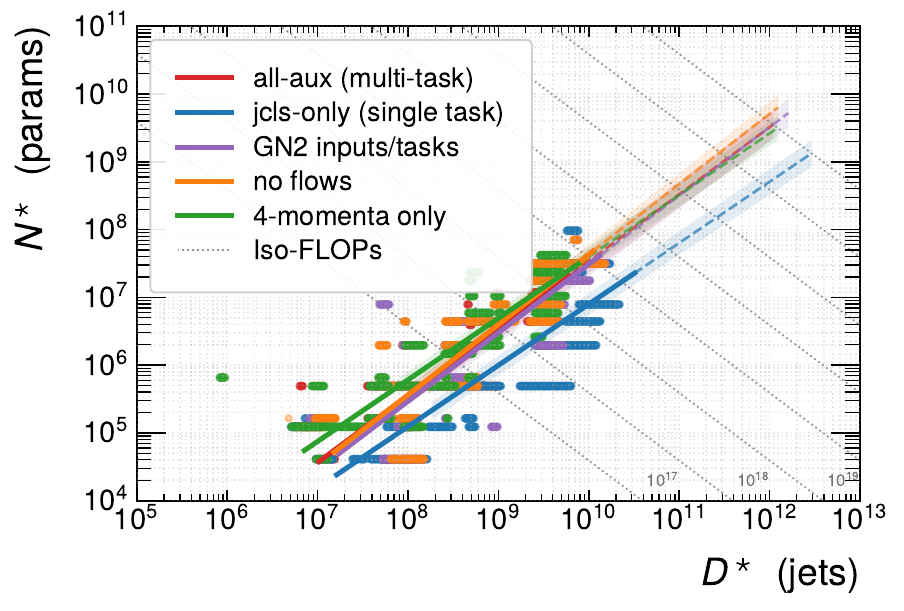}
    \caption{Compute-optimal allocation $(N^\star,D^\star)$ for five settings: two training
    objectives (all-aux, jcls-only) and three reduced input sets, with $1\sigma$ bands from the
    fitted allocation exponents. Dotted diagonals are lines of constant compute at decades
    from $C = 10^{17}$ to $10^{24}$~FLOP. The jet-classification-only setting is the clear outlier, preferring more data
    and a smaller model at every budget; the input level does not appear to shift the allocation.}
    \label{fig:nd_alloc_5settings}
\end{figure}
\begin{figure}[!htbp]
    \centering
    \includegraphics[width=0.8\textwidth]{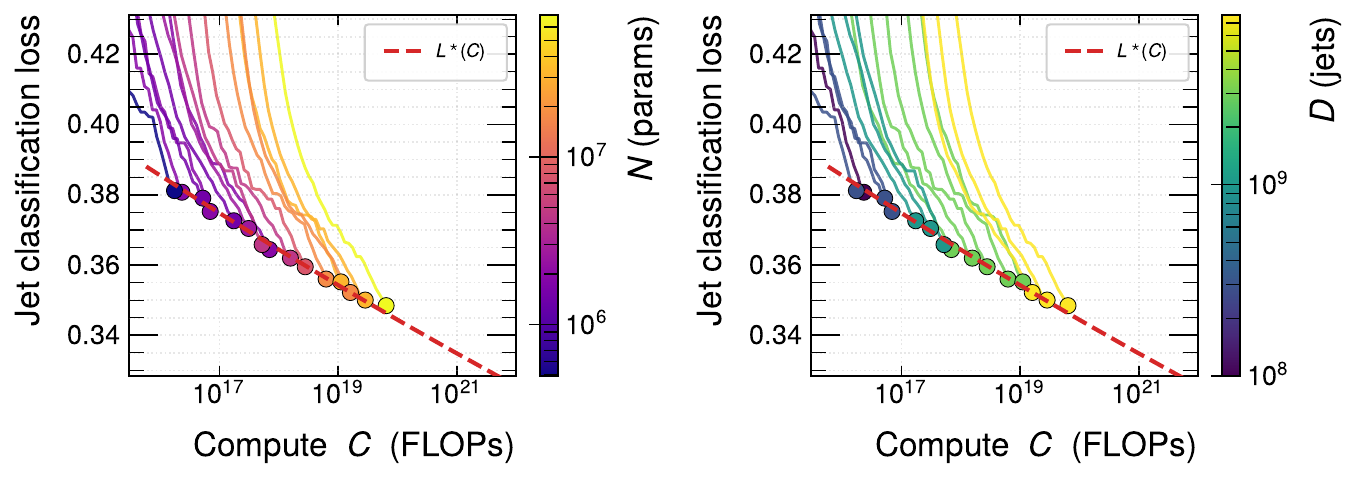}
    \caption{Jet-classification loss against compute for the GN2 input set.}
    \label{fig:variant_gn2}
\end{figure}
\begin{figure}[!htbp]
    \centering
    \includegraphics[width=0.8\textwidth]{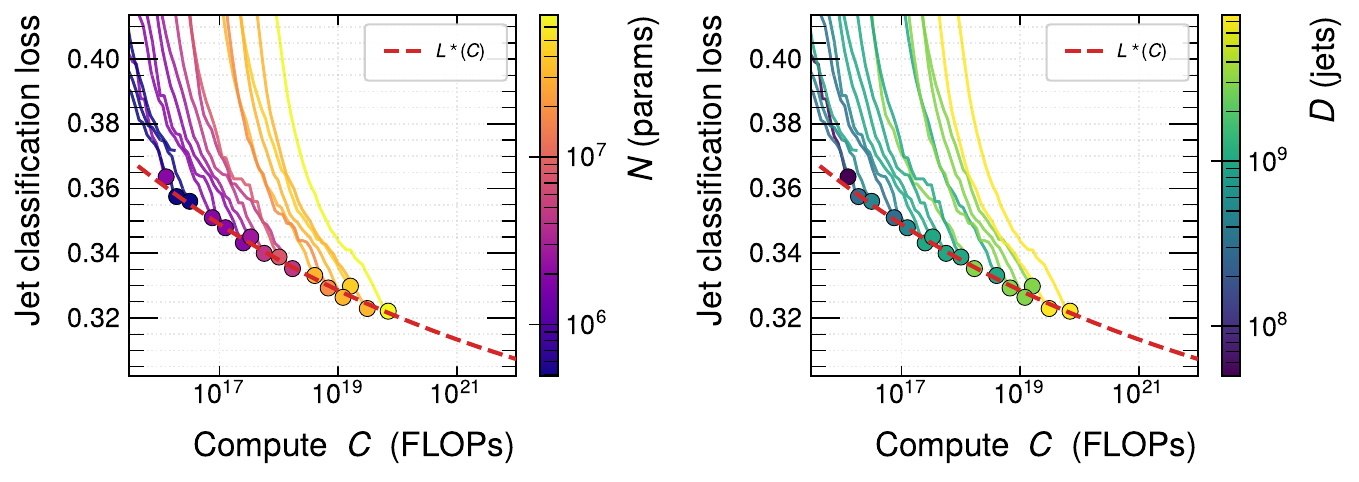}
    \caption{Jet-classification loss against compute for the no-flows input set.}
    \label{fig:variant_noflows}
\end{figure}
\begin{figure}[!htbp]
    \centering
    \includegraphics[width=0.8\textwidth]{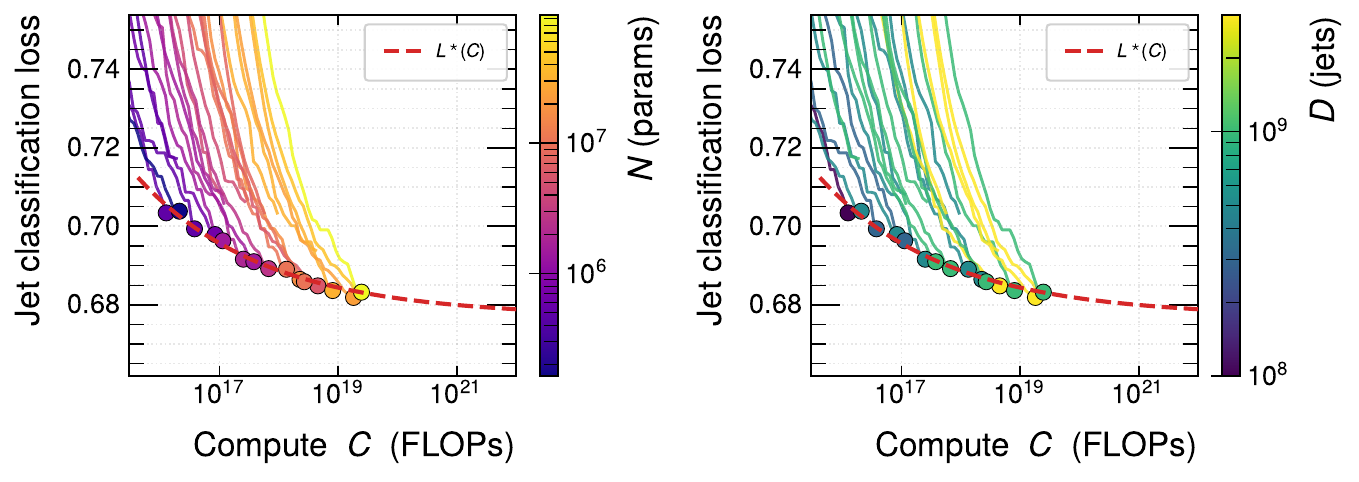}
    \caption{Jet-classification loss against compute for the four-momenta-only input set.}
    \label{fig:variant_fourmom}
\end{figure}

The per-variant jet-classification loss-versus-compute runs behind the envelope overlay of \cref{fig:variant_frontier:loss} are collected here, one panel per reduced input set: the GN2 set (\cref{fig:variant_gn2}), the no-flows variant (\cref{fig:variant_noflows}), and the four-momenta-only model (\cref{fig:variant_fourmom}).

% =============================================================================
\section{Validating the joint-optimum transfer}
\label{app:validation}
% =============================================================================

\begin{figure}[!htbp]
    \centering
    \begin{subfigure}[b]{0.296\textwidth}
        \centering
        \includegraphics[width=\linewidth]{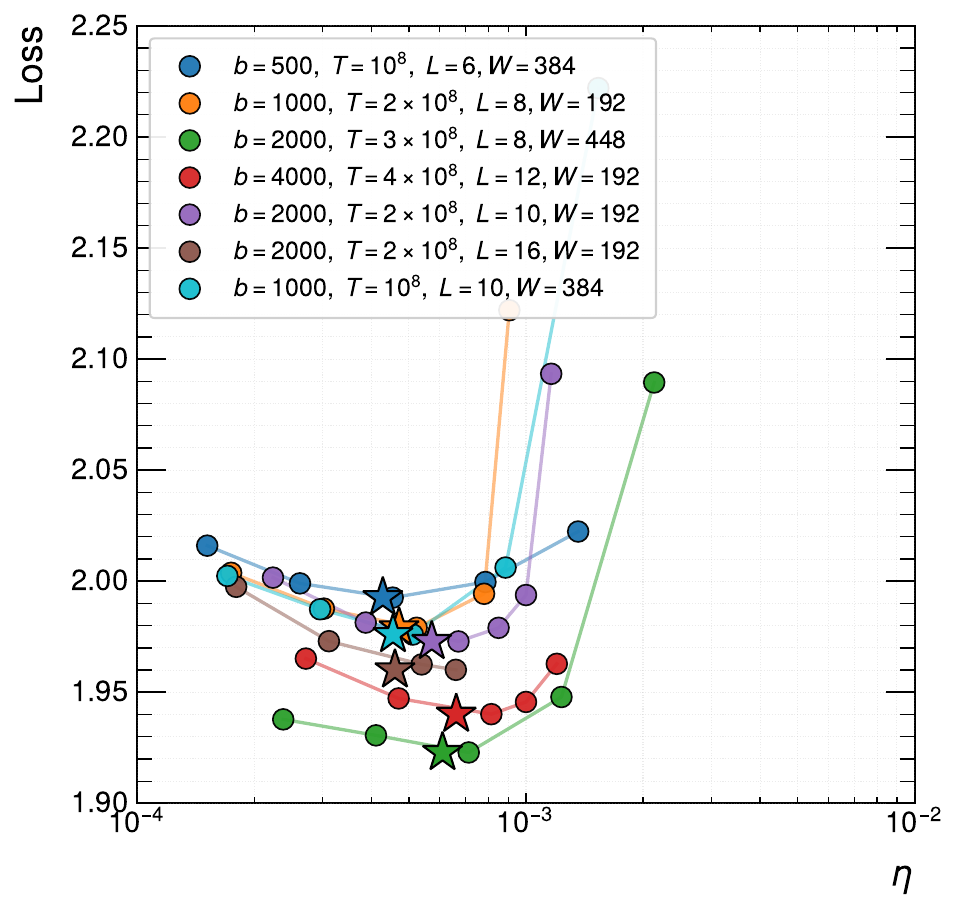}
        \caption{Predicted $\etastar$ on held-out configurations.}
        \label{fig:transfer:eta}
    \end{subfigure}\hfill
    \begin{subfigure}[b]{0.684\textwidth}
        \centering
        \includegraphics[width=\linewidth]{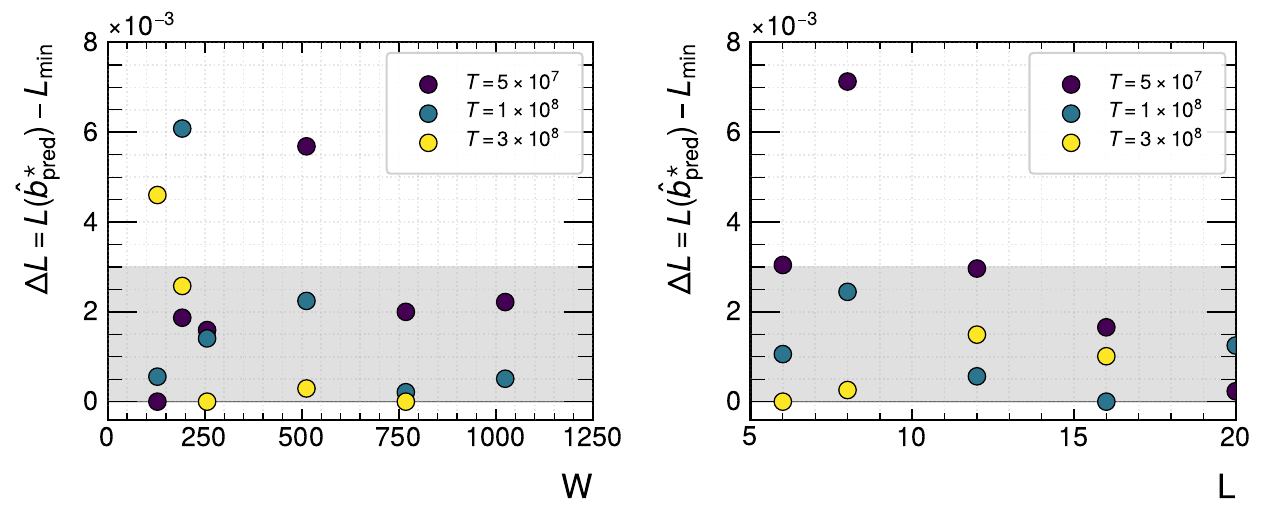}
        \caption{Loss penalty from the predicted $\bstar$.}
        \label{fig:transfer:b}
    \end{subfigure}
    \caption{Validation of joint-optimum transfer. \textbf{(a)} final validation loss against $\eta$ for $(b, T, w, l)$ combinations not used in the multiplicative fit. The star marks the predicted $\etastar$, which sits at or near every minimum, so $\etastar$ transfers across batch size, token budget, and model shape. \textbf{(b)} loss penalty $\Delta L = L(\bstar_{\mathrm{pred}}) - L_{\min}$ from using the predicted batch size rather than the lowest-loss $b$ in the sweep ($L_{\min}$), as a function of width (left sub-panel) and depth (right sub-panel) at three token budgets $T$. The gray band marks the run-to-run spread, showing that the possible residual $\bstar$ shift from the prediction can be neglected once $\etastar$ is tuned.}
    \label{fig:transfer}
\end{figure}

\cref{fig:transfer:eta} validates the $\etastar$ transfer on held-out $(b, T, w, l)$ combinations, where the predicted $\etastar$ lands at the bottom of each loss-versus-$\eta$ curve. The joint optimum $(\etastar, \bstar)$ transfers up to a possible shift $\bstar(N;\etastar)$. For jet classification the shift is trivially absent, the loss being flat in $b$ at $\etastar$. For the multi-task objective $\bstar$ grows as $\sqrt{T}$, leaving that shift as the only possible residual freedom as $N$ varies from the reference $\Nref$. \cref{fig:transfer:b} shows that using the predicted $\bstar$ instead of the lowest-loss $b$ in a sweep costs at most a few $\times 10^{-3}$ in loss, a per-mille-level penalty comparable to run-to-run noise. The joint optimum transfers in practice without needing a further $N$ dependence, possibly helped by the flatness of $\Lstar(\etastar, b, T)$ in $b$ around $\bstar$.

% =============================================================================
\section{A toy scaling-law tutorial}
\label{app:tutorial}
% =============================================================================

The companion tutorial\footnote{Code and step-by-step instructions: \url{https://github.com/mvigl/scaling-laws-demo}.} reproduces the three extraction methods of \cref{sec:scaling-laws} on a synthetic, laptop-scale problem. A fixed, randomly initialized MLP (the \emph{teacher}) defines the target; students of growing width are trained to imitate it on Gaussian inputs $x \sim \mathcal{N}(0, I)$ with label noise, $y = \mathrm{teacher}(x) + \sigma\varepsilon$. Data is drawn fresh at every step, so the run is single-pass and $D$ counts examples seen, with MLP compute $C = (6N - 2\,d\,w)\,D$ ($d$ the input dimension, $w$ the first hidden width). The loss follows the additive form $L \approx L_\infty + A/N^{\alpha} + B/D^{\beta}$, and because the task is synthetic the floor is known exactly, $L_\infty = \sigma^2$, the label noise no model can predict, which lets the fitted floor of Approach~3 be graded against the truth.

\begin{figure}[!htbp]
    \centering
    \includegraphics[width=0.6\textwidth]{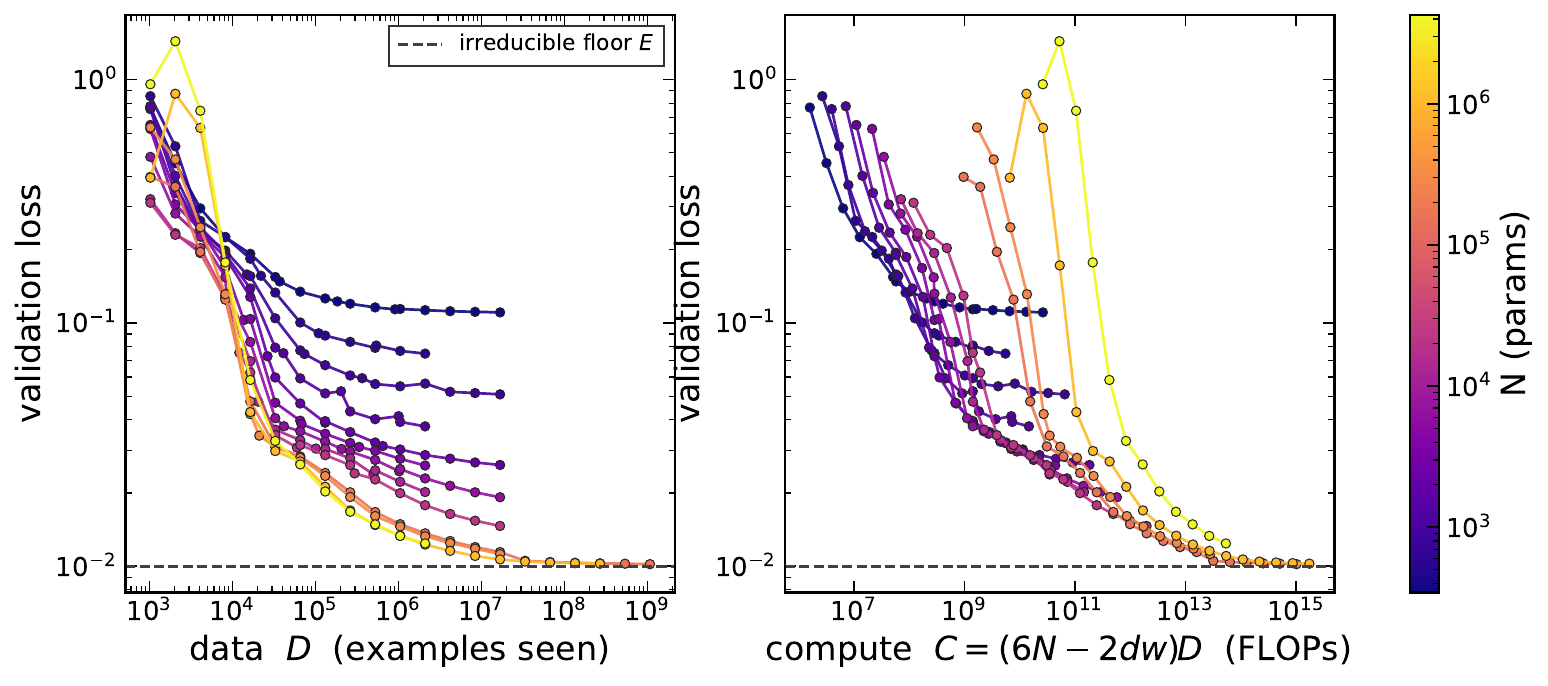}
    \caption{Tutorial scaling laws on the teacher--student toy problem: validation loss against data $D$ \textbf{(left)} and compute $C$ \textbf{(right)}, one curve per student size $N$. Smaller students win at low compute and larger ones take over, with the envelope approaching the known floor $L_\infty = \sigma^2 = 0.01$ (dotted).}
    \label{fig:scaling_demo_overview}
\end{figure}

A grid over $(N, D)$, each cell trained at its own tuned learning rate, recovers the compute-optimal frontier $L^\star(C)$ and allocation $(N^\star, D^\star)$ by all three approaches (\cref{fig:scaling_demo_overview}): all three give near-$\sqrt{C}$ scaling and broadly agree on the allocation.

\begin{figure}[!htbp]
    \centering
    \includegraphics[width=0.6\textwidth]{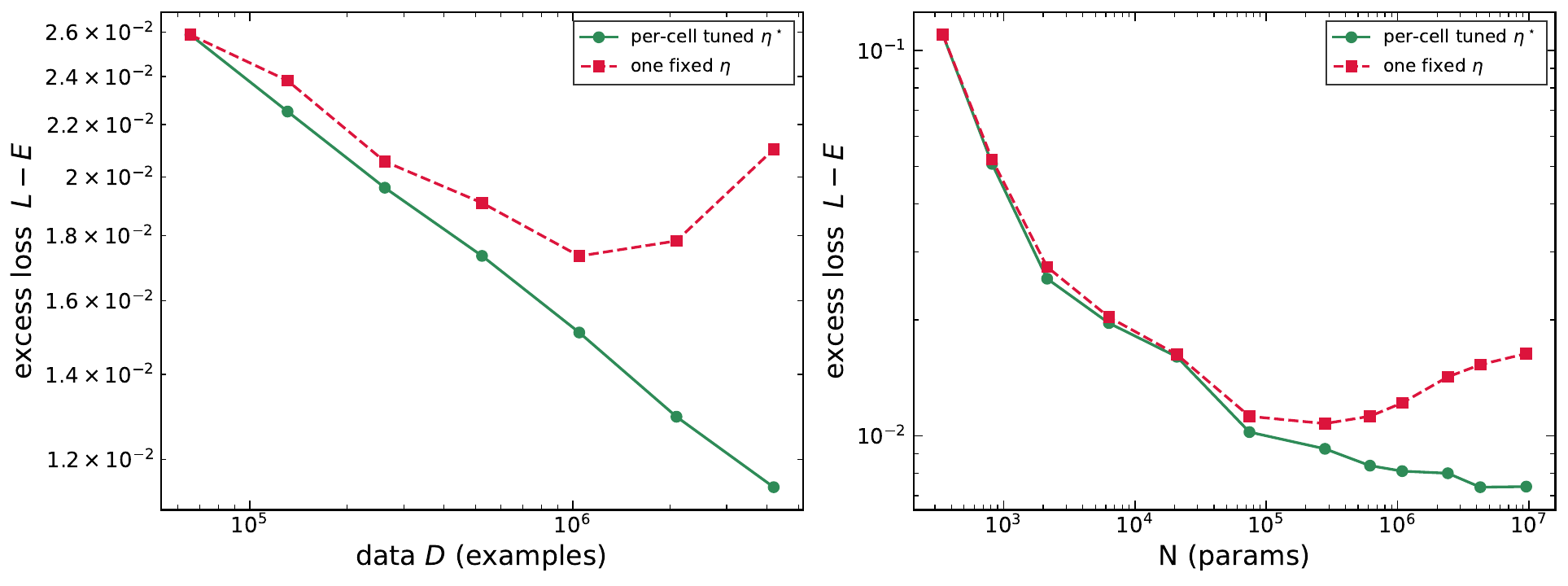}
    \caption{Cost of not re-tuning the learning rate on the teacher--student toy (\cref{app:tutorial}), whose floor $E$ is known exactly. Excess loss $L - E$ for the per-cell tuned $\etastar(w, T)$ (green) against a single learning rate tuned once at the small end and then held fixed (red), growing the data $D$ at fixed width \textbf{(left)} and the model size $N$ at fixed data \textbf{(right)}. The mis-tuned arm does not sit at a constant offset: it departs from the tuned power law and turns upward at scale, so a grid trained this way biases the fitted exponents and can hide the scaling entirely.}
    \label{fig:demo_lr_ablation}
\end{figure}

The tutorial also repeats the grid with the learning rate tuned once and held fixed. The resulting excess-loss comparison is shown in \cref{fig:demo_lr_ablation}.

\begin{figure}[!htbp]
    \centering
    \includegraphics[width=0.6\textwidth]{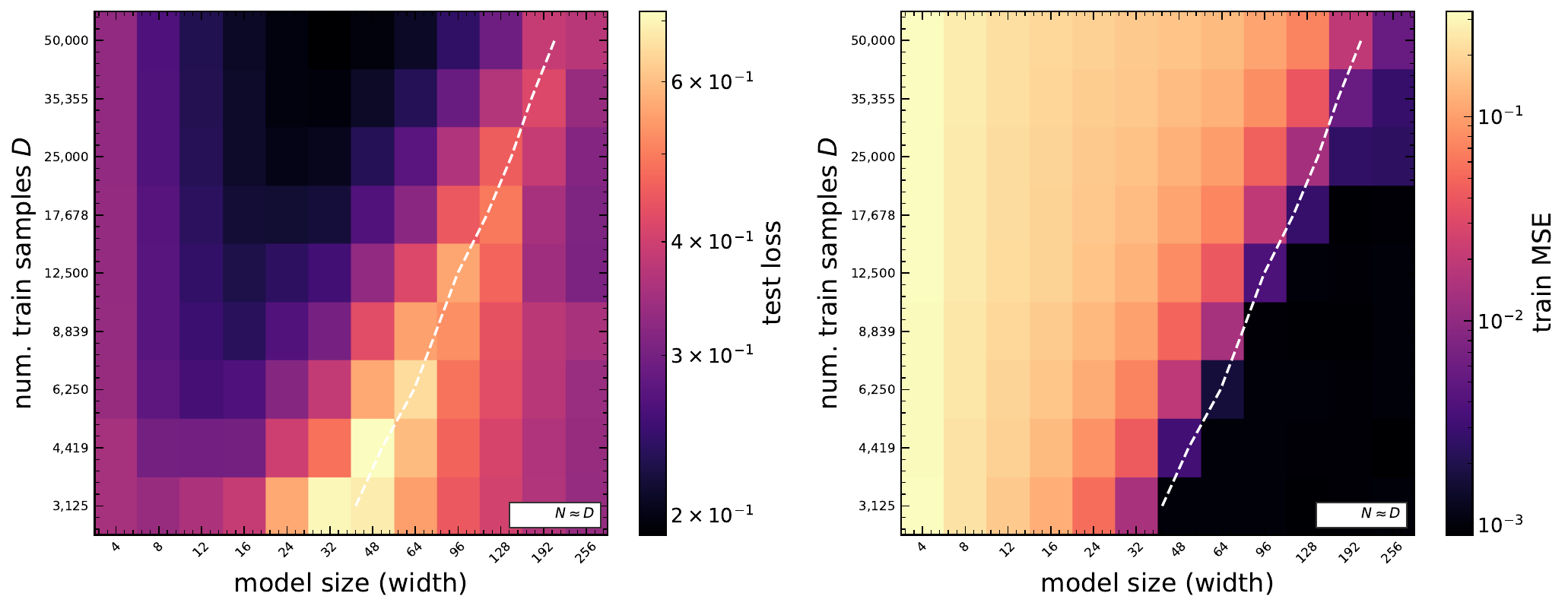}
    \caption{Double descent on the teacher--student toy problem: test loss \textbf{(left)} and train MSE \textbf{(right)} over model width and number of training samples $D$. The test-loss ridge follows the interpolation threshold $N \sim D$ (dashed), where the train MSE drops to zero; beyond it the test loss second-descends.}
    \label{fig:scaling_demo_double_descent}
\end{figure}

A second notebook switches to the repeated-data regime on the same teacher--student setup and reproduces the three double descent cases~\cite{nakkiran2019deepdoubledescentbigger}: sample-wise (a test-loss peak near $D \sim N$), model-wise (a peak at $N \sim D$), and epoch-wise. The same phenomenology was observed on particle-physics data~\cite{vigl2025doubledescentoverparameterizationparticle}. \cref{fig:scaling_demo_double_descent} shows the model-versus-data phase diagram, with the test-loss ridge tracking the interpolation threshold $N \sim D$.

% =============================================================================
\section{Compressed-representation scaling details}
\label{app:simple_scaling}
% =============================================================================

The compressed-representation study of \cref{sec:warmup} freezes a small set-transformer autoencoder ($d_{\mathrm{model}} = 256$, $\sim 8.3$M parameters) trained for one pass over $5$M JetClass jets~\cite{qu_2022_6619768}, and runs the scaling study on one-hidden-layer MLP probes over its latents for $10^8$ jets the encoder never saw, on a $10$-way boosted-jet classification target.

Each jet enters as a set of up to $40$ $\pt$-sorted constituents with $17$ features, which the encoder compresses into one latent vector of dimension $\ell$ and a transformer decoder expands
back into the constituents; no labels enter the objective. The objective is
\begin{equation}
    \mathcal{L} = \mathcal{L}_{\mathrm{cont}}
    + 0.2\,\mathcal{L}_{\mathrm{charge}}
    + 0.2\,\mathcal{L}_{\mathrm{PID}}
    + 0.5\,\mathcal{L}_{\mathrm{mult}} ,
\end{equation}
with $\mathcal{L}_{\mathrm{cont}}$ a masked Huber loss ($\delta = 1$) on the $11$ standardized
kinematic features, $\mathcal{L}_{\mathrm{charge}}$ and $\mathcal{L}_{\mathrm{PID}}$ masked
cross-entropies on the charge and on five particle-identification
classes, and $\mathcal{L}_{\mathrm{mult}}$ a cross-entropy over constituent-multiplicity bins
predicted from the latent directly. The continuous term carries the largest weight, so the latent is
shaped mostly by kinematic reconstruction, with the categorical heads acting as auxiliary
supervision. Training uses AdamW (weight decay $0.01$) with a one-cycle schedule (peak learning
rate $3\times10^{-4}$, $5\%$ warmup, single cosine anneal) at batch size $256$, with only the
latent dimension $\ell$ varied across the series.

\begin{figure}[!htbp]
    \centering
    \includegraphics[width=0.8\textwidth]{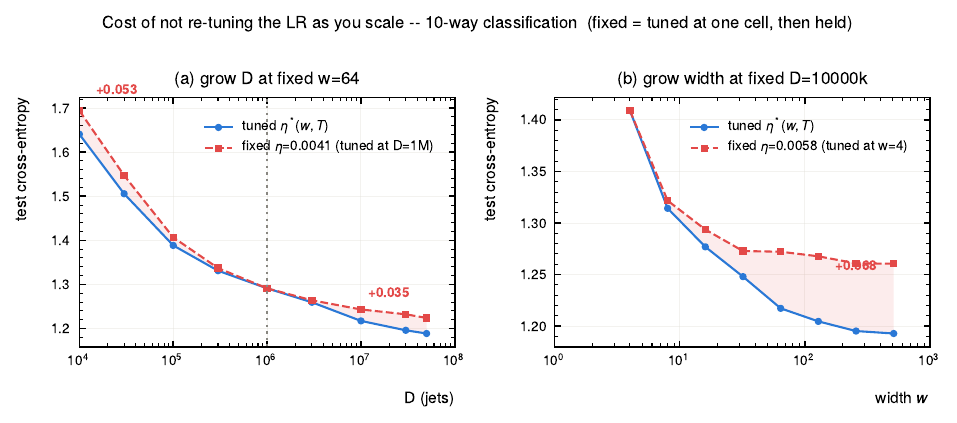}
    \caption{Cost of not re-tuning the learning rate on the compressed-representation system ($10$-way classification). \textbf{(a)} growing $D$ at fixed width $w = 64$: per-cell tuned $\etastar(w, T)$ (blue) against a learning rate tuned at $D = 1$M and then held fixed (red). \textbf{(b)} growing width at fixed $D$. A grid trained this way refits to visibly different scaling-law parameters.}
    \label{fig:simple_lr_ablation}
\end{figure}

\cref{fig:simple_lr_ablation} isolates the cost of not re-tuning the learning rate of the MLP probes when scaling. Against the per-cell tuned baseline, a learning rate tuned once at a single cell and then held fixed leaves a loss gap that grows toward both the large-$D$ and large-$w$ ends of the grid. Refitting the $L(N, D)$ surface on the fixed-LR grid returns a different law altogether, with a floor estimate above the best measured cell, so the mis-tuned grid misreports both the exponents and where scaling ends.

% =============================================================================

\end{document}